\documentclass[prb,onecolumn,superscriptaddress]{revtex4}
\usepackage{amsmath,amssymb,graphicx,epsfig,color, float}
\usepackage{epstopdf}

\newcommand \beq{\begin{eqnarray}}
\newcommand \eeq{\end{eqnarray}}

\newcommand{\bfr}{\boldsymbol{r}}
\newcommand{\bfk}{\boldsymbol{k}}

\newcommand{\bfq}{\boldsymbol{q}}

\newcommand{\im}{\mathrm{Im}}
\newcommand{\sgn}{\mathrm{sgn}}

\begin{document}
\title{Spin Diffusion Equations for Magnetized or  Orbital Polarized Systems }

\author{Vincent Sacksteder IV}
\email{vincent@sacksteder.com}
\affiliation{ Department of Physics, Royal Holloway University of London, Egham, United Kingdom}

\date{\today}

\begin{abstract}
Charge and spin transport in spintronics devices can be described by a spin diffusion equation suitable for modelling scales much larger than the scattering and atomic scales.  This work concerns the coarse graining procedure used to compute the coefficients of the diffusion equation, which are sensitive the details of individual atoms and impurities.  We show with two simple examples that in  spintronics devices which have both a spin-orbit interaction and magnetization, standard coarse graining  can easily obtain diffusion equations which fail to conserve electronic charge.  The same failure can occur in systems with  both a spin-orbit interaction and orbital polarization.  We show that linear response theory, coupled with the self-consistent Born approximation and ladder diagrams, offers an improved way of calculating diffusion equations.  We show that the resulting equations satisfy  a Ward-Takahashi identity that guarantees charge conservation.

\end{abstract}


\maketitle

\section{Introduction}
Note: This paper is in a draft status, does not mention all relevant citations, and may still contain errors.  We expect that any errors are relatively minor and do not affect the main message or results, and that  all the major citations are already included. 

\subsection{What we want to model; the reason for this work}
The immediate stimulus for this work was the recent fabrication of magnetic memory elements in which the magnetization switching is caused by an electric current, not a magnetic field.  (See Miron et al, Nature 476, 189-193, and Buhrman et al, Science 336, 555-558.) \cite{miron2011perpendicular,liu2012spin} These devices are bilayers, with a magnetic layer on top of metallic layer with a strong spin-orbit coupling.  For instance,  Miron's device was a Cobalt-Platinum bilayer.   When a electric current flows through these devices, the spin-orbit coupling causes spin currents and accumulations, and switches in the magnetization are observed.  These devices offer possibilities both for new memory technologies and for integrating memory elements   into CPUs.   We emphasize that, unlike some other earlier spintronics devices, spin-orbit coupling and its effect on transport are key to the new devices.

Modeling these memory devices requires a microscopic description of the  atoms in each of the layers, the spin-orbit coupling in the metallic layer, the interface between the metallic and ferromagnetic layers, and the magnetization in the magnetic layer.   Impurity atoms, scattering, and mismatches between the layers are likely to be important.  Along with this atomic scale physics, the modeling must be able to capture the much larger scale of  the entire memory device - and the first devices were micron sized. 

Similar modeling  capabilities are required by many other candidates for new spintronics technologies: antiferromagnetic memories (Science v. 351 p. 587, 2016 \cite{wadley2016electrical}), graphene with adsorbed impurities  producing a very large spin Hall effect (Nature Communications 5, 4748 \cite{balakrishnan2014giant}), magnetic domain wall motion in racetrack memories, and devices involving the movement of magnetic skyrmions.

The results presented here concern the specific technical challenge of modeling devices which include both spin-orbit couplings and one of the following two properties: magnetization and/or orbital polarization.

\subsection{Coarse Graining with Spin Diffusion Equations}
Several modeling techniques have been developed to meet these modeling requirements.  One approach is to build a model of the device which explicitly includes all the atoms in the device, and then to calculate its behavior using the Schodinger equation.  This approach always involves a limit on the size of the device that can be simulated, but device sizes in excess of a billion atoms have been achieved using tight binding Hamiltonian representations of each atom - see for instance Ferreira's KITE software (https://quantum-kite.com/resources/).   Important work in the same direction has been done using ab initio (DFT) software, although the computational expense significantly reduces the system sizes which can be achieved.  Some of the leaders in this direction, i.e. some of the people who have often focused specifically on ab initio calculation of spintronics devices and spin transport and dynamics, have been Paul Kelly, Ingrid Mertig, Stefan Blugel, the Prague research group of Kudrnovsky, Drchal, and Sob, and others.

Our work presented here  concerns a different strategy for modeling spintronics devices, which is often called spin diffusion equations.  Broadly speaking, this is a coarse-graining strategy.  Starting with the microscopic description of the atoms within a sample, we determine a diffusion equation which controls charge and spin transport  at distances much longer than the scattering length.  The diffusion equation does describe the influence of atomic-scale processes on  long-distance physics, but it does not include a description of transport at the atomic scale.  In a model where each atom has only a spin up and a spin down state, and no additional states associated with different atomic orbitals, diffusion equations describe the evolution of four coarse grained variables: the charge density $\rho$ and the spin densities $S_x,\,S_y,\,S_z$.  These are often written as a charge-spin density four-vector $\vec{\rho} =  \left[\rho,\,S_x,\,S_y,\,S_z\right]^T$.  In models with two or more atoms, or with several orbitals per atom, the charge-spin density vector has $4 N^2$ components, where $N$ is the number of atomic orbitals per unit cell.  The diffusion equation is then written as
\begin{align}
\tau \frac{\partial}{\partial t} \vec{\rho} = D \vec{\rho}
\end{align}
The diffusion operator $D$ is a $4N^2 \times 4N^2$ operator, and is derived from the microscopic description of the model's atoms and the disorder.  It includes the physics of magnetic fields, spin precession, spin-charge conversion, and spin relaxation in both the Elliot-Yafet and the Dyakonov-Perel regimes.  $\tau$ is the momentum relaxation time caused by scattering.  Since this is a first-order partial differential equation, the initial value of $\vec{\rho}$ must be determined, and also geometry and boundary conditions of the sample, and then the spin diffusion equation can be solved straightforwardly.

Spin diffusion equations have four principal advantages: (1) For simple models the spin diffusion equation can be derived analytically, (2) the coefficients of the spin diffusion equation invite theoretical interpretation and analysis, (3) solution of the coarse-grained spin diffusion equation requires much fewer computational resources and therefore allows calculation of samples with realistic sizes, and (4)  the coarse-grained charge and spin  flow determined by a spin diffusion equation is  easier to examine and comprehend than results describing spin and charge at each atomic site.  There are also some disadvantages: (1)  the spin diffusion equation approach involves approximations which typically include omitting certain quantum interference processes and also many-body electronic correlations, and  (2)  the process of starting from a microscopic description of a sample's atoms and then deriving the correct spin diffusion equation requires the use of perturbation theory and  Feynman diagrams, and therefore requires considerable care and expertise.   The correct derivation of spin diffusion equations is the subject of ongoing research and investigation.  In particular the new results presented here are about how to derive correct spin diffusion equations in materials which include both spin-orbit coupling and one of two following effects: magnetization and/or orbital polarization.

\subsection{The Standard Techniques for Calculating the Spin Diffusion Equation}
There are two well-known procedures for calculating the diffusion operator $D$: 
\begin{itemize}
\item The Burkov - Nunez - Macdonald approach (PRB 70 155308 \cite{PhysRevB.70.155308}), which is motivated by linear response theory.
\item The Mishchenko - Shytov - Halperin  approach (PRL 93 226602 \cite{PhysRevLett.93.226602}), which is noteworthy for starting within a Keldysh formalism.  
\end{itemize}
While these two techniques have some significant differences, they have many points in common in their philosophy, development, approximations, and actual computational steps.  Both approaches obtain identical results for   some simple 2-D models with rotational in-plane symmetry, as was demonstrated already in the original publications.  

The work presented in the present article was stimulated by the fact that both approaches (as originally presented) treat the momentum relaxation time $\tau$ as a single parameter.  This is incorrect in systems with magnetization or orbital polarization, where different orbital or spin states can display different momentum relaxation times.  Moreover, often $\tau$ is treated as an input parameter, without calculation from the microscopic quantities that are responsible for scattering.   We will show that in materials with spin-orbit coupling and with also either magnetization or orbital polarization, this simplified treatment of $\tau$ is incorrect and produces incorrect results; $\tau$ must be calculated from the microscopic data instead of simply being chosen as an input paremeter.  In our view neither the Burkov - Nunez - Macdonald approach nor the  Mishchenko - Shytov - Halperin  approach correctly handle the particular case of systems with both  spin-orbit coupling and with also either magnetization or orbital polarization.   We are not aware of work  generalizing either of the two methods to handle this case.  (In particular Wang and Manchon's PRL 108 117201 on CoPt bilayers \cite{PhysRevLett.108.117201} retained the assumption that $\Sigma$ is proportional to the identity.)

In the remainder of this article we will focus on the Burkov - Nunez - Macdonald approach (PRB 70 155308 \cite{PhysRevB.70.155308}), which we will call "the BNM approach", although admittedly the literature is fairly evenly split between this approach and the Mishchenko - Shytov - Halperin approach \cite{PhysRevLett.93.226602}. Since both approaches share so much in common, and since both display the same methodological failure when handling magnetized and/or orbital-polarized systems, our critique of and improvement on the BNM approach reflects also on the Mishchenko - Shytov - Halperin approach.  In the linear response regime the Keldysh formalism, while more powerful for treating interacting systems out of equilibrium, cannot fail to give the same results as linear response theory, if both are calculated exactly. Therefore the only possible source of differences between the two approaches is from the approximations which each employ.

Here we briefly review the equations of the Burkov - Nunez - Macdonald approach , which determine both the self-energy $\Sigma$ and the diffusion operator $D$.  We conflate the spin and orbital degrees of freedom into a single index which we denote by greek letters.  We use the convention that the Green's function is $ G^{A,R}_{\alpha \beta} = ( E - H \mp \Sigma^{A,R})^{-1} $, where $\Sigma^{A,R}$ is the self-energy.  
The $A,R$ superscript notation selects the advanced or retarded Green's function $G$ and corresponding self-energy $\Sigma$.  $\Sigma$ is subtracted for the advanced Green's function and added for the retarded Green's function.
The weak Gaussian disorder in the potential $V$ is described by its second moment  $\Gamma_{\alpha \gamma \delta \beta} \propto \langle V^2 \rangle$, which has four indices because it is the second moment of $V$, which like $H$ has two indices.  The self-consistent Born approximation (SCBA) which we use to approximate the self-energy is:
\begin{align}
\Sigma^R_{\alpha \beta}(E) &= \sum_{\gamma \delta} \Gamma_{\alpha \gamma \delta \beta} \int{d\bfk} G^R_{\gamma \delta} (\bfk, E) 
\end{align}
Often this integral is approximated by keeping only the imaginary part of the Green's function and treating it as a delta function confining the integration ${d\bfk}$ to the Fermi surface.

Both Burkov Nunez and Macdonald \cite{PhysRevB.70.155308} and Mishchenko  Shytov and  Halperin \cite{PhysRevLett.93.226602} assume that $\Sigma_{\alpha \beta}$ is imaginary and proportional to the identity $1$.    Since the imaginary part of $ \pi^{-1} \int{d\bfk} G_{\gamma \delta} (\bfk, E) $ is the single particle density of states, if the disorder has trivial spin structure i.e. $\Gamma \propto 1$, then the assumption that $\Sigma_{\alpha \beta} \propto 1$ is equivalent to assuming that the system has no spin or orbital polarization.   Within this assumption, the self-energy is commonly written as 
\begin{align}
\Sigma = - \imath \hbar / 2 \tau
\end{align}
where $\tau $ is the scattering time.

The SCBA equation is a self-consistent equation, since the Green's function $G$ depends on the self-energy $\Sigma$.  Solving this nonlinear equation requires iterative recalculation of $\Sigma$ from $G$, and $G$ from $\Sigma$, and as a rule must be done numerically rather than analytically.  Therefore theoretical papers often avoid the numerical work and instead use 
\begin{align}
 \hbar / 2 \pi \tau &= \sum_{\gamma \delta} \Gamma_{\alpha \gamma \delta \beta} \;\rho_{\gamma \delta}
\end{align}
It can be tempting to further simplify things by using the bare Green's function $G(\Sigma = 0)$ to evaluate $\rho$, thus avoiding the iterative self-consistency prescribed by the SCBA.  Alternatively, while in principle remembering that $\rho$, $\Sigma$, and $1/\tau$ include self-consistency effects and all powers of $\Gamma$, some authors do not go through the steps of requiring numerical self-consistency, and leave these parameters as basically external parameters.  In either case, the effect is to cut the connection between (A) the scattering processes that determine $\Sigma$ and the single-particle Green's function and (B) the scattering processes that determine the diffusion operator $D$.  When this connection is cut the resulting diffusion operator will fail to conserve charge if $\Sigma$ is not proportional to the identity, as we will show in the next sections.  The only solution is to insist both analytically and numerically that $\Sigma$ be the real numerical solution of the SCBA for the specific values of $\Gamma$ and $G_0$ realized in the system one is modelling.

Burkov Nunez and Macdonald calculated a collision integral $I$ which represents one scattering event where both an electron and a hole scatter off the same impurity:
\begin{align}
I_{\gamma \zeta \lambda \delta}(\bfq, E, \omega) &= \sum_{\alpha \beta}   {\Gamma}_{\gamma \alpha \beta \delta}   \overline{\Phi}_{\alpha \zeta \lambda \beta} (\bfq, \omega, E) \\
\overline{\Phi}_{\alpha \nu \eta \beta} ( \bfq, \omega, E) & = \int {d\bfk} \Phi_{\alpha \nu \eta \beta} (\bfk, \bfq, \omega, E) \\
\Phi_{\alpha \nu \eta \beta}  (\bfk, \bfq, \omega, E) & = G^A_{\alpha \nu }(\bfk + \bfq/2,  E + \omega) G^R_{ \eta \beta}(\bfk-\bfq/2, E)  
\end{align}
Similarly to the integral determining the self-energy,  the collision integral is often approximated by keeping only the imaginary part of one of the Green's functions and treating it as a delta function confining the integration ${d\bfk}$ to the Fermi surface.  

Burkov Nunez and Macdonald determine the diffusion operator $D$ directly from the collision integral:
\begin{align}
\tau \partial_t \vec{\rho}&= D \vec{\rho} \\
\vec{\rho}(t) &= \exp(D t / \tau) \, \vec{\rho}(0) \\
D_{\alpha \gamma \delta \beta}(\bfq, \omega, E) & =        (1- I(\bfq, E, \omega))_{\alpha \gamma \delta \beta}    
\end{align}
Actually one should expand $D$  to second order in  momentum $\bfq$ and zeroth order in energy $\omega$ in order to obtain the diffusion operator.  If one does not do so then it has a non-trivial time dependence and it contains all powers of the spatial derivatives.  There isn't much point in going to higher orders because the main benefit of going to higher order would be to investigate small length scales, but information about small length scales is removed and corrupted by BNM's ladder approximation.  Since this information is absent, attempts to recover it with higher derivatives produce results without any physical meaning.

The evolution operator $\exp(D t / \tau)$  contains all powers of the collision integral $I$, representing repeated scatterings off of impurities.  In perturbation theory the collision integral $I$ looks like one rung of a ladder, so powers of $I$ are called ladder diagrams, and the Burkov Nunez and Macdonald formula  is a ladder approximation. 

 Some recent papers by Ado Dmietriev Ostrovsky and Titov \cite{ado2015anomalous} and by Milletari and Ferreira \cite{PhysRevB.94.134202} have shown that even though the ladder approximation calculates the longitudinal conductivity correctly, it is not sufficient for getting the transverse conductivity and spin Hall physics.  The ladder approximation will give the correct order of magnitude for the Hall coefficient but can get the sign wrong, as well as the numerical coefficient.  The solution to this problem is to renormalize the scattering tensor $\Gamma$ with  a one loop integral. 

\subsection{Failure in Systems which are Magnetized or have Orbital Polarization}
If the self-energy is a proportional to the identity, i.e. $\Sigma = \imath \hbar / 2 \tau$, then the BNM approach produces a diffusion operator $D$ which conserves the electronic charge component of $\vec{\rho}$.  We will prove this statement  in detail later in this article.  The proof relies on $\Sigma$ commuting with the bare Hamiltonian $H_0$ inside the Green's function, allowing $\tau$ it to be pulled outside of the integral over ${d\bfk} $.    A similar trick can be played inside the momentum integral determining $\Sigma$.  Using these tricks, one can show that the collision integral $I$ is independent of $\tau$ and therefore is able to cancel exactly against the $1$ in $D = 1-I$ independently of the value of $\tau$.  The cancellation provides an exact zero eigenvalue in $D$  which means that the inverse of the charge lifetime is zero; i.e.  charge is conserved.  Thus the commutation between $\Sigma$ and $H_0$ provides an elementary mathematical mechanism for guaranteeing charge conservation.

On the other hand, when $\Sigma$ is not proportional to identity, it in general does not commute with $H_0$, and its role in the collision integral $I$ and in $\Sigma$ becomes non-trivial.   In this case the mechanism discussed above for guaranteeing charge conservation does not apply, and deeper physics/math must be used to obtain charge conservation.  If a calculation of the diffusion operator $D$ omits or truncates the necessary physics, then the resulting $D$ will not conserve charge.  We will show two very simple magnetized models which, when one uses the BNM approach and retains the fact that $\Sigma$ is not proportional to the identity, produce $D$'s which exhibit charge non-conservation.  When the spin splittings in $H_0$ and in $\Sigma$ are comparable, charge is lost at a rate which is similar to the scattering rate $\tau^{-1}$.  In this case the diffusion operator completely loses its validity.

\subsection{The Correct Formalism}
The solution to this difficulty is to ground the BNM approach more firmly in  linear response theory.  Rather than attempting to provide a general evolution equation describing the behavior of an arbitrary charge and spin distribution, we restrict our focus to a system's response to 
 small external perturbations $V$.  Linear response theory allows one to compute  $\rho$, the change in charge and spin density induced by $V$: 
\begin{align}
((\hat{ \chi})^{-1})_{\alpha \gamma \delta \beta} \; \rho_{\gamma \delta} &= V_{\alpha \beta} \\
\end{align}
Here  we are using the Einstein convention where repeated indices have implicit sums.  
 The key machinery in this equation is the susceptibility  $\hat{ \chi}$.
The inverse susceptibility $(\hat{ \chi})^{-1}$, if calculated exactly, would contain a complete description of the linear response at both short and long scales, including ballistic physics, diffusive motion, and Anderson localization.   One can obtain from $(\hat{\chi})^{-1}$ the diffusion operator $D$ which  governs long time and distance scales  by expanding  to second order in  momentum $\bfq$ and first order in energy $\omega$.  

In the ladder approximation, $(\hat{\chi})^{-1}$ has a simple relation to the diffusion operator $D = 1 - I $ calculated in the BNM approach:
\begin{align}
\hat{ \chi}_{\alpha \gamma \delta \beta}(\bfq, \omega, E) & = \sum_{\nu \eta}      \overline{\Phi}_{\alpha \nu \eta \beta}(\bfq, \omega, E)   ((1- I(\bfq, E, \omega))^{-1})_{\nu \gamma \delta \eta}    
\end{align}
All of the details of calculating the collision integral $I$, $\overline{\Phi}$, and $\Sigma$ are the same as in the BNM approach.  Charge conservation will not be recovered, however, unless the self-energy $\Sigma$ is the self-consistent solution of the SCBA equation.

The advantage of following linear response theory and calculating the susceptibility $\hat{\chi}_0$ is that a powerful Ward-Takahashi identity guarantees charge conservation.  (todo: advanced-retarded, and check the explanation of $\Lambda$) The Ward-Takahashi identity reads:
\begin{align}
\Sigma_{\nu \eta}(E + \omega) - \Sigma_{\nu \eta}(E) &= 
\imath \omega \sum_{\gamma} \int{d\bfk} \Lambda_{\nu \gamma \gamma \eta}(\bfk, \bfq=0, \omega, E) 
\end{align}
Here we are using Matsubara frequencies $\omega$ which are imaginary.  Upon analytic continuation to the real axis the left hand side of the identity becomes $\Sigma^A_{\nu \eta}(E + \omega) - \Sigma^R_{\nu \eta}(E) $, which is in general non-zero when $\omega \rightarrow 0$.
$\Lambda$ is the interacting vertex function.  It can be obtained from susceptibility $\chi$ by pruning two Green's functions and then substracting $1$,  i.e. roughly $\chi = G^A G^R + G^A G^R \Lambda$.  The $\omega$ multiplying the right-hand side of the Ward-Takahashi identity implies that at small $\omega$ (i.e. long times) $\Lambda$ and $\hat{\chi}$ both diverge like $\omega^{-1}$.  This in turn implies that $(\hat{\chi})^{-1}$ has an eigenvalue which is zero when $\omega=0$.  It is this eigenvalue which ensures that both $(\hat{\chi})^{-1}$ and the diffusion operator $D$ which can be obtained from it by using a Taylor series expansion conserve charge.


The Ward-Takahashi identity was proved in Ramazashvili's PRB 66 220503(R), following similar earlier proofs \cite{PhysRevB.66.220503}.  In this present work we generalize that proof to systems with magnetization or orbital polarization by keeping track of the spin/orbital indices at every step in the proof.  The proof is fully rigorous and non-perturbative, requiring only that global charge be conserved.  The price of this rigor is that one must restrict one's scope to linear response theory, i.e. charge and spin response to small perturbing potentials.

Although the Ward-Takahashi identity guarantees that a system's exact inverse susceptibility $(\hat{\chi})^{-1}$ conserves charge, it does not make guarantees about approximations to the susceptibility, including the ladder approximation $\hat{ \chi}_{\alpha \gamma \delta \beta}  = \sum_{\nu \eta}      \overline{\Phi}_{\alpha \nu \eta \beta}  ((1- I)^{-1})_{\nu \gamma \delta \eta} $ which we have just given.  Therefore the present work presents several checks that the ladder approximation, in combination with a full self-consistent SCBA calculation of the self-energy, does obey the Ward-Takahashi identity. Within the ladder approximation to $\hat{\chi}$, the Ward-Takahashi identity is
\begin{align}
\Sigma_{\nu \eta}(E + \omega) - \Sigma_{\nu \eta}(E) &= 
\imath \omega \sum_{\gamma \alpha \beta}  \hat{\Gamma}_{\nu \alpha \beta \eta}(\bfq=0, \omega, E) \overline{\Phi}_{\alpha \gamma \gamma \beta}(\bfq=0, \omega, E)  \\
\hat{\Gamma}_{\nu \zeta \lambda \eta}  (\bfq, \omega, E) &= \sum_{\gamma \delta} ((1- I(\bfq, E, \omega))^{-1})_{\nu \gamma \delta \eta}  {\Gamma}_{\gamma \zeta \lambda \delta} \
\end{align}

\subsection{What can we gain?}

\begin{itemize}
\item Most importantly, we gain a  working coarse-grained theory of spin transport that can handle modern spintronics devices  much larger than the atomic scale.
\item Use of the SCBA to perform a correct calculation of the self-energy results in a  renormalization of the exchange field $b_z$ and the total magnetization. Moreover the Fermi velocity, which figures in the diffusion constants, spin-spin couplings, and charge-spin couplings, is renormalized.
\item Comparing to  Burkov Nunez and Macdonald's diffusion operator $D = 1- \Gamma \overline{\Phi}$, the new diffusion operator  is  $(1- \Gamma \overline{\Phi}) \overline{\Phi}^{-1}$.  In the regime Dyakanov-Perel spin relaxation, where the spin relaxation rates $ \propto (1- \Gamma \overline{\Phi}) \tau^{-1} \ll \tau^{-1}$, the new matrix $\overline{\Phi}^{-1}$ can safely be truncated at zeroth order in $\omega$ and $\bfq$.  Therefore its only effect is to rotate the Burkov Nunez Macdonald kernel $(1- \Gamma \overline{\Phi})$, causing the spin-charge density $\vec{\rho}$ to be rotated.  In the Elliot-Yafet regime this approximation is not valid and the new matrix must retain its $\omega$ and $\bfq$ dependence, causing  changes in the spin diffusion operator (compared to the BNM operator) that go well beyond rotations of $\vec{\rho}$.
\item The new introduced matrix $\overline{\Phi}^{-1}$ may be nontrivial even when there is no magnetization.  Where this is so, past published results on   spin diffusion equations in  the Elliot-Yafet regime will need revision.
\item Where the BNM approach uses a simple time derivative $\tau \frac{\partial}{\partial t} $, the correct treatment requires expanding $(\hat{\chi})^{-1}$ to first order in $\omega$, and  $\frac{d(\hat{\chi})^{-1}}{d\omega}$ multiplies the time derivative, replacing $\tau$.  Therefore the diffusion equation's time derivative acquires a matrix structure which can include physics well beyond that of $\tau$.   This non-trivial matrix structure of the time derivative can occur in   non-magnetized systems, and if so would force revision of previous literature.
\end{itemize}

\subsection{Outline of this Paper}

The remainder of this work is organized as follows:
\begin{itemize}
\item Section II shows that the BNM approach produces diffusion operators which do not conserve charge when applied to two simple models: (a) a minimal model with magnetization and spin-orbit disorder, and (b)  a model of CoPt which can also be applied to magnetized TIs.  
\item Section III contains (a)  a general proof of the  Ward-Takahashi identity, (b) derivation of the transport equation written  in terms of $\hat{\chi}$, both exactly and when using the ladder diagram approximaiton, and (c) notes on how to perform the analytic continuation from Matsubara frequencies to real frequencies and energies.  Much of the analytical work in this paper, and in particular the proof of the Ward-Takahashi identity, is done using Matsubara frequencies.
\item Section IV verifies that the Ward-Takahashi identity is obeyed when using the SCBA plus ladder diagrams. It does so (a) in any model to first and second order in $\Gamma$, (b) to all orders when $\Sigma$ is proportional to the identity, (c) to all orders in a minimal model with magnetization and spin-orbit disorder - the same one which we showed does not conserve charge in the BNM approach, and (d) to second order in $\Gamma$ a model of CoPt which can also be applied to magnetized TIs  - the same one which we showed does not conserve charge in the BNM approach.   The derivations to second order can be extended to all orders by following the logic of Vollhardt and Wolfle's PRB 22 4666 \cite{PhysRevB.22.4666}.
\item Section V has notes toward calculating the coefficients of the diffusion operator for a simple spin $1/2$ model with both magnetization and a spin-orbit interaction.  This model has been used to analyze CoPt bilayers, and is also applicable to magnetized TIs.  The work reported in this section is analytical, and is incomplete because it remains to solve the SCBA self-consistent equation numerically and then evaluate all the spin diffusion coefficients numerically.
\item  Section VI has some of our notes on the Bethe-Salpeter equation, Quantum Kinetic Equation, and charge conservation in this context.  We worked on this topic a lot because of its connection with Mischenko Shytov and Halperin's Keldysh approach to spin diffusion.
\end{itemize}

\section{Two Models Which Exhibit Charge Non-Conservation}
In this section we will present two simple spin 1/2 models, and show that when  treated using the BNM approach they produce spin diffusion equations which do not conserve charge.

\subsection{Minimal Model: Magnetization Plus Spin-Orbit Disorder}
This minimal model has two orbitals (spin up and spin down),  a $\hat{z}$ Zeeman splitting between its energy levels which causes in-plane precession of the spin, a minimal dependence on $\vec{k}$ i.e. no spin-orbit coupling,  a density of states which is different for spin up than for spin down (the asymmetry is controlled by $n_z$, the spin-$z$ component of the density of states), and both  a spin-conserving disorder $\Gamma_0$ and a spin-orbit disorder $\Gamma_x \sigma_x \otimes \sigma_x$. $\Gamma_0$ is not necessary for obtaining the charge non conservation.  What is necessary to produce this pathology is the presence of two non-commuting operators, the spin-orbit disorder $\Gamma_x \sigma_x \otimes \sigma_x$ and the self-energy which has a component proportional to $\sigma_z$.

We present the Hamiltonian, its eigenvalues, and the density of states $n$ (without disorder).  $H$ is the Hamiltonian, $E$ is its eigenvalues, $s=\pm 1$ chooses among two eigenvalues,  $n$ is the density of states, $\Gamma$ is the scattering vertex,  $\Sigma$ is the self-energy, and $G^R$ is the retarded Green's function.
\begin{eqnarray}
H  & = & H_0(k^2) + E_z \sigma_z, \; E(\vec{k}, s)= H_0(k^2) + s E_z,
\nonumber \\
n &=& -\imath \pi^{-1} \int {d\vec{k}} G^R(\vec{k}) =  n_0  + n_z \sigma_z 
\nonumber \\
\Gamma & = & \Gamma_1 1 \otimes 1 - \Gamma_x \sigma_x \otimes \sigma_x
\nonumber \\
\Sigma &=&-  \imath \Gamma \, \int {d\vec{k}} G^R(\vec{k}) = \pi \Gamma n = \Sigma_1 + \sigma_z \Sigma_z, \; \Sigma_1 = \pi n_1 (\Gamma_1 + \Gamma_x), \; \Sigma_z = \pi n_z (\Gamma_1 - \Gamma_x)
\nonumber \\
G^R(\vec{k},s) &=& \sum_s \frac{1 + s \sigma_z}{2 (E_F - E(\vec{k},s) - \imath \Sigma_1 - \imath s\Sigma_z )}
\end{eqnarray}

Note that we have calculated the self-energy in a way that is often used in the literature, i.e. without making sure that it is the self-consistent solution of the SCBA equation.  We  now  follow the BNM approach for calculating the diffusion operator $D^{-1}$ at zero momentum and zero frequency.  This object controls decay and mixing of spin and charge.
\begin{eqnarray}
(D^{-1}(\vec{q}=0,\omega=0))_{ij} &=& 1- I = \delta_{ij} - \frac{1}{2} \int {d\vec{k}} \; {Tr}(\Gamma \;G^A \sigma_i \;  G^R \sigma_j) 
\nonumber \\
 &=& \delta_{ij} - \frac{1}{8} \sum_{s, \acute{s}} \int {d\vec{k}} \{ \Gamma_1 {Tr}(\sigma_i (1 + s \sigma_z) \sigma_j (1 + \acute{s} \sigma_z)) - t \Gamma_x {Tr}(\sigma_i (1 + s \sigma_z) \sigma_x \sigma_j (1 + \acute{s} \sigma_z) \sigma_x) \} 
 \nonumber \\
 & \times & \{ (E_F - E(\vec{k},s) - \imath \Sigma_1 - \imath s\Sigma_z )^{-1} - (E_F - E(\vec{k},\acute{s}) + \imath \Sigma_1+ \imath \acute{s}\Sigma_z )^{-1} \} 
 \nonumber \\
 & \times&  (-(\acute{s} - s)E_z +\imath 2\Sigma_1 + \imath (s + \acute{s})\Sigma_z )^{-1}
\nonumber \\
 &=& \delta_{ij} - \frac{1}{8} \sum_{s, \acute{s}} \int {d\vec{k}} \{ \Gamma_1 {Tr}(\sigma_i (1 + s \sigma_z) \sigma_j (1 + \acute{s} \sigma_z)) + t \Gamma_x {Tr}(\sigma_i (1 + s \sigma_z) \sigma_x \sigma_j (1 + \acute{s} \sigma_z) \sigma_x) \} 
 \nonumber \\
 & \times & \{ \imath \pi \delta(E_F - E(\vec{k},s) - \imath \Sigma_1 - \imath s\Sigma_z ) + \imath \pi \delta(E_F - E(\vec{k},\acute{s}) + \imath \Sigma_1 + \imath \acute{s}\Sigma_z ) \} 
 \nonumber \\
 & \times&  (-(\acute{s} - s)E_z +\imath 2 \Sigma_1 + \imath (s + \acute{s})\Sigma_z )^{-1}
\nonumber \\
(D^{-1}(\vec{q}=0,\omega=0))_{ij} &=& \delta_{ij} - \frac{1}{8} \sum_{s, \acute{s}} \{ \Gamma_1 {Tr}(\sigma_i (1 + s \sigma_z) \sigma_j (1 + \acute{s} \sigma_z)) + t \Gamma_x {Tr}(\sigma_i (1 + s \sigma_z) \sigma_x \sigma_j (1 + \acute{s} \sigma_z) \sigma_x) \} 
 \nonumber \\
 & \times & \{ \pi (2n_0 +( s + \acute{s}) n_z)\} 
\times (\imath(\acute{s} - s)E_z +2 \Sigma_1 + (s + \acute{s})\Sigma_z )^{-1}
 \end{eqnarray}
 The sign $t=\pm 1$ is $+1$ if the $\Gamma_x$ scattering comes from spin-orbit impurities, and $-1$ if it comes from local magnetizations - see Hikami Larkin and Nagaoka's original paper \cite{hikami1980spin}.
 
If we remove the spin-orbit disorder $\Gamma_x$, then we get
\begin{eqnarray}
(D^{-1}(\vec{q}=0,\omega=0))_{ij} & = & \frac{ E_z}{ E_z^2 + \Sigma_1^2} \begin{bmatrix} 0 & 0 & 0 & \\ 0 &  E_z & -\Sigma_1 & 0 \\ 0 & \Sigma_1 &  E_z & 0 \\ 0 & 0 & 0 & 0 \end{bmatrix}
\end{eqnarray}
This clearly conserves charge.  There is no coupling between charge and spin, i.e. the upper row and left column are filled with zeros.  Moreover charge does not decay - the upper left corner of the matrix is zero.

On the other hand if we remove the ordinary disorder $\Gamma_0$ and retain only the spin-orbit term $\Gamma_x$, then we get two terms:
\begin{eqnarray}
(D^{-1}(\vec{q}=0,\omega=0))_{ij} & = & -\frac{1}{n_1^2 - n_z^2} \begin{bmatrix} 0 & 0 & 0 & \\ 0 & (t-1)n_1^2 + (t+1)n_z^2 & \imath 2t n_1 n_z  & 0 \\ 0 & \imath 2t n_1 n_z  &- (t+1)n_1^2 + (t-1)n_z^2 & 0 \\ 0 & 0 & 0 & 0 \end{bmatrix}
\nonumber \\
 & + & \frac{1}{ E_z^2 + \Sigma_1^2} \begin{bmatrix}  E_z^2 + (1 - t) \Sigma_1^2 & 0 & 0 & \imath  t E_z \Sigma_1 & \\ 0 & 0 & 0 & 0 \\ 0 & 0 & 0 & 0 \\ -\imath  t E_z \Sigma_1 & 0 & 0 & E_z^2 + (1 + t) \Sigma_1^2 \end{bmatrix}
\end{eqnarray}
The second term clearly does not conserve charge: the number in the upper left corner indicates that charge decays, and the numbers in the upper right and lower left corners indicate that charge mixes with the $z$ component of spin $S_z$.  The charge non-conservation effect is biggest, of order $O(1)$, when $\Sigma_1 = \pi n_1 \Gamma_z$ is comparable to $E_z$, i.e. when the self-energy caused by the spin-orbit term is comparable to the level splitting caused by the magnetization.  The $O(1)$ magnitude obtained in this case means that charge decays at the same time scale as the momentum relaxation time $\tau$.

Mathematica allows us to do anything in between  limits of $\Gamma_0=0$ and $\Gamma_x=0$. At lowest order in $\Gamma_x/ \Gamma_0$, we find 
\begin{eqnarray}
(D^{-1})_{00} &=& \frac{\Gamma_x ((E_z^2 (n_1^2 + n_z^2)) + (\Sigma_1^2 ((1-t) n_1^2 + (t+1)n_z^2)))}{\Gamma_0 (n_1^2 - n_z)^2 (E_z^2 + \Sigma_1^2)}
\end{eqnarray}

This result shows that  the BNM approach produces charge non-conservation even without a polarized density of states $n_z $, a polarized lifetime $\Sigma_z $, or spin-conserving disorder $\Gamma_0$.   It is enough to have (a) a non-zero Zeeman splitting $E_z$ and (b) spin-non-conserving disorder $\Gamma_x$.

\subsection{CoPt Model: Magnetization with Spin-Orbit Coupling}
In the minimal model we used spin-orbit disorder but no spin-orbit interaction.  Here we do the reverse, i.e. no spin-orbit disorder and instead a spin-orbit interaction.  The spin non-conservation is caused in this case by the non-commutation of the magnetization with the spin-orbit term.  This model is directly applicable to Cobalt-Platinum bilayers, and has been studied by Pesin and Macdonald's PRB 86 014416 \cite{PhysRevB.86.014416} and by Wang and Manchon's PRL 108 117201 \cite{PhysRevLett.108.117201}.  However previous works did not notice or address the fact that the self-energy is spin dependent, and that correct treatment of the self-energy results is charge non-conservation.

This model can also be used to study topological insulators with magnetization, by setting the $k^2/2m$ term in the Hamiltonian and the eigenvalues to zero.  

In the following we often write the  magnetic field as $\vec{b}$  thus allowing it to point in any direction, but we also at times assume that it is a perpendicular magnetic field $\vec{b} = b_z \hat{z}$ because this is simpler.  We include a spin-orbit interaction $\alpha (k_y \sigma_x - k_x \sigma_y)$.  Because of the magnetic field, the density of states $\hat{\rho}$ and the self-energy $ \Sigma$ are not proportional to the identity.   Significant spin-polarization occurs only near the bottom of the dispersion, where only one of the two spin-split states coincides with the Fermi energy.

We will show that that the BNM approach results in charge non-conservation and a coupling between charge and spin.

The self-energy is:
\begin{align}
\Sigma^A &= \vec{\Sigma}^A \cdot \sigma\\
\Sigma^R &=  \vec{\Sigma}^R \cdot \sigma
\end{align}
We are violating convention here since we have not written the $\imath $ in $\Sigma$ explicitly.  In general $\Sigma$ is expected to have an important imaginary part, and the real part is usually neglected.  
Also note that if $b_x = b_y=0$, then also $\Sigma_x = \Sigma_y=0$.

The Hamiltonian plus self-energy is:
\begin{align}
H^{A,R} &= k^2/2m + \Delta_{SO} (k_y \sigma_x - k_x \sigma_y) +  (\vec{b}+\vec{\Sigma}^{A,R}) \cdot \vec{\sigma} \\
 &= \frac{k^2}{2m}  + (\vec{b}+\vec{\Sigma})  \cdot \vec{\sigma} + \Delta_{SO} k \begin{bmatrix}0& e^{-\imath \phi_k} \\  e^{\imath  \phi_k} & 0\end{bmatrix} \\
&=\frac{k^2}{2m}  + (\vec{b}+\vec{\Sigma})  \cdot \vec{\sigma}  + \Delta_{SO} \begin{bmatrix}0& (k_x-\imath k_y) \\  (k_x+\imath k_y) & 0\end{bmatrix}\\
\end{align}

Since $\vec{b} $ always occurs in combination with $\vec{\Sigma}^{A,R}$, we write the combination as $\vec{B}^{A,R} = \vec{b} + \vec{\Sigma}^{A,R}$.

The energies are as follows. Note that if there is a self energy and it has an imaginary part, then these energies are, in principal, complex.
\begin{align}
 E^{A,R}(s,\vec{k}) &=  \frac{k^2}{2m}+\Sigma_1^{A,R}  + s\sqrt{(\Delta_{SO} k_y + B_x^{A,R})^2 + (-\Delta_{SO} k_x + B_y^{A,R})^2 +( B_z^{A,R})^2}\\
 E^{A,R,z}(s,\vec{k}) &=  \frac{k^2}{2m} +\Sigma_1^{A,R} + s\sqrt{(\Delta_{SO} |k|)^2 + (B_z^{A,R})^2}\\
\end{align}

The Green's functions are:
\begin{align}
 G^{A,R}(\vec{k}, E) &= \frac{1}{2}\sum_s \frac{1 + s \vec{X}^{A,R}  (\vec{k} ) \cdot \vec{\sigma} } {E -E^{A,R}(s,\vec{k}) \mp \imath \epsilon},  \; \vec{\sigma} = \left[\sigma_x, \,\sigma_y,\,\sigma_z\right]^T \\
 \vec{G}^{A,R}(\vec{k}, E) &= \frac{1}{2}\sum_s \frac{  \left[1, sX_1^{A,R}, sX_2^{A,R}, sX_3^{A,R} \right]^T
}  {E -E^{A,R}(s,\vec{k}) \mp \imath \epsilon},  \\
\vec{G}^{A,R,z}(\vec{k}, E) &= \frac{1}{2}\sum_s \frac{  \left[1, s \cos \theta^{A,R}_B \sin \theta_k , -s\cos \theta^{A,R}_B \cos \theta_k, s \sin \theta^{A,R}_B \right]^T
}  {E -\frac{k^2}{2m} -\Sigma_1^{A,R} - s\sqrt{(\Delta_{SO} |k|)^2 + (B^{A,R})_z^2} \mp \imath \epsilon}\\
  X_1^{A,R} &= \frac{\Delta_{SO} k_y + B_x^{A,R}}{ \sqrt{(\Delta_{SO} k_y + B_x^{A,R})^2 + (-\Delta_{SO} k_x + B_y^{A,R})^2 + (B_z^{A,R})^2}},  \; X_1^z = \cos \theta^{A,R}_B \sin \theta_k \\
   X_2^{A,R} &= \frac{-\Delta_{SO} k_x + B_y^{A,R}}{ \sqrt{(\Delta_{SO} k_y + B_x^{A,R})^2 + (-\Delta_{SO} k_x + B_y^{A,R})^2 + (B_z^{A,R})^2}},  \; X_2^z = -\cos \theta^{A,R}_B \cos \theta_k \\
   X_3^{A,R} &= \frac{B_z^{A,R}}{ \sqrt{(\Delta_{SO} k_y + B_x^{A,R})^2 + (-\Delta_{SO} k_x + B_y^{A,R})^2 + (B_z^{A,R})^2}} , \; X_3^z =  \sin \theta_B^{A,R} \\
\cos \theta_B^{A,R} &=   (\Delta_{SO} |k| / \sqrt{\Delta_{SO}^2 |k|^2 + (B^{A,R})^2}), \; \sin \theta_B^{A,R} = (B^{A,R} / \sqrt{\Delta_{SO}^2 |k|^2 + (B^{A,R})^2}) 
 \end{align}

The above formula for the Green's functions is based on the following inversion formula: 
 \begin{align}
&\; (E - (k^2/2m + \Delta_{SO} (k_x \sigma_x + k_y \sigma_y) + B_z \sigma_z ))^{-1} = \frac{1}
{2}\sum_s \frac{1 + s \vec{X}  (\vec{k} ) \cdot \vec{\sigma} } {E -E(s,\vec{k}) } \\
\end{align}

we prove the inversion formula as follows:
\begin{align}
 &\; (E - (k^2/2m + \Delta_{SO} (k_x \sigma_x + k_y \sigma_y) + B_z \sigma_z ))\frac{1}{2}\sum_s \frac{1 + s \vec{X}  (\vec{k} ) \cdot \vec{\sigma} } {E -E(s,\vec{k}) } \\
&= \frac{1}{2}\sum_s \frac{1}{E -E(s,\vec{k}) } ((E - k^2/2m) -s \Delta_{SO} (k_x  X_1+k_y X_2) -sB_z X_3 ) =1\\
&+ \frac{1}{2}\sum_s \frac{1}{E -E(s,\vec{k}) } (s\frac{(E - k^2/2m)}{\sqrt{\Delta_{SO}^2 |k_F|^2 + B_z ^2}}-1) ( \Delta_{SO} (k_x \sigma_x + k_y \sigma_y) + B_z \sigma_z ) =0\\
&+ \frac{1}{2}\sum_s \frac{1}{E -E(s,\vec{k}) } \sigma_z (-\Delta_{SO} k_x  s X_2 + \Delta_{SO} k_y s X_1) =0\\
&+ \frac{1}{2}\sum_s \frac{1}{E -E(s,\vec{k}) } \sigma_y (-B_z  s X_1 + \Delta_{SO} k_x s X_3)=0 \\
&+ \frac{1}{2}\sum_s \frac{1}{E -E(s,\vec{k}) } \sigma_x (-\Delta_{SO} k_y  s X_3 + B_z s X_2) =0
\end{align}

Now we introduce the scattering tensor, which describes nonmagnetic white noise disorder:
\begin{align}
\Gamma_{\alpha \nu, \beta \xi} &= \gamma  \delta_{\alpha \nu} \delta_{\beta \xi} \\
\end{align}

We write the SCBA equation for calculating the self-energy:
\begin{align}
\Sigma^{A,R}_{\alpha \beta}(E) &= \Gamma_{\alpha \nu, \xi \beta} \int {dk} \;G^{A,R}_{\nu \xi}(E,k)= \gamma \int {dk} \;G^{A,R}_{ \alpha \beta}(E,k), \; \vec{\Sigma}^{A,R} = \gamma \int {dk} \;\vec{G}^{A,R}\\
\end{align}

If $\vec{b} = b_z \hat{z}$ then it is trivial to prove that $\Sigma_x = \Sigma_y = 0$, and the remaining integrals have rotational invariance resulting in one dimensional integrals:
\begin{align}
\vec{\Sigma}^{A,R,z} &= 2 \pi \gamma \int_0^{\infty} {d|k|} \; |k|  \; \frac{1}{2}\sum_s \frac{  \left[1, 0, 0, s \sin \theta^{A,R}_B \right]^T
}  {E -\frac{k^2}{2m} -\Sigma_1^{A,R} - s\sqrt{(\Delta_{SO} |k|)^2 + (B^{A,R})_z^2} \mp \imath \epsilon},  \\
\vec{B}^{A,R} &= \vec{b} + \vec{\Sigma}^{A,R} \\
\end{align}

This SCBA equation should be solved self-consistently.  However, many authors do not obtain self-consistency and instead do the following   zeroth order approximation, using the bare Green's function. We follow this practice to illustrate the consequences:
\begin{align}
\vec{\Sigma}^{A,R,0} &= \gamma \int {dk} \;\vec{G}^A(E,k, \vec{\Sigma}=0) = \gamma \int {dk} \; \frac{1}{2}\sum_s \frac{  \left[1, sX_1, sX_2, sX_3 \right]^T
}  {E -E(s,\vec{k}, \Sigma = 0) \mp \imath \epsilon} \\
& \approx \pm \imath \pi \gamma \int {dk}   \; \frac{1}{2}\sum_s  \; \delta (E -E(s,\vec{k}, \Sigma = 0)) \left[1, sX_1, sX_2, sX_3 \right]^T \\
& = \pm \imath \frac{1}{2} \gamma \int {d\phi_k}  \; \frac{1}{2}\sum_s   \; \rho(\phi_k, s) \left[1, sX_1(k_F (s,\phi_k)\hat{\phi}_k), sX_2(k_F(s,\phi_k) \hat{\phi}_k), sX_3(k_F (s,\phi_k)\hat{\phi}_k) \right]^T \\
E &\equiv  E(s,k_F(s,\phi_k) \hat{\phi}_k), \Sigma = 0)\\
\rho(\phi_k,s) &=2 \pi \int_0^{\infty} {d|k|} \; |k|  \; \delta (E -E(s,|k|\hat{\phi}_k, \Sigma = 0))
\end{align}

We begin calculating the collision integral $I$:
\begin{align}
I_{\alpha \gamma, \beta \delta}(E,\omega,q) &=  \gamma \int {dk} G^A_{\alpha \gamma}(E+\omega/2, k+q/2) G^R_{\beta \delta}(E-\omega/2, k-q/2) \\
I_{ij}(E,\omega,q) &=   \gamma \frac{1}{2} \int {dk} {Tr}(G^A_{\alpha \gamma}(E+\omega/2, k+q/2) \sigma_i G^R_{\beta \delta}(E-\omega/2, k-q/2)\sigma_j)  \\
I_{ij}(E,\omega,q) &=   \frac{1}{2} \sum_{l,m} {Tr}(\sigma_l \sigma_i \sigma_m\sigma_j) I_{lm}(E,\omega,q) , \\
I_{lm}(E,\omega,q)  &=  \gamma \int {dk}\; \vec{G}^A_{l}(E+\omega/2, k+q/2)\; \vec{G}^R_{m}(E-\omega/2, k-q/2)  \\
 \vec{G}^{A,R}(\vec{k}, E) &= \frac{1}{2}\sum_s \frac{  \left[1, sX_1^{A,R}, sX_2^{A,R}, sX_3^{A,R} \right]^T
}  {E -E^{A,R}(s,\vec{k}) \mp \imath \epsilon},  \\
E^{A,R}(s,\vec{k}) &=  \frac{k^2}{2m}+\Sigma_1^{A,R}  + s\sqrt{(\Delta_{SO} k_y + B_x^{A,R})^2 + (-\Delta_{SO} k_x + B_y^{A,R})^2 +( B_z^{A,R})^2}\\
  X_1^{A,R} &= \frac{\Delta_{SO} k_y + B_x^{A,R}}{ \sqrt{(\Delta_{SO} k_y + B_x^{A,R})^2 + (-\Delta_{SO} k_x + B_y^{A,R})^2 + (B_z^{A,R})^2}}, \\
   X_2^{A,R} &= \frac{-\Delta_{SO} k_x + B_y^{A,R}}{ \sqrt{(\Delta_{SO} k_y + B_x^{A,R})^2 + (-\Delta_{SO} k_x + B_y^{A,R})^2 + (B_z^{A,R})^2}},    \\
   X_3^{A,R} &= \frac{B_z^{A,R}}{ \sqrt{(\Delta_{SO} k_y + B_x^{A,R})^2 + (-\Delta_{SO} k_x + B_y^{A,R})^2 + (B_z^{A,R})^2}}  \\
\vec{G}^{A,R,z}(\vec{k}, E) &= \frac{1}{2}\sum_s \frac{  \left[1, s \cos \theta^{A,R}_B \sin \theta_k , -s\cos \theta^{A,R}_B \cos \theta_k, s \sin \theta^{A,R}_B \right]^T
}  {E -\frac{k^2}{2m} -\Sigma_1^{A,R} - s\sqrt{(\Delta_{SO} |k|)^2 + (B^{A,R})_z^2} \mp \imath \epsilon}\\
 \cos \theta_B^{A,R} &=   (\Delta_{SO} |k| / \sqrt{\Delta_{SO}^2 |k|^2 + (B^{A,R})^2}), \;  \sin \theta_B^{A,R} = (B^{A,R} / \sqrt{\Delta_{SO}^2 |k|^2 + (B^{A,R})^2}) 
\end{align}

Since we want to show that charge is not conserved in the BNM approach, we choose to focus on $I_{ij}(E,\omega=0,q=0)$. $I_{00}$ determines the charge lifetime, and $I_{0j}$ determines the coupling between charge and spin.  We also insert the Green's functions:
\begin{align}
I_{ij}(E,\omega=0,q=0) &=   \gamma \frac{1}{8} \sum_{s,\acute{s}} \int {dk} {Tr}(\frac{1 + s \vec{X}_A  (\vec{k} ) \cdot \vec{\sigma} } {E -E_A(s,\vec{k}) }   \sigma_i \frac{1 + \acute{s} \vec{X}_R  (\vec{k} ) \cdot \vec{\sigma} } {E -E_R(\acute{s},\vec{k}) }  \sigma_j)  \\
&=   \gamma \frac{1}{8} \sum_{s,\acute{s}} \int {dk} {Tr}((1 + s \vec{X}_A  (\vec{k} ) \cdot \vec{\sigma} )  \sigma_i (1 + \acute{s} \vec{X}_R  (\vec{k} ) \cdot \vec{\sigma} )  \sigma_j) \\
& \times ( \frac{1}{E -E_A(s,\vec{k}) }-\frac{1} {E -E_R(\acute{s},\vec{k}) } ) \frac{1} {E_A(s,\vec{k}) -E_R(\acute{s},\vec{k}) } \\
\end{align}

Following standard practice, we now approximate the denominator, taking the imaginary part, and treating it as a Dirac delta function.  
\begin{align}
I_{ij}&\approx  \imath \pi  \gamma \frac{1}{8} \sum_{s,\acute{s}} \int {dk} \delta(E -Re(E_A(s,\vec{k}))) {Tr}((1 + s \vec{X}_A  (\vec{k} ) \cdot \vec{\sigma} )    \sigma_i (1 + \acute{s} \vec{X}_R  (\vec{k} ) \cdot \vec{\sigma} )  \sigma_j)  \frac{1}{E_A(s,\vec{k}) -E_R(\acute{s},\vec{k}) } \\ 
&+\imath \pi  \gamma \frac{1}{8} \sum_{s,\acute{s}} \int {dk} \delta(E -Re(E_R(\acute{s},\vec{k}))) {Tr}((1 + s \vec{X}_A  (\vec{k} ) \cdot \vec{\sigma} )  \sigma_i (1 + \acute{s} \vec{X}_R  (\vec{k} ) \cdot \vec{\sigma} )  \sigma_j) \frac{1}{E_A(s,\vec{k})-E_R(\acute{s},\vec{k}) }   
\end{align}

Now we require that the magnetic field be perpendicular to the plane, i.e. $\vec{b} = b_z \hat{z}$,  and take advantage of the resulting rotational symmetry:
\begin{align}
I_{ij}&\approx  \sum_{s,\acute{s}} \frac{1}{16}  \frac{  \imath \gamma \rho_A(s) } { E_A(s,\vec{k}_F(s,\phi)) -E_R(\acute{s},\vec{k}_F(s,\phi)) }   \int {d\phi} {Tr}((1 + s \vec{X}_A  (\vec{k}_F(s,\phi) ) \cdot \vec{\sigma} )    \sigma_i (1 + \acute{s} \vec{X}_R  (\vec{k}_F(s,\phi) ) \cdot \vec{\sigma})   \sigma_j)  \\
&+\sum_{s,\acute{s}} \frac{1}{16} \frac{\imath \gamma \rho_R(\acute{s})} {E_A(s,\vec{k}_F(\acute{s},\phi) ) -E_R(\acute{s},\vec{k}_F(\acute{s},\phi) )}    \int {d\phi}  {Tr}((1 + s \vec{X}_A  (\vec{k}_F(\acute{s},\phi) ) \cdot \vec{\sigma})    \sigma_i (1 + \acute{s} \vec{X}_R  (\vec{k}_F(\acute{s},\phi) ) \cdot \vec{\sigma} )  \sigma_j)  \\
\end{align}

Now we specialize to calculating the charge lifetime, i.e. $I_{00}$. If $I_{00}$ is not equal to $1$, then charge has a lifetime.
\begin{align}
I_{00} &\approx  \sum_{s,\acute{s}} \frac{1}{8}  \frac{ \imath \gamma \rho_A(s) } {E_A(s,\vec{k}_F({s},\phi) ) -E_R(\acute{s},\vec{k}_F({s},\phi) )}   \int {d\phi} \; (1 + s \acute{s}  \vec{X}_A  (\vec{k}_F(s,\phi) )\cdot \vec{X}_R  (\vec{k}_F(s,\phi) ))\\
&+  \sum_{s,\acute{s}} \frac{1}{8}  \frac{ \imath \gamma \rho_R(s) } {E_A(\acute{s},\vec{k}_F({s},\phi) ) -E_R(s,\vec{k}_F({s},\phi) )}   \int {d\phi} \; (1 + s \acute{s}  \vec{X}_A  (\vec{k}_F(s,\phi) )\cdot \vec{X}_R  (\vec{k}_F(s,\phi) ))
\end{align}
Now we analyze the deviation of $\vec{X_A} \cdot \vec{X_R}$ from $1$, which is its value in a system without spin polarization.  We also break the sum $\sum_{s,\acute{s}} $ into contributions with $s = \acute{s}$ and contributions with $s = -\acute{s}$.  (Here we are assuming $\rho_A = \rho_R$ to simplify the algebra, and later also $\Sigma^R = - \Sigma^A$, but the same argument goes through without these assumptions.)
\begin{align}
1 + \delta XX &= \vec{X}_A  (\vec{k}_F(s,\phi) )\cdot \vec{X}_R  (\vec{k}_F(s,\phi) ) \\
(1 + s \acute{s}) &= 2 \delta(s,\acute{s}) \\
I_{00} &\approx   \pi \sum_s \frac{ \imath \gamma \rho(s) }{E_A(s,\vec{k}_F({s},\phi) ) -E_R(s,\vec{k}_F({s},\phi) )}  \\
&+ \sum_{s} \frac{1}{4}  \frac{ \imath  \gamma \rho(s) } {E_A(s,\vec{k}_F({s},\phi) ) -E_R(s,\vec{k}_F({s},\phi) )}   \int {d\phi}\;  \delta XX(s,\phi)\\
&-  \sum_{s} \frac{1}{8}  \frac{ \imath \gamma \rho(s) } {E_A(s,\vec{k}_F({s},\phi) ) -E_R(-s,\vec{k}_F({s},\phi) )}   \int {d\phi} \; \delta XX(s,\phi)\\
&-  \sum_{s} \frac{1}{8}  \frac{ \imath \gamma \rho(s) } {E_A(-s,\vec{k}_F({s},\phi) ) -E_R(s,\vec{k}_F({s},\phi) )}   \int {d\phi} \;\delta XX(s,\phi) \\
\end{align}
$\delta X X$ is independent of $\phi$ and the angular integral is trivial. The first of the four terms in $I_{00}$ would be exactly $1$ in systems without spin polarization.  Here it deviates from $1$ meaning that charge is not conserved, because
\begin{align}
 \Sigma^A_1 &=  \pi \gamma  \;\frac{1}{2}\sum_s  \rho(s,E), \Sigma^A_z =\pi \gamma  \;\frac{1}{2}\sum_s  \rho(s,E)\;s X_3 (|\vec{k}(E,s) |) \\
E_{A,R}(s) &= k^2/2m +\imath \Sigma^{A,R}_1 +s \sqrt{\Delta_{SO}^2 |k|^2 + (b_z + \imath \Sigma^{A,R}_z)^2} 
 \end{align}

 Next we evaluate $\delta XX$ and show that it is not zero, implying that charge is not conserved. 
\begin{align}
 X_{1A,R} &= \frac{\Delta_{SO}\, k_x}{ \sqrt{(\Delta_{SO} |k|)^2 + (b_z \pm \imath \Sigma^{A}_z)^2}}, \; X_{2A,R} = \frac{\Delta_{SO}\,  k_y}{ \sqrt{(\Delta_{SO} |k|)^2 + (b_z \pm \imath \Sigma^{A}_z)^2}}, \; X_{3A,R} = \frac{b_z \pm \imath \Sigma^{A}_z}{ \sqrt{(\Delta_{SO} |k|)^2 + (b_z \pm \imath \Sigma^{A}_z)^2}} \\
 \delta XX(s,\phi) &= -1 + \frac{(\Delta_{SO} |k(s,\phi)|)^2 +b_z^2 + (\Sigma^A_z )^2}{ \sqrt{(\Delta_{SO} |k(s,\phi)|)^2 + (b_z + \imath \Sigma^{A}_z)^2} \sqrt{(\Delta_{SO} |k(s,\phi)|)^2 + (b_z - \imath \Sigma^{A}_z)^2}}\\
  &= -1 + \sqrt{ \frac{(\Delta_{SO} |k(s,\phi)|)^4 +b_z^4 + (\Sigma^A_z )^4 + 2 b_z^2 (\Sigma^A_z )^2 + 2b_z^2(\Delta_{SO} |k(s,\phi)|)^2 +2  (\Sigma^A_z )^2(\Delta_{SO} |k(s,\phi)|)^2}{(\Delta_{SO} |k(s,\phi)|)^4 +b_z^4 + (\Sigma^A_z )^4   + 2 b_z^2 (\Sigma^A_z )^2+2b_z^2(\Delta_{SO} |k(s,\phi)|)^2-2  (\Sigma^A_z )^2(\Delta_{SO} |k(s,\phi)|)^2} } \\
 &= -1 + \sqrt{1+ \frac{4  (\Sigma^A_z )^2(\Delta_{SO} |k(s,\phi)|)^2}{(\Delta_{SO} |k(s,\phi)|)^4 +b_z^4 + (\Sigma^A_z )^4   + 2 b_z^2 (\Sigma^A_z )^2+2b_z^2(\Delta_{SO} |k(s,\phi)|)^2-2  (\Sigma^A_z )^2(\Delta_{SO} |k(s,\phi)|)^2} } \\
\end{align}
This expression's deviation from zero controls charge non-conservation.  Its numerator (inside the square root) is proportional to $\Sigma^A_z \Delta_{SO} k_F$.   Overall the deviation from zero is of order $1$ when  $\Sigma_z$ and $\Delta_{SO} |k(s,\phi)|$ are of the same magnitude, i.e. when the spin-orbit splitting $\Delta_{SO}$ is of the same magnitude as the magnetization energy $\Sigma_z$.  In this case we find that the decay constant is of order $O(1)$, meaning that the charge decay time is of the same order as the momentum relaxation time.

In summary, charge is not conserved, i.e. $I_{00}$ is not $1$, both because $ \pi \sum_s \frac{  \gamma \rho(s) } { 2\, Im(E_A(s)) }$ is not one, and because $\delta X X $ is not zero.

Lastly, we consider the charge-spin coupling, which should be zero. 
\begin{align}
I_{0j} &\approx  \sum_{s,\acute{s}} \frac{1}{16}  \frac{ \imath  \gamma \rho_A(s) } { E_A(s,\vec{k}_F(s,\phi)) -E_R(\acute{s},\vec{k}_F(s,\phi)) }  \int {d\phi} {Tr}((1 + s \vec{X}_A  (\vec{k}_F(s,\phi) ) \cdot \vec{\sigma} )    (1 + \acute{s} \vec{X}_R  (\vec{k}_F(s,\phi) ) \cdot \vec{\sigma})   \sigma_j)  \\
&+\sum_{s,\acute{s}} \frac{1}{16} \frac{ \imath \gamma \rho_R(\acute{s})}{E_A(s,\vec{k}_F(\acute{s},\phi) ) -E_R(\acute{s},\vec{k}_F(\acute{s},\phi) )}   \int {d\phi}  {Tr}((1 + s \vec{X}_A  (\vec{k}_F(\acute{s},\phi) ) \cdot \vec{\sigma})     (1 + \acute{s} \vec{X}_R  (\vec{k}_F(\acute{s},\phi) ) \cdot \vec{\sigma} )  \sigma_j)  \\
I_{0j} &\approx  \sum_{s,\acute{s}} \frac{1}{8}  \frac{ \imath \gamma \rho_A(s) } { E_A(s,\vec{k}_F(s,\phi)) -E_R(\acute{s},\vec{k}_F(s,\phi)) }  \int {d\phi} (sX_{Aj}  (\vec{k}_F(s,\phi) ) + \acute{s}X_{Rj}  (\vec{k}_F(s,\phi) ))\\
&+\sum_{s,\acute{s}} \frac{1}{8} \frac{\imath \gamma \rho_R(\acute{s})} {E_A(s,\vec{k}_F(\acute{s},\phi) ) -E_R(\acute{s},\vec{k}_F(\acute{s},\phi) )}  \int {d\phi} (sX_{Aj}  (\vec{k}_F(\acute{s},\phi) ) + \acute{s}X_{Rj}  (\vec{k}_F(\acute{s},\phi) ))  \\
\end{align}
Now we decompose $X_A, X_R$ into their mean value and difference.  In a system without spin polarization, the difference $\delta X$ is zero.  Because of the accompanying factor of $s+\acute{s}$, the mean value $X_+$ contributes only when $s = \acute{s}$.  Similarly, the difference contributes only when $s=-\acute{s}$.  
\begin{align}
X_{Aj} &= X_{+j} + \delta X_{j}, X_{Rj} = X_{+j} - \delta X_j \\
I_{0j} &\approx  \sum_{s,\acute{s}} \frac{1}{8}  \frac{ \imath \gamma \rho_A(s) } { E_A(s,\vec{k}_F(s,\phi)) -E_R(\acute{s},\vec{k}_F(s,\phi)) }  \int {d\phi} ( (s + \acute{s}) X_{+j}(\vec{k}_F(s,\phi) ) + (s - \acute{s}) \delta X_{j}(\vec{k}_F(s,\phi) )) \\
&+\sum_{s,\acute{s}} \frac{1}{8} \frac{ \imath \gamma \rho_R(\acute{s})} {E_A(s,\vec{k}_F(\acute{s},\phi) ) -E_R(\acute{s},\vec{k}_F(\acute{s},\phi) )}  \int {d\phi}( (s + \acute{s}) X_{+j}(\vec{k}_F(\acute{s},\phi) ) + (s - \acute{s}) \delta X_{j}(\vec{k}_F(\acute{s},\phi) ))  \\
&= \sum_{s} \frac{1}{4}  \frac{ s \gamma \rho(s) } { Im(E_A(s)) }   \int {d\phi} (  X_{+j}(\vec{k}_F(s,\phi) ))  \\
&+ \sum_{s=-\acute{s}} \frac{1}{8}  \frac{ \imath \gamma \rho_A(s) (s - \acute{s})} { E_A(s,\vec{k}_F(s,\phi)) -E_R(\acute{s},\vec{k}_F(s,\phi)) }  \int {d\phi}   \delta X_{j}(\vec{k}_F(s,\phi) ) \\
&+\sum_{s=-\acute{s}} \frac{1}{8} \frac{ \imath \gamma \rho_R(\acute{s})(s - \acute{s})} {E_A(s,\vec{k}_F(\acute{s},\phi) ) -E_R(\acute{s},\vec{k}_F(\acute{s},\phi) )}  \int {d\phi}   \delta X_{j}(\vec{k}_F(\acute{s},\phi) )  \\
\end{align}
 In the first term, the angular integral evaluates to zero when $j=1,2$, but not when $j=3$; the coupling between charge and $S_z$ spin is non-zero. In the second and third terms $\delta X_j$ also integrates to zero when $j=1,2$, but not when $j=3$. In all terms $X_3$ is independent of $\phi$ so the angular integral is trivial.

In summary, the $j=3$ terms in $I_{0j}$ are non-zero, which means that charge mixes with $S_z$, which again means that charge is not conserved.

\section{Ward-Takahashi Identity and Diffusion Equations }

In this section we  derive a Ward-Takahashi identity which ensures charge conservation.  Our work is inspired by two papers:
\begin{itemize}
\item Ward-Takahashi identities for disordered metals and superconductors, by Revaz Ramazashvili, PRB 66 220503(R), 2002 \cite{PhysRevB.66.220503}.
\item Ward-Takahashi identity for nonequilibrium Fermi systems, by Velicky Kalvova and Spicka, PRB 77 041201(R), 2008 \cite{PhysRevB.77.041201}.
\end{itemize}
We prove the Ward-Takahashi identity using very strong and general non-perturbative arguments which are articulated in these two papers. The proof applies to any interacting theory which commutes with the global charge operator.  The main value-added of our presentation is that we are explicitly emphasizing spin (and orbital) indices, and their role in the Ward-Takahashi identity, where the other works did not.  Our development was important for understanding how charge conservation can be guaranteed in spin-polarized or orbital-polarized systems.  The Ward-Takahashi identity which we will prove is:
\begin{align}
\Sigma_{\alpha \beta}(E+\omega,p)-\Sigma_{\alpha \beta}(E, p)  &= -\omega \sum_\gamma \Lambda_{\alpha \beta, \gamma \gamma}(E + \omega,E,p,p)
\end{align}
$\Sigma$ is the self-energy. $ \Lambda $ is a four-point function, and is dimensionless.  This Ward-Takahashi identity applies to several variants of $\Lambda$: the advanced-retarded variant $\Lambda^{AR}$, the retarded-retarded variant $ \Lambda^{RR}$, and  the advanced-advanced variant $\Lambda^{AA}$. 

Ramazashvilli's PRB 66 220503(R) \cite{PhysRevB.66.220503}  showed that  the Ward-Takahashi identity for $\Lambda^{AR}$ is equivalent to a statement that $\Lambda^{AR}_{\alpha \beta, \gamma \gamma}(E + \omega,E,p,p)$ has a pole at $\omega = 0$, i.e. 
\begin{align}
\Lambda^{AR}_{\alpha \beta, \gamma \gamma}(E + \omega,E,p,p) &= 2 \Sigma / \omega \propto (\tau \omega)^{-1}
\end{align}
Moreover, if we keep the $k$ momentum in $\Lambda^{AR}$,  we will get something like
\begin{align}
\Lambda^{AR}_{\alpha \beta, \gamma \gamma}(E + \omega,E,p+k,p) &  \propto (\tau \omega + Dk^2)^{-1}
\end{align}
This is recognizable as a diffuson pole.  The absence of a constant term in this expression, i.e. the absence of a lifetime, indicates that charge is conserved.
From this form written as a pole one can derive  a diffusion equation governing transport.  The pole guarantees that the diffusion equation will satisfy charge conservation.

The Ward-Takahashi identity is very powerful for three reasons: (1) it concerns $\Lambda$ which is the unique operator that controls charge and spin transport (in the linear response approximation), (2) $\omega$'s  explicit role in the Ward-Takahashi identity makes the existence of the pole $\Lambda$ unavoidable, and (3) the Ward-Takahashi identity is rigorous for every system which conserves global charge.

There is another Ward-Takahashi identity which has been proved  by Vollhardt and Wolfle's PRB 22 4666 \cite{PhysRevB.22.4666}: 
\begin{align}
\Sigma^R_{\alpha \beta}(k,E+\epsilon)-\Sigma^A_{\alpha \beta}(k,E)  &= \int {d\acute{k}}\; U_{\alpha \beta, \gamma \delta}(k,\acute{k}, E, \epsilon) \; (G^R_{\gamma \delta}(\acute{k}, E+\epsilon)\otimes 1-1\otimes G^A_{\gamma \delta}(\acute{k}, E))
\end{align}
We will call this Ward II, in contrast to the more powerful one which we favor in this text. Here  $U$ is the irreducible four point vertex, and has units of energy squared.  Vollhardt and Wolfle's Ward II identity is proved perturbatively, and does not contain an explicit factor of $\omega$.  We have not found it directly useful for proving charge conservation in the general case of spin-polarized or orbital-polarized systems.  However, Vollhardt and Wolfle's PRB 22 4666 Appendix A  is very useful for understanding that in disordered systems the a ladder-diagram approximation for $\Lambda$, combined with the CPA/SCBA approximation for $\Sigma$, does satisfy charge conservation \cite{PhysRevB.22.4666}. The perturbative expansion and manipulation used by Vollhardt and Wolfle's PRB 22 4666 Appendix A to prove their Ward II identity transfers exactly to showing that CPA/SCBA + the ladder-diagram approximation satisfies charge conservation.  

After proving the Ward-Takahashi identity, we derive transport equations.  First we discuss the general case of a charge-conserving system, and then we present the equations for disordered systems with the CPA/SCBA + the ladder-diagram approximation.  Lastly there is an unfinished section discussing how these results, derived using Matsubara frequencies, should be analytically continued to real frequencies.

\subsection{The Ward-Takahashi identity}
We will derive the Ward-Takahashi identity for the vertex correction under the imaginary-time (Matsubara) formalism,
with $\hbar =1$.
The action $S[\psi^\dag,\psi]$ is uniquely obtained from the Hamiltonian $H(\tau)[\psi^\dag,\psi]$ as
\begin{align}
S[\psi^\dag,\psi] &= \sum_\alpha \int_0^\beta d\tau \int d\bfr \ \psi_\alpha^\dag(\bfr,\tau) \partial_\tau \psi_\alpha(\bfr,\tau) + \int_0^\beta d\tau \ H(\tau)[\psi^\dag(\tau),\psi(\tau)],
\end{align}
with the fermionic field variables (Grassmann numbers) $\psi_\alpha^{(\dag)}(\bfr,\tau)$.
The Hamiltonian $H[\psi^\dag,\psi]$ should be normal-ordered
so that it can be appropriately converted to the path-integral formalism.
The integral over the imaginary time $\tau$ is taken from $0$ to $\beta = 1/T$.
With this action,
the partition function is given by the path integral
\begin{align}
Z = \int \mathcal{D}\psi^\dag \mathcal{D}\psi \ e^{-S[\psi^\dag,\psi]},
\end{align}
where we suppress the source terms for the field variables.

The quantum and thermal average of the observable $\mathcal{A}$ is given by the path integral
\begin{align}
\langle \mathcal{A} \rangle = \frac{1}{Z} \int \mathcal{D}\psi^\dag \mathcal{D}\psi \ \mathcal{A} \ e^{-S[\psi^\dag,\psi]}.
\end{align}
The two-point Green's function for the fermion is given by
\begin{align}
G_{\alpha \beta}(\bfr,\tau;\bfr',\tau') &= \langle \psi_\alpha(\bfr,\tau) \psi^\dagger_\beta(\bfr',\tau') \rangle = \frac{1}{Z} \int \mathcal{D}\psi^\dag \mathcal{D}\psi \ \psi_\alpha(\bfr,\tau) \psi^\dagger_\beta(\bfr',\tau') \ e^{-S[\psi^\dag,\psi]}.
\end{align}

For the spinless non-interacting fermions, the action can be written in general as
\begin{align}
S[\psi^\dag,\psi] = \sum_{\omega_n,\bfk} \psi^\dag(\bfk,\omega_n) \left[ -i\omega_n + \epsilon(\bfk) \right] \psi(\bfk,\omega_n) \equiv \sum_{\omega_n,\bfk} \psi^\dag(\bfk,\omega_n) D(\bfk,\omega_n) \psi(\bfk,\omega_n),
\end{align}
with the Fourier transformed field variables
\begin{align}
\psi(\bfr,\tau) = \frac{1}{\sqrt{\beta V}}  \sum_{\omega_n,\bfk} e^{-i\omega_n \tau -i \bfk\cdot\bfr} \psi(\bfk,\omega_n), \\
\psi(\bfk,\omega_n) = \frac{1}{\sqrt{\beta V}} \int_0^\beta {d\tau}  \sum_{\bfr} e^{i\omega_n \tau +i \bfk\cdot\bfr} \psi(\bfr,\tau),
\end{align}
where $\omega_n = (2\pi/\beta)(n+1/2)$ is the fermionic Matsubara frequency.
Thus the non-interacting partition function $Z$ can be straightforwardly obtained,
\begin{align}
Z = \int \mathcal{D}\psi^\dag \mathcal{D}\psi \ e^{-S[\psi^\dag,\psi]} &= \prod_{\omega_n,\bfk} \left[\int d\psi^\dag(\bfk,\omega_n) d\psi(\bfk,\omega_n) \left(1 -\psi^\dag(\bfk,\omega_n) D(\bfk,\omega_n) \psi(\bfk,\omega_n)\right) \right] \\
 &= \prod_{\omega_n,\bfk} D(\bfk,\omega_n),
\end{align}
where we have used the definition that Grassmann integral is equivalent to the derivative,
and that the Grassmann derivatives also anticommute each other.

Similarly, the averaging over $\psi \psi^\dag$ is given by
\begin{align}
Z \langle \psi(\bfk,\omega_n) \psi^\dag(\bfk,\omega_n) \rangle
 &= \int d\psi^\dag(\bfk,\omega_n) d\psi(\bfk,\omega_n) \left[ -\psi^\dag(\bfk,\omega_n) \psi(\bfk,\omega_n) \left(1-\psi^\dag(\bfk,\omega_n) D(\bfk,\omega_n) \psi(\bfk,\omega_n)\right)\right] \\
 & \quad \times \prod_{(\omega'_n,\bfk') \neq (\omega_n,\bfk)} \left[\int d\psi^\dag(\bfk',\omega'_n) d\psi(\bfk',\omega'_n) \left(1-\psi^\dag(\bfk',\omega'_n) D(\bfk',\omega'_n) \psi(\bfk',\omega'_n)\right) \right] \nonumber \\
 &= \prod_{(\omega'_n,\bfk') \neq (\omega_n,\bfk)} D(\bfk',\omega'_n).
\end{align}
As a result, the non-interacting Green's function can be written in a simple form,
\begin{align}
G_0(\bfk,\omega_n) = D^{-1}(\bfk,\omega_n) = \frac{1}{-i\omega_n + \epsilon(\bfk)},
\end{align}
There is only one momentum index and one frequency index because we are exploiting momentum conservation and energy conservation.  In real space the Green's function is
\begin{align}
G_0(\bfr,\tau;\bfr',\tau') = \frac{1}{\beta V} \sum_{\omega_n,\bfk} e^{-i\omega_n(\tau-\tau') -i\bfk\cdot(\bfr-\bfr')} G_0(\bfk,\omega_n).
\end{align}

We now generalize from scalar variables to a multi-component system, which is similar.  The action, the partition function, and the Green's function are given by
\begin{align}
S[\psi^\dag,\psi] &= \sum_{\omega_n,\bfk} \psi_\alpha^\dag(\bfk,\omega_n) \left[ -i\omega_n + H_{\alpha\beta}(\bfk) \right] \psi_\beta(\bfk,\omega_n) \equiv \sum_{\omega_n,\bfk} \psi_\alpha^\dag(\bfk,\omega_n) D_{\alpha\beta}(\bfk,\omega_n) \psi_\beta(\bfk,\omega_n), \\
Z &= \prod_{\omega_n,\bfk} \det D(\bfk,\omega_n), \\
G_{0,\alpha\beta}(\bfk,\omega_n) &= \langle \psi_\alpha(\bfk,\omega_n) \psi_\beta^\dag(\bfk,\omega_n) \rangle = [D^{-1}(\bfk,\omega_n)]_{\alpha\beta} = \left[\frac{1}{-i\omega_n + H(\bfk)}\right]_{\alpha\beta},
\end{align}
respectively.

As long as the original Hamiltonian $H[\psi^\dag,\psi]$ conserves the total particle number,
the action $S[\psi^\dag,\psi]$ is invariant under the \textbf{time-independent global U(1) gauge transformation}
\begin{align}
\psi_\alpha(\bfr,\tau) \mapsto e^{i\theta} \psi_\alpha(\bfr,\tau), \quad \psi_\alpha^\dag(\bfr,\tau) \mapsto \psi_\alpha^\dag(\bfr,\tau) e^{-i\theta}.
\end{align}
The conservation of total charge
\begin{align}
Q(\tau) =\sum_\alpha \int d\bfr \ \psi_\alpha^\dag(\bfr,\tau) \psi_\alpha(\bfr,\tau)
\end{align}
is equivalent to this global gauge symmetry.

We can also define the \textbf{time-dependent U(1) gauge transformation}
\begin{align}
\tilde{\psi}_\alpha(\bfr,\tau) = e^{i\theta(\tau)} \psi_\alpha(\bfr,\tau), \quad
\tilde{\psi}_\beta^\dag(\bfr',\tau') = \psi^\dag_\beta(\bfr',\tau') e^{-i\theta(\tau')},
\end{align}
with a time-dependent scalar function $\theta(\tau)$.
Under this transformation, the variation of the action $S[\psi^\dag,\psi]$ comes only from a ``Berry phase'' term:
\begin{align}
\delta_\theta S = S[\tilde{\psi}^\dag,\tilde{\psi}] - S[\psi^\dag,\psi] &= i \int d\tau'' d\bfr'' \ \psi_\gamma^\dag(\bfr'',\tau'')\dot{\theta}(\tau'') \psi_\gamma(\bfr'',\tau'') \label{eq:gauge-artifact} \\
 &= i \int d\tau'' \ \dot{\theta}(\tau'') Q(\tau'').
\end{align}

We now derive the Ward-Takahashi identity related to the U(1) gauge invariance, namely  total charge conservation.
Since the path integral $\int \mathcal{D}\psi^\dag \mathcal{D}\psi$ is invariant under the local (time-dependent) U(1) gauge transformation,
the following equality is justified:
\begin{align}
G_{\alpha\beta}(\bfr,\tau;\bfr',\tau') &= \frac{1}{Z} \int \mathcal{D}\psi^\dag \mathcal{D}\psi \ \psi_\alpha(\bfr,\tau)\psi_\beta^\dag(\bfr',\tau') e^{-S[\psi^\dag,\psi]} \\
 &= \frac{1}{Z} \int \mathcal{D}\tilde{\psi}^\dag \mathcal{D}\tilde{\psi} \ \tilde{\psi}_\alpha(\bfr,\tau)\tilde{\psi}_\beta^\dag(\bfr',\tau') e^{-S[\tilde{\psi}^\dag,\tilde{\psi}]} \\
 &= \frac{1}{Z} \int \mathcal{D}\psi^\dag \mathcal{D}\psi \ \tilde{\psi}_\alpha(\bfr,\tau)\tilde{\psi}_\beta^\dag(\bfr',\tau') e^{-S[\tilde{\psi}^\dag,\tilde{\psi}]} \\
 &= \frac{1}{Z} \int \mathcal{D}\psi^\dag \mathcal{D}\psi \ e^{i[\theta(\tau)-\theta(\tau')]} \psi_\alpha(\bfr,\tau) \psi_\beta^\dag(\bfr',\tau') e^{-S[\psi^\dag,\psi]-\delta_\theta S[\psi^\dag,\psi]}.
\end{align}
Comparing the first and the fourth lines and taking $\theta(\tau)$ infinitesimal, we obtain
\begin{align}
0 &= \frac{1}{Z} \int \mathcal{D}\psi^\dag \mathcal{D}\psi \ \psi_\alpha(\bfr,\tau) \psi_\beta^\dag(\bfr',\tau') \left[1- e^{i[\theta(\tau)-\theta(\tau')]} e^{-\delta_\theta S[\psi^\dag,\psi]} \right] e^{-S[\psi^\dag,\psi]} \\
 &= \frac{1}{Z} \int \mathcal{D}\psi^\dag \mathcal{D}\psi \ \psi_\alpha(\bfr,\tau) \psi_\beta^\dag(\bfr',\tau') \left[-i[\theta(\tau)-\theta(\tau')] +\delta_\theta S[\psi^\dag,\psi] \right] e^{-S[\psi^\dag,\psi]}.
\end{align}
Using the form of $\delta_\theta S$ in Eq.~(\ref{eq:gauge-artifact}),
we obtain the Ward-Takahashi identity relating the two-point and four-point functions,
\begin{align}
[\theta(\tau)-\theta(\tau')] G_{\alpha\beta}(\mathbf{x}; \mathbf{x}') &=  \int d\tau'' d\bfr'' \ \dot{\theta}(\tau'')\chi_{\beta\gamma\gamma\alpha}(\mathbf{x}; \mathbf{x}''; \mathbf{x}''; \mathbf{x}'). \label{eq:WI-real-space}
\end{align}
Here we have introduced  the shorthand notation $\mathbf{x} = (\boldsymbol{r},\tau)$ and $\mathbf{K} \equiv (\bfk,\omega_n)$.

Next we Fourier transform the Ward-Takahashi identity from position-time coordinates $ \mathbf{x} = (\boldsymbol{r},\tau)$ to momentum-frequency coordinates  $\mathbf{K} \equiv (\bfk,\omega_n)$.  Taking the specific form
\begin{align}
\theta(\tau'') = \bar{\theta} e^{-i\bar{\omega}_m \tau''}, \;\mathbf{W} \equiv (0,\mathbf{\bar{\omega}}_m)
\end{align}
with $\bar{\omega}_m = (2\pi/\beta)m$ the bosonic Matsubara frequency,
the left-hand and right-hand sides of Eq.~(\ref{eq:WI-real-space}) read
\begin{align}
\mathrm{[LHS]} &= [\theta(\tau)-\theta(\tau')] G_{\alpha\beta}(\bfr,\tau;\bfr',\tau') \\
& =    \frac{\bar{\theta}}{\beta V}   \sum_{\mathbf{K}}   e^{-i\mathbf{K}\cdot(\mathbf{x}-\mathbf{x}') } \left[ e^{-i\mathbf{W}\cdot \mathbf{x}} - e^{-i\mathbf{W}\cdot \mathbf{x}'} \right] G_{\alpha\beta}(\mathbf{K})\\
& =    \frac{\bar{\theta}}{\beta V}  e^{-i \mathbf{W} \cdot (\mathbf{x}+\mathbf{x}')/2} \sum_{\mathbf{K}}   e^{-i\mathbf{K}\cdot(\mathbf{x}-\mathbf{x}') } \left[ e^{-i\mathbf{W}\cdot ( \mathbf{x} - \mathbf{x}')/2} - e^{i\mathbf{W}\cdot ( \mathbf{x} - \mathbf{x}')/2} \right] G_{\alpha\beta}(\mathbf{K})\\
\mathrm{[RHS]} &=\int d\mathbf{x}'' \ \dot{\theta}(\tau'')  \chi_{\beta \gamma \gamma \alpha}(\mathbf{x};\mathbf{x}'';\mathbf{x}'';\mathbf{x}')
  \\
 &= -i \bar{\omega}_m \bar{\theta} \int d\mathbf{x}'' \ e^{-i\mathbf{W}\cdot\mathbf{x}''}  \frac{1}{(\beta V)^3} \sum_{\mathbf{K},\mathbf{Q},\mathbf{K}''} e^{i(\mathbf{K} + \mathbf{Q}/2)\cdot \mathbf{x} - i(\mathbf{K}'' + \mathbf{Q}/2)\cdot\mathbf{x}''} e^{i(\mathbf{K}'' - \mathbf{Q}/2) \cdot \mathbf{x}'' - i(\mathbf{K} - \mathbf{Q}/2)  \cdot \mathbf{x}'} \\
 & \times \chi_{\beta \gamma \gamma \alpha}(\mathbf{K};\mathbf{Q};\mathbf{K}'') \\
&= -i \bar{\omega}_m   \  \frac{\bar{\theta}}{(\beta V)^2}  \sum_{\mathbf{K},\mathbf{K}''} e^{i(\mathbf{K} - \mathbf{W}/2)\cdot \mathbf{x} } e^{ - i(\mathbf{K} + \mathbf{W}/2)  \cdot \mathbf{x}'} \chi_{\beta \gamma \gamma \alpha}(\mathbf{K};-\mathbf{W};\mathbf{K}'') \\
&= -i \bar{\omega}_m   \  \frac{\bar{\theta}}{(\beta V)^2} e^{-i \mathbf{W} \cdot (\mathbf{x}+\mathbf{x}')/2} \sum_{\mathbf{K},\mathbf{K}''}  e^{  i\mathbf{K}   \cdot (\mathbf{x}- \mathbf{x}')}G_{\beta_1 \beta}(\mathbf{K}- \mathbf{W}/2) G_{\alpha \alpha_1}(\mathbf{K} + \mathbf{W}/2) \\
& \times  \left[ (\beta V) \delta_{\beta_1 \gamma } \delta_{\gamma\alpha_1} \delta(\mathbf{K} - \mathbf{K}'')+  \Lambda_{\beta_1\gamma\gamma\alpha_1}(\mathbf{K};-\mathbf{W};\mathbf{K}'') \right] \\
\end{align}
We have defined the vertex function $\Lambda$, a four point function, which we will discuss in more detail  in the next section.

As a result, the Fourier transformed Ward-Takahashi identity is 
\begin{align}
G_{\alpha\beta}\mathbf{K}-\mathbf{W}/2) - G_{\alpha\beta}(\mathbf{K}+\mathbf{W}/2) = -i\bar{\omega}_m &[ G_{\alpha\gamma}(\mathbf{K}+\mathbf{W}/2) G_{\gamma\beta}(\mathbf{K}-\mathbf{W}/2) \\
&+ G_{\alpha\alpha_1}(\mathbf{K}+\mathbf{W}/2)  \frac{1}{\beta V} \sum_{\mathbf{K}'} {\Lambda}_{\beta_1 \gamma \gamma \alpha_1}(\mathbf{K};\mathbf{W};\mathbf{K}') G_{\beta_1\beta}(\mathbf{K}-\mathbf{W}/2) ].
\end{align}

Multiplying $[G^{-1}(\mathbf{K}+\mathbf{W}/2)]_{\alpha'\alpha}$ from the left and $[G^{-1}(\mathbf{K}-\mathbf{W}/2)]_{\beta\beta'}$ from the right, we obtain
\begin{align}
[G^{-1}(\mathbf{K}+\mathbf{W}/2)]_{\alpha'\beta'} - [G^{-1}(\mathbf{K}-\mathbf{W}/2)]_{\alpha'\beta'} &= -i\bar{\omega}_m \left[ \delta_{\alpha'\beta'} + \frac{1}{\beta V} \sum_{\mathbf{K}'}{\Lambda}_{\beta'\gamma\gamma\alpha'}(\mathbf{K};\mathbf{W};\mathbf{K}') \right] 
\end{align}
Using the specific form
\begin{align}
[G^{-1}(\bfk,\omega_n)]_{\alpha\beta} = -i\omega_n \delta_{\alpha\beta} + [H(\bfk) -\Sigma(\bfk,\omega_n)]_{\alpha\beta},
\end{align}
\begin{align}
G^{-1} &= -\imath E + H - \Sigma 
\end{align}

we finally obtain
\begin{align}
[\Sigma(\mathbf{K}+\mathbf{W}/2) - \Sigma(\mathbf{K}-\mathbf{W}/2)]_{\alpha\beta} & = i\bar{\omega}_m \frac{1}{\beta V} \sum_{\mathbf{K}'}{\Lambda}_{\beta\gamma\gamma\alpha}(\mathbf{K};\mathbf{W};\mathbf{K}'), \; \mathbf{W} \equiv (0,\bar{\omega}_m) \\
 & = i\bar{\omega}_m \frac{1}{\beta V} \sum_{\mathbf{K}}{\Lambda}_{\gamma \alpha \beta \gamma}(\mathbf{K};\mathbf{W};\mathbf{K}'),
\end{align}
which relates the self energy $\Sigma$ and the vertex correction $\Lambda$.  This is the final form of the Ward-Takahashi identity.  In the second line we have written the companion identity which applies to the $\Lambda$'s outer indices.

Note that the Ward-Takahashi identity proves that $\Lambda$ contains a pole at $\bar{\omega}_m = 0$.   This is true because the left hand side of the Ward-Takahashi identity does not go to zero when $\bar{\omega}_m \rightarrow 0$ - there is a branch cut in $\Sigma$ at $\bar{\omega}_m = 0$.  Therefore the right side of the Ward-Takahashi identity must also be non-zero.    And therefore some component of $\Lambda$ must scale as $(\bar{\omega}_m)^{-1}$ to cancel the explicit factor of $\bar{\omega}_m$.

%
%
If the disorder is static, independent of time, then the energy components of $\mathbf{K}, \mathbf{K}''$ must be equal: $\omega = \omega''$.  Therefore the sum $\omega$ or $\omega''$ in the above expressions is extraneous, and we need only a sum over $\mathbf{k}$ or $\mathbf{k}''$.  The Ward-Takahashi identity is now
\begin{align}
\sum_{\mathbf{k}''}{\Lambda}_{\beta\gamma\gamma\alpha}(\omega,\mathbf{k};\omega',\mathbf{q}=0;\mathbf{k}'') &= \frac{\beta V}{i\bar{\omega}_m} [\Sigma(\mathbf{k},\omega + \omega'/2) - \Sigma(\mathbf{k},\omega - \omega'/2)]_{\alpha\beta} 
\end{align}

These equations also prove  that the susceptibility tensor $\chi$  contains a pole at $\bar{\omega}_m = 0$, since $\chi$ is linearly related to $\Lambda$. Please refer to the next section which defines $\chi^{(0)}$ and $\chi^{(vc)}$.  With those definitions in place, we obtain
\begin{align}
\chi_{\alpha\beta\gamma\delta}(\mathbf{K};\mathbf{Q};\mathbf{K}'') &=  \chi^{(0)}+\chi^{(vc)}= G_{\alpha_1 \alpha}(\mathbf{K} + \mathbf{Q}/2) G_{\delta \delta_1}(\mathbf{K} - \mathbf{Q}/2) \\
& \times \left[ (\beta V) \delta_{\alpha_1 \beta } \delta_{\gamma\delta_1} \delta(\mathbf{K} + \mathbf{K}'')+  \Lambda_{\alpha_1\beta\gamma\delta_1}(\mathbf{K};\mathbf{Q};\mathbf{K}'') \right] \\
\sum_{\mathbf{K}''} \chi^{(vc)}_{\alpha\gamma\gamma\delta}(\mathbf{K};\mathbf{W};\mathbf{K}'') &=G_{\alpha_1 \alpha}(\mathbf{K}+\mathbf{W}/2) G_{\delta \delta_1}(\mathbf{K}-\mathbf{W}/2) \sum_{\mathbf{K}''} \Lambda_{\alpha_1\gamma\gamma\delta_1}(\mathbf{K};\mathbf{W};\mathbf{K}'') \\
&= \frac{\beta V}{i\bar{\omega}_m} G_{\alpha_1 \alpha}(\mathbf{K}+\mathbf{W}/2) G_{\delta \delta_1}(\mathbf{K}-\mathbf{W}/2)  [\Sigma(\mathbf{K}+\mathbf{W}/2) - \Sigma(\mathbf{K}-\mathbf{W}/2)]_{\alpha_1\delta_1} 
\end{align}
A similar identity shows that  that there is also a pole in $\sum_{\mathbf{K}} \chi^{(vc)}_{\alpha \beta \gamma \alpha}(\mathbf{K};\mathbf{W};\mathbf{K}'') $, i.e. the Ward-Takahashi identity applies both to $\chi$'s internal indices and to its external indices. Therefore we can conclude that the inverse of $\chi$ conserves charge, and is (up to an inconsequential multiplicative constant) the  charge transport operator.

Both ${\chi}$ and ${\Lambda}$ are fully symmetric under interchange of inner and outer indices and retain the same pole.

This concludes our proof of the Ward-Takahashi identity.

\subsection{Preparation for the Transport Equation}

We define a local spin-charge density $\rho_{\alpha \beta}(\mathbf{x}, \tau)$ and a perturbing local potential $V_{\alpha \beta}(\mathbf{x}, \tau)$.  The charge transport equation governing $\rho$ is
\begin{align}
\rho_{\alpha \beta}(\mathbf{x}) = \int{d\mathbf{x}'} \; \hat{ \chi}_{\alpha \gamma \delta \beta}(\mathbf{x},\mathbf{x}') \;   V_{\gamma \delta}(\mathbf{x}')
\end{align}

The transport kernel is $\hat{ \chi}$, which is related to the four-point function, namely the susceptibility tensor, by
\begin{align}
\hat{\chi}_{\alpha\beta\gamma\delta}(\mathbf{x}'; \mathbf{x}) & = \chi_{\alpha\beta\gamma\delta}(\mathbf{x}; \mathbf{x}'; \mathbf{x}'; \mathbf{x}) \\
\chi_{\alpha\beta\gamma\delta}(\mathbf{x}_1; \mathbf{x}_2; \mathbf{x}_3; \mathbf{x}_4) & \equiv \langle \psi_\alpha^\dag(\mathbf{x}_1) \psi_\beta(\mathbf{x}_2) \psi_\gamma^\dag(\mathbf{x}_3) \psi_\delta(\mathbf{x}_4) \rangle.
\end{align}
$\hat{\chi}^{-1} \rho$  has the same units as $V$, and $\chi$ has the meaning of $d \rho / d V$.   

The diffusion operator, i.e. the equation which controls transport at scales larger than the scattering length/time, can be extracted by expanding $(\hat{\chi})^{-1}$ to linear order in $\omega'$ and second order in $q$.  After doing this expansion, if one Fourier transforms from $\omega'$ to $t$ then one obtains a simple exponential evolution controlled by the kernel of  $\hat{\chi}$.


We now discuss the Fourier transform of  $\hat{\chi}$ in some detail, to get a good understanding of how the transport equations work.  We require translational invariance and energy conservation, which makes the description in the reciprocal space efficient. The Fourier transform of the transport function $\hat{\chi}$, and of any other two point function such as the Green's function $G_{\alpha \beta}$, is given by
\begin{align}
\hat{\chi}_{\alpha\beta}(\mathbf{x};\mathbf{x}') &= \hat{\chi}_{\alpha\beta}(\mathbf{x}-\mathbf{x}') = \frac{1}{\beta V} \sum_{\mathbf{K}} e^{-i\mathbf{K}\cdot(\mathbf{x}-\mathbf{x}')} \hat{\chi}_{\alpha\beta}(\mathbf{K}) \\
\hat{\chi}_{\alpha\beta}(\mathbf{K})  & =\int {d(\mathbf{x}-\mathbf{x}')} e^{i\mathbf{K}\cdot(\mathbf{x}-\mathbf{x}')} \hat{\chi}_{\alpha\beta}(\mathbf{x};\mathbf{x}')
\end{align}
The Fourier transform  of the susceptibility tensor, and of any other four point function, is
\begin{align}
\chi_{\alpha\beta\gamma\delta}(\mathbf{x};\mathbf{x}';\mathbf{x}'';\mathbf{x}''') 
 &= \frac{1}{(\beta V)^3} \sum_{\mathbf{K},\mathbf{Q},\mathbf{K}''} e^{i(\mathbf{K} + \mathbf{Q}/2)\cdot \mathbf{x} - i(\mathbf{K}'' + \mathbf{Q}/2)\cdot\mathbf{x}'} e^{i(\mathbf{K}'' - \mathbf{Q}/2) \cdot \mathbf{x}'' - i(\mathbf{K} - \mathbf{Q}/2)  \cdot \mathbf{x}'''} \chi_{\alpha\beta\gamma\delta}(\mathbf{K};\mathbf{Q};\mathbf{K}'') \\
\chi_{\alpha\beta\gamma\delta}(\mathbf{K};\mathbf{Q};\mathbf{K}'')
 &=  \int { d(\mathbf{x}'-\mathbf{x}) d(\mathbf{x}''-\mathbf{x}) d(\mathbf{x}'''-\mathbf{x})}  e^{-i(\mathbf{K} + \mathbf{Q}/2)\cdot \mathbf{x} + i(\mathbf{K}'' + \mathbf{Q}/2)\cdot\mathbf{x}'} e^{-i(\mathbf{K}''- \mathbf{Q}/2) \cdot \mathbf{x}'' + i(\mathbf{K} - \mathbf{Q}/2)  \cdot \mathbf{x}'''} \\
 & \times   \chi_{\alpha\beta\gamma\delta}(\mathbf{x};\mathbf{x}';\mathbf{x}'';\mathbf{x}''') 
\end{align}

We now separate the  response function into two parts.  The first part $\chi^{(0)}$ gives the complete response function of a non-interacting theory, and can be expressed as a product of two Green's functions using  Wick's theorem. In  interacting systems there is a  vertex correction $\chi^{(vc)}$. We write $\chi^{(vc)}$ in terms of the  vertex function $\Lambda$, a four point function.  This is a definition of $\Lambda$:
\begin{align}
\chi & = \chi^{(0)} + \chi^{(vc)} \\
\chi^{(0)}_{\alpha\beta\gamma\delta}(\mathbf{x}; \mathbf{x}'; \mathbf{x}''; \mathbf{x}''') & = G_{\beta \alpha}(\mathbf{x}'; \mathbf{x}) G_{\delta \gamma}(\mathbf{x}'''; \mathbf{x}'') \\
\chi^{(vc)}_{\alpha\beta\gamma\delta}(\mathbf{x}; \mathbf{x}'; \mathbf{x}''; \mathbf{x}''') &=  \int d\mathbf{x}_1 d\mathbf{x}'_1 \ G_{\alpha_1 \alpha}(\mathbf{x}_1; \mathbf{x}) G_{\delta \delta_1}(\mathbf{x}'''; \mathbf{x}'_1) \Lambda_{\alpha_1\beta\gamma\delta_1}(\mathbf{x}_1; \mathbf{x}'; \mathbf{x}''; \mathbf{x}'_1) 
\end{align}

The Fourier transforms of these quantities are
\begin{align}
\chi^{(0)}_{\alpha\beta\gamma\delta}(\mathbf{K};\mathbf{Q};\mathbf{K}'') &= G_{\beta \alpha}(\mathbf{K} + \mathbf{Q}/2) G_{\delta \gamma}(\mathbf{K} - \mathbf{Q}/2) \times (\beta V) \delta(\mathbf{K} - \mathbf{K}'')\\
\chi^{(vc)}_{\alpha\beta\gamma\delta}(\mathbf{K};\mathbf{Q};\mathbf{K}'') &=  G_{\alpha_1 \alpha}(\mathbf{K} + \mathbf{Q}/2) G_{\delta \delta_1}(\mathbf{K} - \mathbf{Q}/2)  \Lambda_{\alpha_1\beta\gamma\delta_1}(\mathbf{K};\mathbf{Q};\mathbf{K}'')\\
\chi_{\alpha\beta\gamma\delta}(\mathbf{K};\mathbf{Q};\mathbf{K}'') &=  G_{\alpha_1 \alpha}(\mathbf{K} + \mathbf{Q}/2) G_{\delta \delta_1}(\mathbf{K} - \mathbf{Q}/2) \left[ (\beta V) \delta_{\alpha_1 \beta } \delta_{\gamma\delta_1} \delta(\mathbf{K} - \mathbf{K}'')+  \Lambda_{\alpha_1\beta\gamma\delta_1}(\mathbf{K};\mathbf{Q};\mathbf{K}'') \right] \\
\hat{\chi}^{(0)}_{\alpha\beta\gamma\delta}(\mathbf{Q}) & =  \frac{1}{(\beta V)^2} \sum_{\mathbf{K},\mathbf{K}'}\chi^{(0)}_{\alpha\beta\gamma\delta}(\mathbf{K};\mathbf{Q};\mathbf{K}''), \; \hat{\chi}^{(vc)}_{\alpha\beta\gamma\delta}(\mathbf{Q})  =  \frac{1}{(\beta V)^2} \sum_{\mathbf{K},\mathbf{K}'}\chi^{(vc)}_{\alpha\beta\gamma\delta}(\mathbf{K};\mathbf{Q};\mathbf{K}'') \\
\end{align}

The Ward-Takahashi identity is
\begin{align}
[\Sigma(\mathbf{K}+\mathbf{W}/2) - \Sigma(\mathbf{K}-\mathbf{W}/2)]_{\alpha\beta} & =  i\bar{\omega}_m \frac{1}{\beta V} \sum_{\mathbf{K}'}{\Lambda}_{\gamma \alpha \beta \gamma}(\mathbf{K};\mathbf{W};\mathbf{K}')
\end{align}
where the self-energy is defined by
\begin{align}
[G^{-1}(\bfk,\omega_n)]_{\alpha\beta} = -i\omega_n \delta_{\alpha\beta} + [H(\bfk) -\Sigma(\bfk,\omega_n)]_{\alpha\beta}.
\end{align}

We have neglected the diamagnetic contribution. 

\begin{align}
\hat{\chi}^{(0)}_{\alpha\beta\gamma\delta}(\mathbf{x}'; \mathbf{x}) & = \hat{\chi}^{(0)}_{\alpha\beta\gamma\delta}(\mathbf{x}; \mathbf{x}'; \mathbf{x}'; \mathbf{x})  = G_{\beta \alpha}(\mathbf{x}'; \mathbf{x}) G_{\delta \gamma}(\mathbf{x}; \mathbf{x}') \\
&= \frac{1}{(\beta V)^2} \sum_{\mathbf{K}, \mathbf{K}'} e^{i(\mathbf{K}-\mathbf{K}')\cdot(\mathbf{x}-\mathbf{x}')} G_{\beta \alpha}(\mathbf{K}) G_{\delta \gamma}(\mathbf{K}')\\
\hat{\chi}^{(0)}_{\alpha\beta\gamma\delta}(\mathbf{Q}) &= \int {d(\mathbf{x}'-\mathbf{x})}   \frac{1}{(\beta V)^2} \sum_{\mathbf{K}, \mathbf{K}'} e^{i\mathbf{Q}\cdot(\mathbf{x}'-\mathbf{x})}  e^{i(\mathbf{K}-\mathbf{K}')\cdot(\mathbf{x}-\mathbf{x}')} G_{\beta \alpha}(\mathbf{K}) G_{\delta \gamma}(\mathbf{K}')\\
&=    \frac{1}{\beta V} \sum_{\mathbf{K}}  G_{\beta \alpha}(\mathbf{K}) G_{\delta \gamma}(\mathbf{K}-\mathbf{Q}) = \frac{1}{(\beta V)^2} \sum_{\mathbf{K},\mathbf{K}'}\chi^{(0)}_{\alpha\beta\gamma\delta}(\mathbf{K};\mathbf{Q};\mathbf{K}'') \\
\end{align}

\subsection{The Diffusive Transport Equations - the Ladder diagram susceptibility and the SCBA}

Now we introduce the ladder approximation for $\chi, \Lambda$.  We also introduce the SCBA for $\Sigma$, which requires an assumption that disorder is weak and governed by Gaussian statistics.   This is combination of approximations is exactly the same as the RPA with a local  $\phi^4(x)$ interaction and an interaction strength set by $\Gamma$, with a single modification: the disorder is static, so most of the integrations over energy $\omega$ are removed.   As long as the impurity density is small enough, we can easily treat strong impurities and non-Gaussian statistics by (a) moving from the SCBA to the CPA, and (b) adding T-matrices to the ladder approximation.   See work by Aires Ferreira and collaborators for development of T-matrices \cite{PhysRevLett.112.066601,PhysRevB.90.035444}.

Within these approximations, transport equations are:
\begin{align}
 \Lambda_{\alpha \beta \gamma \delta} (\omega, \bfk; \omega', \bfq; \omega'', \bfk'') &=     ((1- I(\omega,\omega',\mathbf{q}) )^{-1})_{\nu \gamma \delta \eta}  \Gamma_{\alpha \nu \eta \beta} G_{\zeta \alpha}(\omega + \omega'/2, \mathbf{k} + \mathbf{q}/2) G_{\beta \lambda}(\omega - \omega'/2,\mathbf{k} - \mathbf{q}/2) \delta(\omega - \omega'') \\
I_{\gamma \zeta \lambda \delta}(\omega,\omega',\mathbf{q})  &=    {\Gamma}_{\gamma \alpha \beta \delta} \Phi, \\
\Phi &=   \frac{1}{V} \sum_{\mathbf{k}} G_{\zeta \alpha}(\omega + \omega'/2, \mathbf{k} + \mathbf{q}/2) G_{\beta \lambda}(\omega - \omega'/2,\mathbf{k} - \mathbf{q}/2) \\
\Sigma_{\alpha \beta}(\omega) &= \sum_{\gamma \delta} \Gamma_{\alpha \gamma \delta \beta} \int{d\bfk} G_{\gamma \delta} (\omega, \bfk) 
\end{align}
$I$ is the collision integral. The diffusion operator can be extracted by expanding $(\hat{\chi})^{-1}$ to linear order in $\omega'$ and second order in $q$.  After doing this expansion, if one Fourier transforms from $\omega'$ to $t$ then one obtains a simple exponential evolution controlled by the kernel of  $\hat{\chi}$.

The Ward-Takahashi identity is
\begin{align}
[\Sigma(\omega+\bar{\omega}_m /2) - \Sigma(\omega-\bar{\omega}_m /2)]_{\alpha\beta} & =  i\bar{\omega}_m  \frac{1}{V} \sum_{\omega'', \mathbf{k}''} {\Lambda}_{\gamma \alpha \beta \gamma}(\omega,\mathbf{k};\bar{\omega}_m,\mathbf{q}=0;\omega'',\mathbf{k}'') \\
& =  i\bar{\omega}_m \; I(\omega,\bar{\omega}_m,\mathbf{q}=0)  ((1- I(\omega,\bar{\omega}_m,\mathbf{q}=0) )^{-1})_{\nu \gamma \delta \eta}
\end{align}


Comparing to Burkov Nunez and Macdonald \cite{PhysRevB.70.155308}, their transport kernel $((\hat{ \chi}^{transport})^{-1})$ is $1- I(\bfq, E, \omega)$, where $I$ is a collision operator.  In other words, they have, roughly,  $\hat{ \chi}^{transport} = (1- I(\bfq, E, \omega))^{-1} $.  
The biggest differences between their result and ours  are that (A) we have an additional factor of $\Gamma G G$ multiplying their kernel, and (B) we require that the  equation for the self-energy $\Sigma$ be solved self-consistently. 

\subsection{Some notes on the Matsubara formalism}
This section is not complete yet.

The goal of this section is to perform the transition from Matsubara frequencies to real frequencies.  The discussion mostly follows the discussion in ``Condensed Matter Field Theory'' by Altland and Simons \cite{altland2010condensed}.

If we want to analytically continue to the real axis then the $\imath \omega$ prefactor turns into $\omega$, the quantities at $E$ become retarded quantities, and the quantities at $E + \omega$ become advanced quantities.  One obtains:
\begin{align}
 \Sigma^R_{\epsilon \xi}(k,E) -  \Sigma^A_{\epsilon \xi}(k,E+\omega)  &= \omega \sum_{ \gamma }   \int  { dk_5} \Lambda^{RA}_{\epsilon \gamma \gamma \xi}(k, k_5, k_5, k, E + \omega, E+\omega, E, E) \\
 \Sigma^A_{\epsilon \xi}(k,E) -  \Sigma^A_{\epsilon \xi}(k,E+\omega)  &=   \omega \sum_{ \gamma }   \int  { dk_5} \Lambda^{AA}_{\epsilon \gamma \gamma \xi}(k, k_5, k_5, k, E + \omega, E+\omega, E, E) \\
 \Sigma^R_{\epsilon \xi}(k,E) -  \Sigma^R_{\epsilon \xi}(k,E+\omega)  &=   \omega \sum_{ \gamma }   \int  { dk_5} \Lambda^{RR}_{\epsilon \gamma \gamma \xi}(k, k_5, k_5, k, E + \omega, E+\omega, E, E) \\
\end{align}

Suppose the Hamiltonian for the electrons is given by the matrix form,
\begin{align}
\hat{H}_0 = \int d\bfr \ \psi_\alpha^\dag(\bfr) \hat{h}_{\alpha\beta} \psi_\beta(\bfr) = \frac{1}{V} \sum_{\bfk} \psi_\alpha^\dag(\bfk) h_{\alpha\beta}(\bfk) \psi_\beta(\bfk).
\end{align}
Then the imaginary-time action is given by
\begin{align}
S_0[\psi^\dag,\psi] &= \int d\tau d\bfr \ \psi_\alpha^\dag(\bfr,\tau)\left[ \partial_\tau + \hat{h}_{\alpha\beta} \right]\psi_\beta(\bfr,\tau) = \frac{1}{\beta V} \sum_{\omega_n,\bfk} \psi_\alpha^\dag(\bfk,\omega_n)\left[-i\omega_n + h_{\alpha\beta}(\bfk) \right]\psi_\beta(\bfk,\omega_n),
\end{align}
where $\beta = (k_B T)^{-1}$ and $\omega_n = (2\pi/\beta)(n+1/2)$.
We also add a source term coupled to generalized density,
\begin{align}
\delta S[\psi^\dag,\psi;\phi] &= \int d\tau d\bfr \ \psi_\alpha^\dag(\bfr,\tau) \phi_{\alpha\beta}(\bfr,\tau) \psi_\beta(\bfr,\tau) = \frac{1}{\beta V} \sum_{\bar{\omega}_m,\bfq} \phi_{\alpha\beta}(\bfq,\bar{\omega}_m) \rho_{\alpha\beta}(\bfq,\bar{\omega}_m),
\end{align}
where $\bar{\omega}_m = (2\pi/\beta)m$ and
\begin{align}
\rho_{\alpha\beta}(\bfq,\bar{\omega}_m) = \frac{1}{\beta V} \sum_{\omega_n,\bfk} \psi_\alpha^\dag(\bfk+\tfrac{\bfq}{2},\omega_n) \psi_\beta(\bfk+\tfrac{\bfq}{2},\omega_n+\bar{\omega}_m).
\end{align}
The overall partition function is defined by the path integral,
\begin{align}
Z[\phi] = \int \mathcal{D}\psi^\dag \mathcal{D}\psi \ e^{-S_0[\psi^\dag,\psi]-\delta S[\psi^\dag,\psi;\phi]} .
\end{align}

The expectation value of the charge is given by differentiating the partition function by $\phi$,
\begin{align}
\langle \rho_{\alpha\beta}(\bfr,\tau) \rangle &= -\frac{\delta}{\delta \phi_{\alpha\beta}(\bfr,\tau)} \ln Z[\phi] \bigr{|}_{\phi=0} \simeq -\int d\tau' d\bfr' \frac{\delta^2}{\delta \phi_{\alpha\beta}(\bfr,\tau) \delta \phi_{\alpha'\beta'}(\bfr',\tau')} \ln Z[\phi] \bigr{|}_{\phi=0} \phi_{\alpha'\beta'}(\bfr',\tau').
\end{align}
Thus the response function $\chi$ is given by the density-density correlation,
\begin{align}
\chi_{\alpha\beta,\alpha'\beta'}(\bfr,\tau|\bfr',\tau') &= -\int d\tau' d\bfr' \frac{\delta^2}{\delta \phi_{\alpha\beta}(\bfr,\tau) \delta \phi_{\alpha'\beta'}(\bfr',\tau')} \ln Z[\phi] \bigr{|}_{\phi=0} = -\langle \rho_{\alpha\beta}(\bfr,\tau) \rho_{\alpha'\beta'}(\bfr',\tau') \rangle.
\end{align}
In the Fourier space, we have the familiar structure from the Wick's theorem,
\begin{align}
\chi_{\alpha\beta,\alpha'\beta'}(\bfq,\bar{\omega}_m) &= -\langle \rho_{\alpha\beta}(\bfq,\bar{\omega}_m) \rho_{\alpha'\beta'}(-\bfq,-\bar{\omega}_m) \rangle \\
 &= -\frac{1}{(\beta V)^2} \sum_{\omega_n,\omega_n',\bfk,\bfk'} \langle \psi_\alpha^\dag(\bfk-\tfrac{q}{2},\omega_n) \psi_\beta(\bfk+\tfrac{\bfq}{2},\omega_n+\bar{\omega}_m) \psi_{\alpha'}^\dag(\bfk'+\tfrac{q}{2},\omega_n'+\bar{\omega}_m) \psi_{\beta'}(\bfk'-\tfrac{\bfq}{2},\omega_n') \rangle \\
 &= \frac{1}{\beta V} \sum_{\omega_,\bfk} G_{\alpha\beta'}(\bfk-\tfrac{\bfq}{2},\omega_n) G_{\alpha'\beta}(\bfk+\tfrac{\bfq}{2},\omega_n+\bar{\omega}_m).
\end{align}

The disorder-averaged Green's function is given by
\begin{align}
\left[G^{-1}(\bfk,\omega_n)\right]_{\alpha\beta} &= i\omega_n - h_{\alpha\beta}(\bfk) +i\gamma_{\alpha\beta}(\bfk,\omega_n) \sgn\omega_n.
\end{align}
As $h$ and $\gamma$ are Hermitian matrices, the Green's function can be diagonalized by a certain unitary transformation $U$,
\begin{align}
G_{\alpha\beta}(\bfk,\omega_n) &\equiv \sum_{\mu} U_{\alpha\mu}^\dag(\bfk,\omega_n) \frac{1}{i\omega_n -\epsilon_\mu(\bfk) +i\gamma_\mu(\bfk,\omega_n) \sgn\omega_n} U_{\mu\beta}(\bfk,\omega_n).
\end{align}

The summation over Matsubara frequencies can be transformed into the loop integral in the complex plane using the Fermi-Dirac function $f(z) = (1 + \exp(\beta z))^{-1}$,
\begin{align}
\chi_{\alpha\beta,\alpha'\beta'}(\bfq,\bar{\omega}_m) &= \frac{1}{\beta V}\sum_{\omega_n,\bfk} \sum_{\mu\nu} U_{\alpha\mu}^\dag U_{\mu\beta'} U_{\alpha'\nu}^\dag U_{\nu\beta} \frac{1}{i\omega_n-\epsilon_\mu+i\gamma_\mu\sgn\omega_n} \frac{1}{i(\omega_n+\bar{\omega}_m)-\epsilon_\nu+i\gamma_\nu\sgn(\omega_n+\bar{\omega}_m)} \\
 &= -\frac{1}{ 2\pi i  V} \sum_{\bfk}\sum_{\mu\nu} \oint dz \ U_{\alpha\mu}^\dag U_{\mu\beta'} U_{\alpha'\nu}^\dag U_{\nu\beta} \frac{f(z)}{z-\epsilon_\mu+i\gamma_\mu\sgn\im z} \frac{1}{(z+i\bar{\omega}_m)-\epsilon_\nu+i\gamma_\nu\sgn\im(z+i\bar{\omega}_m)} \\
 \end{align}
 The contour must contain all the Matsubara frequencies $\imath \omega_n$ but it may not contain the branch cuts at  $\im z = 0$ and $\im(z+i\bar{\omega}_m)=0$.  Moreover parts of the contour at infinity are zero.  The remaining integrals sum to
 \begin{align}
 \chi &= -\frac{1}{2\pi i V} \sum_{\bfk}\sum_{\mu\nu} \int_{-\infty}^{\infty} d\epsilon \Biggl\{ f(\epsilon) U_{\alpha\mu}^\dag U_{\mu\beta'} U_{\alpha'\nu}^\dag U_{\nu\beta}\\
 & \quad \quad \quad \quad \quad \quad \quad\quad\quad 
\times  \left[g_\mu^R(\epsilon)g_\nu^R(\epsilon+i\bar{\omega}_m) - g_\mu^A(\epsilon)g_\nu^R(\epsilon+i\bar{\omega}_m) + g_\mu^A(\epsilon-i\bar{\omega}_m)g_\nu^R(\epsilon) - g_\mu^A(\epsilon-i\bar{\omega}_m)g_\nu^A(\epsilon) \right]\Biggr\} \nonumber \\
 &= -\frac{1}{2\pi i V} \sum_{\bfk} \int_{-\infty}^{\infty} d\epsilon \ f(\epsilon) \Bigl[ G_{\alpha\beta'}^R(\epsilon)G_{\alpha'\beta}^R(\epsilon+i\bar{\omega}_m) - G_{\alpha\beta'}^A(\epsilon)G_{\alpha'\beta}^R(\epsilon+i\bar{\omega}_m) \\
 & \quad \quad \quad \quad \quad \quad \quad\quad\quad + G_{\alpha\beta'}^A(\epsilon-i\bar{\omega}_m)G_{\alpha'\beta}^R(\epsilon) - G_{\alpha\beta'}^A(\epsilon-i\bar{\omega}_m)G_{\alpha'\beta}^A(\epsilon)\Bigr] \nonumber
\end{align}
By the analytic continuation $i\bar{\omega}_m \rightarrow \omega$, and then reorganizing, we obtain
\begin{align}
\chi_{\alpha\beta,\alpha'\beta'}(\bfq,\omega) &=-\frac{1}{2\pi i V} \sum_{\bfk} \int_{-\infty}^{\infty} d\epsilon \ \Bigl\{ f(\epsilon) \left[ G_{\alpha\beta'}^R(\epsilon)G_{\alpha'\beta}^R(\epsilon+\omega) - G_{\alpha\beta'}^A(\epsilon)G_{\alpha'\beta}^R(\epsilon+\omega)\right] \\
 & \quad \quad \quad \quad \quad \quad \quad\quad\quad +f(\epsilon+\omega) \left[ G_{\alpha\beta'}^A(\epsilon)G_{\alpha'\beta}^R(\epsilon+\omega) - G_{\alpha\beta'}^A(\epsilon)G_{\alpha'\beta}^A(\epsilon+\omega) \right] \Bigr\}. \nonumber \\
  &=\frac{1}{2\pi i V} \sum_{\bfk} \int_{-\infty}^{\infty} d\epsilon \ \Bigl\{( f(\epsilon) -  f(\epsilon+\omega)) \;  G_{\alpha\beta'}^A(\epsilon)G_{\alpha'\beta}^R(\epsilon+\omega) \\
 & \quad \quad \quad \quad \quad \quad \quad\quad\quad -\left[ f(\epsilon)  G_{\alpha\beta'}^R(\epsilon)G_{\alpha'\beta}^R(\epsilon+\omega)   - f(\epsilon+\omega)  G_{\alpha\beta'}^A(\epsilon)G_{\alpha'\beta}^A(\epsilon+\omega) \right] \Bigr\}
\end{align}

At this point one can take the imaginary part of the  $\chi$ and after a few short steps obtain the Kubo formula for the conductivity, as shown in Bruus and Flensberg \cite{bruus2004many}.

\section{Checks that our formalism combining the SCBA with  a susceptibility built from ladder diagrams obeys the Ward-Takahashi identity} 
Although any field theory that conserves global charge must obey the Ward-Takahashi identity and charge conservation, approximations made to that theory may result in charge non-conservation if the approximation used to calculate the self energy $\Sigma$ does not match the approximation used to calculate the susceptibility $\chi$.  We have proposed to use the SCBA for the self energy and a sum of ladder diagrams (very similar to RPA) for the susceptibility.  In this section we perform many checks that  the proposed approximations do satisfy the Ward-Takahashi identity.

Within the combination of SCBA and ladder diagrams, the Ward-Takahashi identity is:
\begin{align}
\Sigma^A_{\alpha \beta}(E+\omega) - \Sigma^R_{\alpha \beta}(E) &= -\imath \omega (I(E,\omega,q=0) (1- I(E,\omega,q=0) )^{-1})_{\alpha \gamma, \beta \gamma} \\
I_{\alpha \gamma, \beta \delta}(E,\omega,q) &= \Gamma_{\alpha \nu, \beta \xi} \int {dk} G^A_{\nu \gamma}(E+\omega, k+q/2) G^R_{\xi \delta}(E, k-q/2) \\
\Sigma^A_{\alpha \beta}(E) &=\Gamma_{\alpha \gamma,  \beta \delta} \int {dk} \;G^A_{\gamma \delta}(E,k)
\end{align}
Sums are implied when we repeat indices.

In the next subsections discuss (A)  checks of the Ward-Takahashi identity for all hamiltonians $H$ and disorder $\Gamma$, order by order in $\Gamma$. Next we show (B) that when the self-energy $\Sigma$ is proportional to the identity, the Ward-Takahashi identity is always satisfied.  Next we show that (C) the minimal model  (magnetization plus spin-orbit disorder) always satisfies the Ward-Takahashi identity. Lastly we develop (D) the CoPt model (magnetization with spin-orbit coupling) and show how to check the Ward-Takahashi identity order by order.  In an earlier section we have showed that both the minimal model and the CoPt model result in equations which  violate charge conservation when treated with the BNM approach, so the results shown here confirm that SCBA + ladder diagrams is superior.

\subsection{Check of the Ward-Takahashi identity at first and second order in the disorder $\Gamma$}


At first order in $\Gamma$ the Ward-Takahashi identity is:
\begin{align}
 & \sum_{\gamma \delta} \Gamma_{\nu \gamma \delta \eta} \int{d\bfk} ( G^0_{\gamma \delta} (\bfk, E + \omega) - G^0_{\gamma \delta} (\bfk, E) ) \\
 &= \imath \omega \sum_{\gamma \alpha \beta}  {\Gamma}_{\nu \alpha \beta \eta} \int {d\bfk}   G^0_{\alpha \gamma }(\bfk ,  E + \omega) G^0_{ \gamma \beta}(\bfk, E)\\
&= \imath \omega \sum_{ \alpha \beta}  {\Gamma}_{\nu \alpha \beta \eta} \int {d\bfk}  (\frac{1}{-\imath E - \imath \omega + H(\bfk)}  \frac{1}{-\imath E  + H(\bfk)} )_{\alpha \beta} \\
 &=  \sum_{ \alpha \beta}  {\Gamma}_{\nu \alpha \beta \eta} \int {d\bfk}  (\frac{1}{-\imath E - \imath \omega + H(\bfk)}  -\frac{1}{-\imath E  + H(\bfk)} )_{\alpha \beta} \\
 &=  \sum_{ \alpha \beta}  {\Gamma}_{\nu \alpha \beta \eta} \int {d\bfk}  ( G^0_{\alpha \beta }(\bfk ,  E + \omega) -G^0_{ \alpha \beta}(\bfk, E) )
\end{align}
So the Ward-Takahashi identity is satisfied at first order. 

At second order in $\Gamma$ the Ward-Takahashi identity is:
\begin{align}
 & \sum_{\gamma \delta} \Gamma_{\nu \gamma \delta \eta} \int{d\bfk}  ( G^0(\bfk, E + \omega) \Sigma(E + \omega) G^0 (\bfk, E + \omega)- (G^0 (\bfk, E) \Sigma(E) G^0 (\bfk, E))_{\gamma \delta} ) \\
&= \imath \omega \sum_{\gamma \alpha \beta}  {\Gamma}_{\nu \alpha \beta \eta} \int {d\bfk}   G^0_{\alpha \gamma }(\bfk ,  E + \omega) (G^0(\bfk, E) \Sigma(E) G^0(\bfk, E) )_{ \gamma \beta}\\
&+ \imath \omega \sum_{\gamma \alpha \beta}  {\Gamma}_{\nu \alpha \beta \eta} \int {d\bfk}   (G^0(\bfk ,  E + \omega) \Sigma(E + \omega) G^0(\bfk ,  E + \omega))_{\alpha \gamma } G^0_{ \gamma \beta}(\bfk, E) \\
&+  \imath \omega \sum_{\gamma \alpha \beta x_1 y_1 x_2 y_2} {\Gamma}_{\nu x_1  y_1 \eta} \int {d\bfk} {d\bfk'}   G^0_{x_1 x_2}(\bfk' ,  E + \omega) G^0_{ y_2 y_1 }(\bfk', E)  {\Gamma}_{x_2 \alpha \beta y_2 }   G^0_{\alpha \gamma }(\bfk ,  E + \omega) G^0_{ \gamma \beta}(\bfk, E)\\
&=  \sum_{ \alpha \beta}  {\Gamma}_{\nu \alpha \beta \eta} \int {d\bfk}    ((G^0_{\alpha \gamma }(\bfk ,  E + \omega)-G^0(\bfk, E))\Sigma(E) G^0(\bfk, E) )_{ \alpha \beta}\\
&+ \sum_{ \alpha \beta}  {\Gamma}_{\nu \alpha \beta \eta} \int {d\bfk}   (G^0(\bfk ,  E + \omega) \Sigma(E + \omega) (G^0(\bfk ,  E + \omega)-G^0_{ \gamma \beta}(\bfk, E)))_{\alpha \beta }  \\
&+\sum_{ \alpha \beta x_1 y_1 x_2 y_2} {\Gamma}_{\nu x_1  y_1 \eta} \int {d\bfk} {d\bfk'}   G^0_{x_1 x_2}(\bfk' ,  E + \omega) G^0_{ y_2 y_1 }(\bfk', E)  {\Gamma}_{x_2 \alpha \beta y_2 }  ( G^0(\bfk ,  E + \omega) -  G^0(\bfk, E))_{\alpha \beta}\\
&=  \sum_{ \alpha \beta}  {\Gamma}_{\nu \alpha \beta \eta} \int {d\bfk}    ((G^0_{\alpha \gamma }(\bfk ,  E + \omega)-G^0(\bfk, E))\Sigma(E) G^0(\bfk, E) )_{ \alpha \beta}\\
&+ \sum_{ \alpha \beta}  {\Gamma}_{\nu \alpha \beta \eta} \int {d\bfk}   (G^0(\bfk ,  E + \omega) \Sigma(E + \omega) (G^0(\bfk ,  E + \omega)-G^0_{ \gamma \beta}(\bfk, E)))_{\alpha \beta }  \\
&+\sum_{ x_1 y_1} {\Gamma}_{\nu x_1  y_1 \eta} \int {d\bfk'}  ( G^0(\bfk' ,  E + \omega) ( \Sigma( E + \omega) -  \Sigma( E)) G^0(\bfk', E) )_{x_1 y_1}  \\
\end{align}
So the Ward-Takahashi identity is satisfied at second order.

@e believe that similar results can be proved at any order if one uses the strategy described by Vollhardt and Wolfle's PRB 22 4666 Appendix A \cite{PhysRevB.22.4666}. The examples we showed for first and second order already implement that strategy.

\subsection{Charge conservation when $\Sigma$ is proportional to the identity}
Now we look at the simplifying case where $\Sigma$ is proportional to the identity, and prove that probability conservation is obtained trivially.  In this case:
\begin{align}
\Sigma_{\alpha \beta}(E) &= \sum_{\gamma \delta} \Gamma_{\alpha \gamma \delta \beta} \int{d\bfk} G_{\gamma \delta} (\bfk, E)  = 1 \times  \frac{\imath \hbar}{  2 \tau(E)} \\
\overline{\Phi}_{\alpha \nu \eta \beta} & ( \bfq=0, \omega, E) = \int {d\bfk} G_{\alpha \nu }(\bfk  ,  E + \omega) G_{ \eta \beta}(\bfk, E)\\
& = \int {d\bfk} (\frac{1}{-\imath E - \imath \omega + H(\bfk) - \Sigma(E + \omega)})_{\alpha \nu }  (\frac{1}{-\imath E  + H(\bfk) - \Sigma(E )})_{ \eta \beta}\\
\sum_{\gamma} \overline{\Phi}_{\alpha \gamma \gamma \beta} & ( \bfq=0, \omega, E) =  \frac{1}{- \Sigma(E ) +  \Sigma(E+ \omega ) + \imath \omega} \int {d\bfk}( G(\bfk  ,  E + \omega) - G(\bfk, E))_{ \alpha \beta}\\
\sum_{\gamma} I_{\zeta\gamma \gamma \delta}(\bfq=0, E, \omega) &= \sum_{\alpha \beta \gamma}   {\Gamma}_{\zeta \alpha \beta \delta}   \overline{\Phi}_{\alpha \gamma \gamma \beta} (\bfq=0, \omega, E) = \frac{- \Sigma(E ) +  \Sigma(E+ \omega ) }{- \Sigma(E ) +  \Sigma(E+ \omega ) + \imath \omega} \\
\sum_{\gamma} I_{\zeta\gamma \gamma \delta}(\bfq=0, E, \omega=0) & =  \delta_{\zeta \delta}
\end{align}
The last line proves that the operator $1-I$, the diffusion kernel proposed by Burkov Nunez and Macdonald, conserves charge when $\Sigma$ is proportional to the identity.  However it only proves charge conservation when the charge operator acts on one side of $I$, not when it acts on the other side of $I$. We can expect that if $\Gamma$ has spin structure then as a general rule $I$ will conserve charge on only one side.

 In the last line we have assumed that $0 \neq \lim_{\omega \rightarrow 0 } - \Sigma(E ) +  \Sigma(E+ \omega )$, which is reasonable because $\Sigma(E )$ is analytically continued to $\Sigma^R$ and $ \Sigma(E+ \omega )$ is analytically continued to $\Sigma^A$.

Continuing through with the full Ward-Takahashi identity in the case where $\Sigma$ is proportional to the identity:
\begin{align}
& \Sigma_{\nu \eta}(E + \omega) - \Sigma_{\nu \eta}(E) = 
\imath \omega  \sum_{\gamma} \int{d\bfk} \Lambda_{\nu \gamma \gamma \eta}(\bfk, \bfq=0, \omega, E) \\
& = \imath \omega(1 - \hat{I})^{-1} \hat{I}=\imath \omega (\hat{I}^{-1} - 1)^{-1}, \; \hat{I} =  \frac{- \Sigma(E ) +  \Sigma(E+ \omega ) }{- \Sigma(E ) +  \Sigma(E+ \omega ) + \imath \omega} \\
& = \imath \omega (\frac{- \Sigma(E ) +  \Sigma(E+ \omega ) + \imath \omega} {- \Sigma(E ) +  \Sigma(E+ \omega ) } - 1)^{-1}  = - \Sigma(E ) +  \Sigma(E+ \omega )\\
\end{align}
Q.E.D.

 \subsection{Minimal Model: Magnetization Plus Spin-Orbit Disorder }

We already discussed this model and showed that when treated with the BNM approach the resulting spin diffusion equations do not conserve charge.  Here we show that using the SCBA plus ladder diagrams in the correct way one does obtain charge conservation.

In the minimal model there are two orbitals (spin up and spin down), $H$ is the Hamiltonian, $n$ is the density of states, $\Gamma$ is the scattering vertex, $\Sigma$ is the self-energy. This minimal model has a $\hat{z}$ Zeeman splitting between its energy levels which causes in-plane precession of the spin, has minimal dependence on $\vec{k}$, has a density of states which is larger for spin up than for spin down (the asymmetry is controlled by $n_z$), and has a spin-conserving disorder $\Gamma_0$ and a spin-orbit disorder $\Gamma_x \sigma_x \otimes \sigma_x$. $\Gamma_0$ is not necessary for obtaining the probability non conservation.

\begin{align}
H &= k^2/2m + b_z \sigma_z, \Gamma = \Gamma_1 1\otimes1 + \Gamma_x \sigma_x \otimes \sigma_x, \Sigma = \Sigma_1 + \Sigma_z \sigma_z \\
H  + \Sigma & =  H_0(k^2)  + \Sigma^{A,R}_1 + (b_z + \Sigma^{A,R}_z) \sigma_z \\
E^{A,R}(k, s) & = H_0(k^2) + \Sigma^{A,R}_1 +  s (b_z + \Sigma^{A,R}_z), \\
{G}^{A,R}(\vec{k}, E) &= \frac{1}{2}\sum_s \frac{ 1 + s \sigma_z
}  {E -E^{A,R}(s,\vec{k}) \mp \imath \epsilon}
\end{align}

We use the SCBA/CPA equation to calculate $\Sigma$:
\begin{align}
\Sigma^{A,R}_{\alpha \beta}(E) &= \Gamma_{\alpha \nu, \xi \beta} \int {dk} \;G^{A,R}_{\nu \xi}(E,k), \; \\
\Sigma^{A,R}(E) &=( \Gamma_1 - \Gamma_z) \frac{1}{2}  \sum_s \int {dk}  \frac{1}  {E -E^{A,R}(s,\vec{k}) \mp \imath \epsilon} + ( \Gamma_1 + \Gamma_z) \sigma_z  \frac{1}{2}  \sum_s \int {dk} \frac{s}  {E -E^{A,R}(s,\vec{k}) \mp \imath \epsilon}\\
&=( \Gamma_1 + \Gamma_z) \frac{1}{2}  \sum_s   F(E- \Sigma^{A,R}_1 -  s (b_z + \Sigma^{A,R}_z) \mp \imath \epsilon)  + ( \Gamma_1 - \Gamma_z) \sigma_z  \frac{1}{2}  \sum_s  s F(E- \Sigma^{A,R}_1 -  s (b_z + \Sigma^{A,R}_z) \mp \imath \epsilon)\\
\Sigma_s^{A,R}(E) &= \Gamma_1   F(E- \Sigma^{A,R}_s -  s b_z \mp \imath \epsilon)  + \Gamma_z  F(E- \Sigma^{A,R}_{-s} + s b_z  \mp \imath \epsilon)\\
F(u) &= \int {dk} \frac{1} {u - H_0(k^2)} = - 2\pi m \ln ( \frac{u - \Lambda/2m}{u})\\
\end{align}

For ordinary disorder, the collision integral is:
\begin{align}
I_{\alpha \gamma, \beta \delta}(E,\omega,q) &=  \gamma \int {dk} G^A_{\alpha \gamma}(E+\omega/2, k+q/2) G^R_{\beta \delta}(E-\omega/2, k-q/2) \\
{G}^{A,R}(\vec{k}, E) &= \frac{1}{2}\sum_s \frac{ 1 + s \sigma_z}  {E -E^{A,R}(s,\vec{k}) \mp \imath \epsilon}
\end{align}

The $\sigma_x$ disorder  just adds a $\sigma_x$ after (or before?) each of the Green's functions, so for our model we use the equations below.
At this point we introduce  $t$, which is an extra sign  found in the HLN paper. $t=+1$ if the $\Gamma_x$ scattering comes from spin-orbit impurities, and $t=-1$ if it comes from local magnetizations - see Hikami-Larkin-Nagaoka's original paper \cite{hikami1980spin}.

 \begin{align}
 I(E,\omega,q)  &=  \frac{1}{4} \sum_{s,\acute{s}} \left[ \Gamma_1 (1 + s\sigma_z) \otimes (1 + \acute{s} \sigma_z)  + t \Gamma_z (1 + s\sigma_z) \sigma_x \otimes (1 + \acute{s} \sigma_z) \sigma_x  \right] \\
 & \times \int {dk} \frac{ 1 }  {E +\omega/2-E^{A}(s,k+q/2) - \imath \epsilon} \frac{ 1 }  {E -\omega/2-E^{R}(\acute{s},k-q/2) + \imath \epsilon} \\
 I(E,\omega,q)  &=  \frac{1}{4} \sum_{s,\acute{s}} \left[ \Gamma_1 (1 + s\sigma_z) \otimes (1 + \acute{s} \sigma_z)  + t \Gamma_z (\sigma_x + s \imath \sigma_y)  \otimes (\sigma_x  + \acute{s} \imath \sigma_y)   \right] \\
 & \times \int {dk} \left[ \frac{ 1 }  {E +\omega/2-E^{A}(s,k+q/2) - \imath \epsilon} - \frac{ 1 }  {E -\omega/2-E^{R}(\acute{s},k-q/2) + \imath \epsilon}  \right]  \\
 & \times \frac{1}{-\omega - E^{R}(\acute{s},k-q/2) + E^{A}(s,k+q/2)} \\
E^{A,R}(k, s) & = H_0(k^2) + \Sigma^{A,R}_1 +  s (b_z + \Sigma^{A,R}_z), \\
\Sigma_1^{A,R} (E)&=( \Gamma_1 + \Gamma_z)  \frac{1}{2}  \sum_s F( E - \Sigma_1^{A,R} -s b_z - s\Sigma_z^{A,R} \mp \imath \epsilon) \\
\Sigma_z^{A,R}(E)&= ( \Gamma_1 - \Gamma_z)   \frac{1}{2}\sum_s  s F  (E - \Sigma_1^{A,R} -s b_z - s\Sigma_z^{A,R} \mp \imath \epsilon)\\
F(u) &= \int {dk} \frac{1} {u - H_0(k^2)} = - 2\pi m \ln ( \frac{u - \Lambda/2m}{u})\\
\end{align}
 
 To work toward the Ward-Takahashi identity, we set $q=0$:
 \begin{align}
  I(E,\omega,q=0)  &=  \frac{1}{4} \sum_{s,\acute{s}} \left[ \Gamma_1 (1 + s\sigma_z) \otimes (1 + \acute{s} \sigma_z)  + t \Gamma_z (\sigma_x + s \imath \sigma_y)  \otimes (\sigma_x  + \acute{s} \imath \sigma_y)   \right] \\
 & \times \int {dk} \left[ \frac{ 1 }  {E +\omega/2-E^{A}(s,k) - \imath \epsilon} - \frac{ 1 }  {E -\omega/2-E^{R}(\acute{s},k) + \imath \epsilon}  \right]  \\
 & \times \frac{1}{-\omega - E^{R}(\acute{s},k) + E^{A}(s,k)} \\
 I(E,\omega,q)  &=  \frac{1}{4} \sum_{s,\acute{s}} \left[ \Gamma_1 (1 + s\sigma_z) \otimes (1 + \acute{s} \sigma_z)  + t \Gamma_z (\sigma_x + s \imath \sigma_y)  \otimes (\sigma_x  + \acute{s} \imath \sigma_y)   \right] \\
 & \times \frac{-F( E-\omega/2 - \Sigma_{\acute{s}}^{R} - \acute{s}b_z  + \imath \epsilon)+F( E +\omega/2- \Sigma_s^{A} - sb_z  - \imath \epsilon)}{-\omega - \Sigma^{R}_{\acute{s}}(E-\omega/2) + \Sigma^{A}_s(E+\omega/2)} \\
\end{align}

Analyzing the matrix elements in the numerator gives:
 \begin{align}
 ++ \rightarrow ++ &= \Gamma_1 \frac{1}{4} \sum_{s,\acute{s}}(1 + s)(1+ \acute{s}) \\
 ++ \rightarrow -- &= \Gamma_z \frac{1}{4} \sum_{s,\acute{s}} (1 + s)(1+ \acute{s})\\
 -- \rightarrow ++ &= \Gamma_z  \frac{1}{4} \sum_{s,\acute{s}}(1 - s)(1- \acute{s})\\
 -- \rightarrow -- &= \Gamma_1 \frac{1}{4} \sum_{s,\acute{s}}(1 - s)(1- \acute{s})
 \end{align}
 
 Converting to a different basis the identity and $\sigma_z$:
 \begin{align}
 1 \rightarrow 1 &= (\Gamma_1 + t\Gamma_z) \frac{1}{4} \sum_{s,\acute{s}} (1 +s  \acute{s}) =  (\Gamma_1 + t\Gamma_z) \frac{1}{2} \sum_{s = \acute{s}}  \\
 1 \rightarrow z &= (\Gamma_1 - t\Gamma_z)\frac{1}{4} \sum_{s,\acute{s}}  (1 +s  \acute{s})= (\Gamma_1 - t\Gamma_z)\frac{1}{2} \sum_{s=\acute{s}}\\
 z \rightarrow 1 &= (\Gamma_1 + t\Gamma_z) \frac{1}{4} \sum_{s,\acute{s}}(s +\acute{s}) =  (\Gamma_1 + t\Gamma_z) \frac{1}{2} \sum_{s=\acute{s}}s\\
 z \rightarrow z &= (\Gamma_1 - t\Gamma_z)\frac{1}{4} \sum_{s,\acute{s}} (s+  \acute{s})  = (\Gamma_1 - t\Gamma_z)\frac{1}{2} \sum_{s=\acute{s}} s \\
 \end{align}

There is also a $S_x, S_y$ sector, but this sector is decoupled from the $1, S_z$ sector and doesn't affect probability conservation.

In the $1, S_z$ sector we have $s= \acute{s}$, and $I$ simplifies to:
\begin{align}
 I(E,\omega,q=0,s=\acute{s}=1)  &=    \left[ \Gamma_1 ( ++ \rightarrow ++ ) +  \Gamma_z ( ++ \rightarrow --) \right]  d_+\\
 I(E,\omega,q=0,s=\acute{s}=-1)  &=     \left[ \Gamma_1 (-- \rightarrow ++ ) +  \Gamma_z ( -- \rightarrow --) \right] d_- \\
d_s &=  \frac{F(E+\omega/2 -\Sigma^A_s -sb_z -\imath \epsilon)-F(E-\omega/2-\Sigma^R_s - s b_z +\imath \epsilon) } {-\omega  + \Sigma^A_s(E+\omega/2) - \Sigma^R_s(E-\omega/2) } \\
\end{align}

In summary (and we can absorb the $(-2\pi m)$ factors into $\Gamma_1, \Gamma_z$, giving $\Gamma$ units of energy):
\begin{align}
  I(E,\omega,q=0)  &= \begin{bmatrix} d_+ & 0 \\ 0 & d_- \end{bmatrix}  \begin{bmatrix} \Gamma_1 & \Gamma_z \\ \Gamma_z & \Gamma_1 \end{bmatrix} (-2 \pi m) \\
 d_s &=(- 2\pi m )^{-1}  \int {dk} \frac{ 1 }  {E +\omega/2-\Sigma^A_s -sb_z -H_0(k^2)- \imath \epsilon} \frac{ 1 }  {E -\omega/2-\Sigma^R_s - s b_z-H_0(k^2) + \imath \epsilon} \\
 d_s &=  \frac{F(E+\omega/2 -\Sigma^A_s -sb_z -\imath \epsilon)-F(E-\omega/2-\Sigma^R_s - s b_z +\imath \epsilon) } {-\omega  + \Sigma^A_s(E+\omega/2) - \Sigma^R_s(E-\omega/2) } \\
\Sigma_s^{A,R}(E) &=(- 2\pi m ) \Gamma_1   F(E- \Sigma^{A,R}_s +  s b_z \mp \imath \epsilon)  + (- 2\pi m )\Gamma_z  F(E- \Sigma^{A,R}_{-s} -  s b_z  \mp \imath \epsilon)\\
F(u) &=(- 2\pi m )^{-1} \int {dk} \frac{1} {u - H_0(k^2)} = \ln ( \frac{u - U_C}{u}), \; U_C = \Lambda/2m\\
\end{align}
The susceptibilities are:
\begin{align}
\chi_0 &= (-2 \pi m)^{-1} \begin{bmatrix} d_+ & 0 \\ 0 & d_- \end{bmatrix} \\
\chi_{diff} &= I (1- I)^{-1} \chi_0
\end{align}

Note that $1-I$ does not conserve charge, i.e. it is not zero when multiplied by either $\begin{bmatrix} 1 & 1 \end{bmatrix}$ (on the left) or $\begin{bmatrix} 1 \\ 1 \end{bmatrix}$ (on the right).  This shows again that the BNM approach for calculating spin diffusion equations does not satisfy charge conservation.

The Ward-Takahashi identity is:
\begin{align}
&\begin{bmatrix} \Sigma^A_+(E+\omega/2) - \Sigma^R_+(E-\omega/2)  \\ \Sigma^A_-(E+\omega/2) - \Sigma^R_-(E-\omega/2)\end{bmatrix} = -\omega \begin{bmatrix} 1 & 1 \end{bmatrix} I (1- I)^{-1}  
\end{align}


  We check the Ward-Takahashi identity when $\Gamma_z = 0$:
  \begin{align}
  I_s  &= \frac{\Gamma_1 F(E+\omega/2 -\Sigma^A_s -sb_z -\imath \epsilon)- \Gamma_1 F(E-\omega/2-\Sigma^R_s - s b_z +\imath \epsilon) } {-\omega  + \Gamma_1 F(E+\omega/2 -\Sigma^A_s -sb_z -\imath \epsilon)- \Gamma_1 F(E-\omega/2-\Sigma^R_s - s b_z +\imath \epsilon) } \\
 1 -  I_s  &=-\omega  \frac{1 } {-\omega  + \Gamma_1 F(E+\omega/2 -\Sigma^A_s -sb_z -\imath \epsilon)- \Gamma_1 F(E-\omega/2-\Sigma^R_s - s b_z +\imath \epsilon) } \\
 & -\omega  \begin{bmatrix} 1 & 1 \end{bmatrix}  I (1- I)^{-1}   =    \begin{bmatrix} 1 & 1 \end{bmatrix}  \begin{bmatrix} \Sigma^A_+(E+\omega/2) - \Sigma^R_+(E-\omega/2) & 0 \\ 0 &  \Sigma^A_-(E+\omega/2) - \Sigma^R_-(E-\omega/2)  \end{bmatrix} \\
 & = \begin{bmatrix}\Sigma^A_+(E+\omega/2) - \Sigma^R_+(E-\omega/2)\\ \Sigma^A_-(E+\omega/2) - \Sigma^R_-(E-\omega/2) \end{bmatrix} 
  \end{align}
  So the Ward-Takahashi identity is satisfied when $\Gamma_z = 0$.
  
    We check the Ward-Takahashi identity when $\Gamma_1 = 0$:
\begin{align}
   I(E,\omega,q=0)  &=   \begin{bmatrix} 0 & e_+ \\ e_- & 0 \end{bmatrix} , \; I^2 = e_+ e_-\begin{bmatrix}  1 & 0 \\ 0 & 1 \end{bmatrix} , \; I^3 =  e_+ e_- I  \\
 e_s &=  \frac{ \Sigma^A_s(E+\omega/2) - \Sigma^R_s(E-\omega/2) } {-\omega  + \Sigma^A_{-s}(E+\omega/2) - \Sigma^R_{-s}(E-\omega/2) } \\
  I - I^2 &= \begin{bmatrix} -e_+ e_- & e_+ \\ e_- & -e_+ e_- \end{bmatrix} , \; I(1-I)^{-1} = (1 - e_+ e_-)^{-1} \begin{bmatrix} -e_+ e_- & e_+ \\ e_- & -e_+ e_- \end{bmatrix}  \\
&\begin{bmatrix} \Sigma^A_+(E+\omega/2) - \Sigma^R_+(E-\omega/2)  \\ \Sigma^A_-(E+\omega/2) - \Sigma^R_-(E-\omega/2)\end{bmatrix}  =  -\omega  \begin{bmatrix} 1 & 1 \end{bmatrix}  I (1- I)^{-1}  
\end{align}
  So the Ward-Takahashi identity is satisfied when $\Gamma_1 = 0$.
  
  Moreover, using mathematica, we have checked that the Ward-Takahashi identity works for all values of $\Gamma_1, \Gamma_z$. 
  So the Ward-Takahashi identity is satisfied for all $\Gamma_1, \Gamma_z$.
  
  Investigating $1 - I^{-1}$, we find that at $\omega = 0$ one of its eigenvalues is zero.  The matrix is not symmetric.  The right eigenvector corresponding to the zero eigenvalue has a simple form.
  
  We think that a spin diffusion equation does not have to be symmetric, but perhaps its $q=0, \omega=0$ component, i.e. its lifetime matrix, should be symmetric.

\subsection{CoPt Model: Magnetization with Spin-Orbit Coupling}

This model can also be used to study topological insulators with magnetization, by setting the $k^2/2m$ term in the Hamiltonian and the eigenvalues to zero. 

In the following we sometimes work with the magnetic field $\vec{b}$  pointing in any direction, but we also sometimes specialize to  perpendicular magnetic field $\vec{b} = b_z \hat{z}$ because this is simpler.  This model includes a spin-orbit interaction $\alpha (k_y \sigma_x - k_x \sigma_y)$.  The density of states $\hat{\rho}$ and the self-energy $ \Sigma$ are not proportional to the identity.   Significant spin-polarization occurs only near the bottom of the dispersion, where only one of the two spin-split states coincides with the Fermi energy.

The self-energy is:
\begin{align}
\Sigma^A &= \vec{\Sigma}^A \cdot \sigma\\
\Sigma^R &=  \vec{\Sigma}^R \cdot \sigma
\end{align}
We are violating convention here since we have not written the $\imath $ in $\Sigma$ explicitly.  In general $\Sigma$ is expected to have an important imaginary part, and the real part is usually neglected.  
Also note that if $b_x = b_y=0$, then also $\Sigma_x = \Sigma_y=0$.

The Hamiltonian plus self-energy is:
\begin{align}
H^{A,R} &= H_0(k^2) + \Delta_{SO} (k_y \sigma_x - k_x \sigma_y) +  (\vec{b}+\vec{\Sigma}^{A,R}) \cdot \vec{\sigma} \\
 &= H_0(k^2)  + (\vec{b}+\vec{\Sigma})  \cdot \vec{\sigma} + \Delta_{SO} k \begin{bmatrix}0& \imath e^{-\imath \phi_k} \\  -\imath e^{\imath  \phi_k} & 0\end{bmatrix} \\
&=H_0(k^2)  + (\vec{b}+\vec{\Sigma})  \cdot \vec{\sigma}  + \Delta_{SO} \begin{bmatrix}0& (k_y+\imath k_x) \\  (k_y-\imath k_x) & 0\end{bmatrix}\\
\end{align}

Since $\vec{b} $ always occurs in combination with $\vec{\Sigma}^{A,R}$, we write the combination as $\vec{B}^{A,R} = \vec{b} + \vec{\Sigma}^{A,R}$.

The energies are as follows. Note that if there is a self energy and it has an imaginary part, then these energies are, in principal, complex.
\begin{align}
 E^{A,R}(s,\vec{k}) &=  H_0(k^2)+\Sigma_1^{A,R}  + s\sqrt{(\Delta_{SO} k_y + B_x^{A,R})^2 + (-\Delta_{SO} k_x + B_y^{A,R})^2 +( B_z^{A,R})^2}\\
 E^{A,R,z}(s,\vec{k}) &= H_0(k^2) +\Sigma_1^{A,R} + s\sqrt{(\Delta_{SO} |k|)^2 + (B_z^{A,R})^2}\\
\end{align}

The Green's functions are:
\begin{align}
 G^{A,R}(\vec{k}, E) &= \frac{1}{2}\sum_s \frac{1 + s \vec{X}^{A,R}  (\vec{k} ) \cdot \vec{\sigma} } {E -E^{A,R}(s,\vec{k}) \mp \imath \epsilon},  \; \vec{\sigma} = \left[\sigma_x, \,\sigma_y,\,\sigma_z\right]^T \\
  X_1^{A,R} &= \frac{\Delta_{SO} k_y + B_x^{A,R}}{ \sqrt{(\Delta_{SO} k_y + B_x^{A,R})^2 + (-\Delta_{SO} k_x + B_y^{A,R})^2 + (B_z^{A,R})^2}},  \; X_1^z = \cos \theta^{A,R}_B \sin \theta_k \\
   X_2^{A,R} &= \frac{-\Delta_{SO} k_x + B_y^{A,R}}{ \sqrt{(\Delta_{SO} k_y + B_x^{A,R})^2 + (-\Delta_{SO} k_x + B_y^{A,R})^2 + (B_z^{A,R})^2}},  \; X_2^z = -\cos \theta^{A,R}_B \cos \theta_k \\
   X_3^{A,R} &= \frac{B_z^{A,R}}{ \sqrt{(\Delta_{SO} k_y + B_x^{A,R})^2 + (-\Delta_{SO} k_x + B_y^{A,R})^2 + (B_z^{A,R})^2}} , \; X_3^z =  \sin \theta_B^{A,R} \\
\cos \theta_B^{A,R} &=   (\Delta_{SO} |k| / \sqrt{\Delta_{SO}^2 |k|^2 + (B^{A,R})^2}), \; \sin \theta_B^{A,R} = (B^{A,R} / \sqrt{\Delta_{SO}^2 |k|^2 + (B^{A,R})^2}) 
 \end{align}

The above formula for the Green's functions is based on the following inversion formula: 
 \begin{align}
&\; (E - (k^2/2m + \Delta_{SO} (k_x \sigma_x + k_y \sigma_y) + B_z \sigma_z ))^{-1} = \frac{1}
{2}\sum_s \frac{1 + s \vec{X}  (\vec{k} ) \cdot \vec{\sigma} } {E -E(s,\vec{k}) } \\
\end{align}

Now we introduce the scattering tensor, which describes nonmagnetic white noise disorder:
\begin{align}
\Gamma_{\alpha \nu, \beta \xi} &= \gamma  \delta_{\alpha \nu} \delta_{\beta \xi} \\
\end{align}

We compute the self-energy with the SCBA/CPA:
\begin{align}
\Sigma^{A,R}_{\alpha \beta}(E) &= \Gamma_{\alpha \nu, \xi \beta} \int {dk} \;G^{A,R}_{\nu \xi}(E,k)= \gamma \int {dk} \;G^{A,R}_{ \alpha \beta}(E,k), \; \vec{\Sigma}^{A,R} = \gamma \int {dk} \;\vec{G}^{A,R}\\
\end{align}
If $\vec{b} = b_z \hat{z}$ then it is trivial to prove that $\Sigma_x = \Sigma_y = 0$, and the remaining integrals have rotational invariance resulting in one dimensional integrals:
\begin{align}
\vec{\Sigma}^{A,R,z}(E) &=  \gamma  \;\frac{1}{2}\sum_s \int {dk} \frac{ 1 +  s \sin \theta^{A,R}_B (E) \sigma_z}  {E -H_0(k^2) -\Sigma_1^{A,R}(E) - s\sqrt{(\Delta_{SO} |k|)^2 + (b_z + \Sigma_z^{A,R}(E))^2} \mp \imath \epsilon},  \\
{\Sigma}^{A,R}_{\hat{s}}(E) &=  \gamma  \;\frac{1}{2}\sum_s \int {dk} \frac{ 1 +  \hat{s} s \sin \theta^{A,R}_B (E) }  {E -H_0(k^2) -\Sigma_1^{A,R}(E) - s\sqrt{(\Delta_{SO} |k|)^2 + (b_z + \Sigma_z^{A,R}(E))^2} \mp \imath \epsilon},  \\
\sin \theta_B^{A,R}(E) &= \frac{b_z + \Sigma_z^{A,R}(E)}{\sqrt{(\Delta_{SO} |k|)^2 + (b_z + \Sigma_z^{A,R}(E))^2}}  \\
\end{align}

We begin calculating the collision integral $I$:
\begin{align}
I(E,\omega,q=0) &=  \gamma \int {dk} G^A(E+\omega/2, k) \otimes G^R(E-\omega/2, k) \\
 G^{A,R}(E) &= \;\frac{1}{2}\sum_s  \frac{ 1 + s \cos \theta^{A,R}_B(E) (\sin \theta_k \sigma_x - \cos \theta_k \sigma_y) +  s \sin \theta^{A,R}_B (E) \sigma_z}  {E -H_0(k^2) -\Sigma_1^{A,R}(E) - s\sqrt{(\Delta_{SO} |k|)^2 + (b_z + \Sigma_z^{A,R}(E))^2} \mp \imath \epsilon},  \\
\sin \theta_B^{A,R}(E) &= \frac{b_z + \Sigma_z^{A,R}(E)}{\sqrt{(\Delta_{SO} |k|)^2 + (b_z + \Sigma_z^{A,R}(E))^2}}  \\
\end{align}

We now resolve $I$ into its $1,S_x,S_y,S_z$ components by tracing over Pauli matrices.
After doing the angular integral, the $S_x,S_y$ components decouple from the $1,S_z$ components. We don't evalulate the $S_x,S_y$ components here because they do not affect the Ward-Takahashi identity.

Resolved into   $1,S_z$ components, after performing the angular average, the numerator of $I$ is:
\begin{align}
\mathcal{I} &=  \frac{\gamma}{4} \sum_{s,\acute{s}} 
\begin{bmatrix}
1 + s \acute{s} \cos(\theta_B^A(E+\omega/2) - \theta_B^R(E-\omega/2)) & s \sin \theta_B^A(E+\omega/2) + \acute{s} \sin \theta_B^R(E-\omega/2) \\
s \sin \theta_B^A(E+\omega/2) + \acute{s} \sin \theta_B^R(E-\omega/2) &  1 - s\acute{s} \cos(\theta_B^A(E+\omega/2) + \theta_B^R(E-\omega/2))
\end{bmatrix} \\
\end{align}

When $\Sigma = 0$, we obtain $\theta_B^A = \theta_B^R$, and the numerator simplifies enormously.  When $s=\acute{s}$, the numerator is 
\begin{align}
\mathcal{I}  &= \frac{\gamma}{2} \sum_{s}  \begin{bmatrix} 
1 & s \sin \theta_B \\ s  \sin \theta_B &  \sin^2 \theta_B
\end{bmatrix} \\
\end{align}

When $s = -\acute{s}$, the numerator is 
\begin{align}
\mathcal{I} &= \frac{\gamma}{2} \sum_{s}  \begin{bmatrix}
0 & 0 \\ 0 &  \cos^2 \theta_B
\end{bmatrix} \\
\end{align}
The fact that there is no coupling from charge to spin when $s= -\acute{s}$ is crucial for obtaining the Ward-Takahashi identity.

\begin{align}
I(E,\omega, q=0) &= \int {dk} \frac{\mathcal{I} }{E+\omega/2 -H_0(k^2) -\Sigma_1^{A}(E+\omega/2) - s\sqrt{(\Delta_{SO} |k|)^2 + (b_z + \Sigma_z^{A}(E+\omega/2))^2} - \imath \epsilon}\\
&\times \frac{1}{E -\omega/2-H_0(k^2) -\Sigma_1^{R}(E-\omega/2) - \acute{s}\sqrt{(\Delta_{SO} |k|)^2 + (b_z + \Sigma_z^{R}(E-\omega/2))^2} + \imath \epsilon} 
\end{align}

The Ward-Takahashi Identity is:
\begin{align}
&
\begin{bmatrix} 
\Sigma^A_1(E+\omega/2)  -  \Sigma^R_1(E-\omega/2)  \\
 \Sigma^A_z(E+\omega/2) -  \Sigma^R_z(E-\omega/2) 
\end{bmatrix}
= -\omega \begin{bmatrix} 1 &0 \end{bmatrix} I(E,\omega,q=0) (1-I(E,\omega,q=0))^{-1} \\
{\Sigma}^{A,R}(E) &= \frac{ \gamma  }{2}\sum_s \int {dk} \frac{ 1 +  s \sigma_z \sin \theta^{A,R}_B (E) }  {E -H_0(k^2) -\Sigma_1^{A,R}(E) - s\sqrt{(\Delta_{SO} |k|)^2 + (b_z + \Sigma_z^{A,R}(E))^2} \mp \imath \epsilon},  \\
\sin \theta_B^{A,R}(E) &= \frac{b_z + \Sigma_z^{A,R}(E)}{\sqrt{(\Delta_{SO} |k|)^2 + (b_z + \Sigma_z^{A,R}(E))^2}}  \\
&\begin{bmatrix} \Sigma^{A,R}_1 \\ \Sigma^{A,R}_z \end{bmatrix} =
\begin{bmatrix} 
\frac{1}{2} {Tr}( \Sigma^{A,R}(E)  ) \\
\frac{1}{2} {Tr}( \Sigma^{A,R}(E) \sigma_z  ) 
\end{bmatrix}\\
I(E,\omega, q=0) &= \int {dk} \frac{\mathcal{I} }{E+\omega/2 -H_0(k^2) -\Sigma_1^{A}(E+\omega/2) - s\sqrt{(\Delta_{SO} |k|)^2 + (b_z + \Sigma_z^{A}(E+\omega/2))^2} - \imath \epsilon}\\
&\times \frac{1}{E -\omega/2-H_0(k^2) -\Sigma_1^{R}(E-\omega/2) - \acute{s}\sqrt{(\Delta_{SO} |k|)^2 + (b_z + \Sigma_z^{R}(E-\omega/2))^2} + \imath \epsilon} \\
\mathcal{I} &=  \frac{\gamma}{4} \sum_{s,\acute{s}} 
\begin{bmatrix}
1 + s \acute{s} \cos(\theta_B^A(E+\omega/2) - \theta_B^R(E-\omega/2)) & s \sin \theta_B^A(E+\omega/2) + \acute{s} \sin \theta_B^R(E-\omega/2) \\
s \sin \theta_B^A(E+\omega/2) + \acute{s} \sin \theta_B^R(E-\omega/2) &  1 - s\acute{s} \cos(\theta_B^A(E+\omega/2) + \theta_B^R(E-\omega/2))
\end{bmatrix} \\
\chi_0 &= \gamma^{-1} I \\
\chi_{diff}  &= \gamma^{-1} I (1-I)^{-1} I 
\end{align}
These equations are ready to evaluate numerically.  The integrals do not have to be done right, as long as the integral in the self-energy matches the integral in $I$.  More precisely, when $s= \acute{s}$ in $I$ and $\Sigma = 0$ in $I$, then $-\omega I $ should be equal to the zeroth-order approximation to $\Sigma^A(E) - \Sigma^R(E)$.  In particular, probably one could make a Fermi-surface approximation, though some care would have to be given to choosing $\tau$ for each value of $s$, and there would be some ambiguity about interpreting this choice.  There is no numerical reason to make this approximation, it's not any trouble to perform the integration, but in terms of physical interpretation it would allow some interpretation of the total result in terms of a density of states and a lifetime.
 
 It is required that that the self-energy calculation be taken all the way to self-consistency, and that the calculation of $I$ use the correct self-energy.   We also think that this means that it is not OK to approximate the difference in energies in the two denominators as $\omega + \imath / \tau$ - you have to use the correct self energies, and the correct $s, \acute{s}$.
 
 At leading order in $\Gamma$ the Ward-Takahashi Identity is:
\begin{align}
 & -\omega I(E,\omega, q=0) \\
& =\int {dk} \; -\omega \gamma G_0^A(E+\omega/2) G_0^R(E-\omega/2) = \\
 &= \int {dk} \; \frac{-\omega \gamma}{2}\sum_s  \frac{ 1 + s \cos \theta_B (\sin \theta_k \sigma_x - \cos \theta_k \sigma_y) +  s \sin \theta_B  \sigma_z}  { (E +\omega/2 -H_0(k^2)  - s\sqrt{(\Delta_{SO} |k|)^2 + b_z^2} - \imath \epsilon)(E -\omega/2 -H_0(k^2)  - s\sqrt{(\Delta_{SO} |k|)^2 + b_z^2 + \imath \epsilon} + \imath \epsilon)}  \\
 &= \int {dk} \; \frac{\gamma}{2}\sum_s (1 + s \cos \theta_B (\sin \theta_k \sigma_x - \cos \theta_k \sigma_y) +  s \sin \theta_B  \sigma_z) \\
 & \times \frac{ 1}  { E +\omega/2 -H_0(k^2)  - s\sqrt{(\Delta_{SO} |k|)^2 + b_z^2} - \imath \epsilon} \frac{1} {E -\omega/2 -H_0(k^2)  - s\sqrt{(\Delta_{SO} |k|)^2 + b_z^2 + \imath \epsilon} + \imath \epsilon}  \\
 & = \int {dk} \;  \gamma (G_0^A(E+\omega/2,k) - G_0^R(E-\omega/2,k)) \\
& = \Sigma^A(E+\omega/2) - \Sigma^R(E-\omega/2) \\
\sin \theta_B &= \frac{b_z}{\sqrt{(\Delta_{SO} |k|)^2 + b_z^2}}  \\
\end{align}

At second order in $\gamma$  the Ward-Takahashi identity is again satisfied:
\begin{align}
 & -\omega I(E,\omega, q=0) (1 -  I(E,\omega, q=0))^{-1} \\
& =\int {dk} \; -\omega \gamma G_0^A(E+\omega/2,k) \Sigma^A(E+\omega/2) G_0^A(E+\omega/2,k) G_0^R(E-\omega/2,k)  \\
& +\int {dk} \; -\omega \gamma G_0^A(E+\omega/2,k) G_0^R(E-\omega/2,k) \Sigma^R(E-\omega/2) G_0^R(E-\omega/2,k)  \\
& +\int {dk} {d\acute{k}} \; -\omega \gamma^2 G_0^A(E+\omega/2,\acute{k})  G_0^A(E+\omega/2,k) G_0^R(E-\omega/2,\acute{k})  G_0^R(E-\omega/2,k) \\
&=\int {dk} {d\acute{k}} \;  \gamma^2 G_0^A(E+\omega/2,k) G_0^A(E+\omega/2,\acute{k}) (G_0^A(E+\omega/2,k) -G_0^R(E-\omega/2,k)  ) \\
& +\int {dk} {d\acute{k}} \;  \gamma^2 (G_0^A(E+\omega/2,k) - G_0^R(E-\omega/2,k)) G_0^R(E-\omega/2,\acute{k}) G_0^R(E-\omega/2,k)  \\
& +\int {dk} {d\acute{k}} \;  \gamma^2 G_0^A(E+\omega/2,\acute{k})  (G_0^A(E+\omega/2,k)- G_0^R(E-\omega/2,k))  G_0^R(E-\omega/2, \acute{k})  \\
&=\int {dk} {d\acute{k}} \;  \gamma^2( G_0^A(E+\omega/2,k) G_0^A(E+\omega/2,\acute{k}) G_0^A(E+\omega/2,k)   - G_0^R(E+\omega/2,k) G_0^R(E+\omega/2,\acute{k}) G_0^R(E+\omega/2,k)) \\
& = \Sigma^A(E+\omega/2) - \Sigma^R(E-\omega/2) 
\end{align}

We believe that similar results can be proved at any order if one uses the strategy described by Vollhardt and Wolfle's PRB 22 4666 Appendix A \cite{PhysRevB.22.4666}. The examples we showed for first and second order already implement that strategy.


 \section{Spin Diffusion Equations for the CoPt Model: Magnetization with Spin-Orbit Coupling}

This section has notes toward calculating the coefficients of the diffusion operator for a simple spin $1/2$ model with both magnetization and a spin-orbit interaction.  This model has been used to analyze CoPt bilayers, and is also applicable to magnetized TIs by setting the $k^2/2m$ term in the Hamiltonian and the eigenvalues to zero.  The work reported in this section is analytical, and is incomplete because it remains to solve the SCBA self-consistent equation numerically and then evalulate all the spin diffusion coefficients numerically.

In the following we work with $\vec{b}$  pointing in any direction, but we also keep the equations for perpendicular magnetic field $\vec{b} = b_z \hat{z}$ because this is simpler.  We include a spin-orbit interaction $\alpha (k_y \sigma_x - k_x \sigma_y)$.  The density of states $\hat{\rho}$ and the self-energy $ \Sigma$ are not proportional to the identity.

The self-energy is:
\begin{align}
\Sigma^A (E) &= \vec{\Sigma}^A \cdot \sigma\\
\Sigma^R(E)  &=  \vec{\Sigma}^R \cdot \sigma
\end{align}
We are violating convention here since we have not written the $\imath $ in $\Sigma$ explicitly.  In general $\Sigma$ is expected to have an important imaginary part, and the real part is usually neglected.  
Also note that if $b_x = b_y=0$, then also $\Sigma_x = \Sigma_y=0$.

The Hamiltonian minus self-energy is:
\begin{align}
H^{A,R} &= H_0(\bfk^2) + \Delta_{SO} (\bfk_y \sigma_x - \bfk_x \sigma_y) +  (\vec{b}-\vec{\Sigma}^{A,R}(E)) \cdot \vec{\sigma} 
\end{align}
In the case of a TI we set $H_0(\bfk^2) = 0$.  In the case of a 2-D electron gas we set $H_0(\bfk^2) = h_0 \bfk^2/2k_0^2$.

The eigenvalues are as follows. Note that if there is a self energy and it has an imaginary part, then these eigenvalues are, in principal, complex.  The first equation gives the general case for any $\vec{b}, \Sigma$, and the second equation gives the eigenvalues when $\vec{b}, \Sigma$ are aligned with the $z$ axis.
\begin{align}
 \lambda^{A,R}(s,\vec{k},E) &=  H_0(k^2)-\Sigma_1^{A,R}  + s\sqrt{(\Delta_{SO} k_y + b_x - \Sigma_x^{A,R})^2 + (-\Delta_{SO} k_x + b_y -  \Sigma_y^{A,R})^2 +( b_z - \Sigma_z^{A,R})^2}\\
 \lambda^{A,R,z}(s,\vec{k},E) &= H_0(k^2) -\Sigma_1^{A,R} + s\sqrt{(\Delta_{SO} |k|)^2 + (b_z -\Sigma_z^{A,R}(E))^2}\\
\end{align}

In the following expressions for the Green's functions, we use the following inversion formula: 
\begin{align}
(E + \vec{A} \cdot \vec{\sigma})^{-1} &= \frac{1}{2} \sum_s \frac{1 + s \vec{A} \cdot \vec{\sigma} /|\vec{A}|}{E + \lambda(\vec{A} \cdot \vec{\sigma},s)}, \lambda = s |\vec{A}|\\
1 &?=?  \frac{1}{2} \sum_s \frac{1 + s \vec{A} \cdot \vec{\sigma} /|\vec{A}|}{E + s|\vec{A}|} (E + \vec{A} \cdot \vec{\sigma}) \\
& = \frac{1}{2} \sum_s \frac{E + E s\vec{A} \cdot \vec{\sigma} /|\vec{A}| + \vec{A} \cdot \vec{\sigma}+ s|\vec{A}| }{E + s|\vec{A}|} \\
& = \frac{1}{2} \sum_s (1 + s \vec{A} \cdot \vec{\sigma}/|\vec{A}| ) \\
&= 1
\end{align}

The Green's functions are:
\begin{align}
  G^{A,R}(\bfk, E) &= \frac{1}{E + H(\bfk) - \Sigma^{A,R}(\bfk,E)\mp \imath \epsilon}\\
 G^{A,R}_{\alpha \beta}(\bfk, E) &=  \frac{1}{2}\sum_s \frac{\delta_{\alpha \beta} + s (S^{A,R}(\bfk^2,E))^{-1} (\Delta_{SO} \bfk_y   \sigma_x - \Delta_{SO} \bfk_x   \sigma_y + b_z \sigma_z- \Sigma_z^{A,R}(E) \sigma_z)_{\alpha \beta}} {E + H_0(\bfk^2) - \Sigma_1^{A,R} (E) + s S^{A,R}(\bfk^2,E) \mp \imath \epsilon},  \\
  S^{A,R}(\bfk^2,E) &= \sqrt{\Delta_{SO}^2 \bfk^2 + (b_z - \Sigma_z^{A,R}(E))^2} \\
 G^{A,R}_{\alpha \beta}(\bfk \pm \bfq/2, E \pm \omega/2) &\approx  \frac{1}{2}\sum_s \frac{\delta_{\alpha \beta} + s (S^{A,R}(\bfk^2,E \pm \omega/2))^{-1} (\Delta_{SO} \bfk_y   \sigma_x - \Delta_{SO} \bfk_x   \sigma_y + b_z \sigma_z- \Sigma_z^{A,R}(E \pm \omega/2) \sigma_z)_{\alpha \beta}} {E  \pm \omega/2 + H_0(\bfk^2) - \Sigma_1^{A,R} (E \pm \omega/2) + s S^{A,R}(\bfk^2,E \pm \omega/2) \mp \imath \epsilon  \pm \hbar \vec{V}(E \pm \omega/2) \cdot \bfq/2} \\
\vec{V}^{A,R}(\bfk,E \pm \omega/2,s) & = \frac{d}{\hbar d\bfk} (H_0(\bfk^2)  + s S^{A,R}(\bfk^2,E \pm \omega/2)) \\
V^{A,R}(\bfk,E\pm \omega/2,s) & = \frac{d}{\hbar d|\bfk|} (H_0(\bfk^2)  + s S^{A,R}(\bfk^2,E\pm \omega/2))\\
V &= (V^R(\bfk,E,\acute{s}) + V^A(\bfk,E,s))/2 \\
\frac{d S^{A,R}(\bfk^2,E\pm \omega/2)}{\hbar d|\bfk|}  & = \frac{\Delta_{SO}^2 |\bfk|}{\hbar \; S^{A,R}(\bfk^2,E\pm \omega/2)} 
\end{align}

The approximation in the last line is justified because if we write $\bfk + \bfq/2$, or $E + \omega/2$, these have order of magnitude $q/k_F=l_F / L$, or $E/E_F = t_F / T$, which are small in the diffusive regime.  It's only when these are divided by $\Sigma / E_F = \hbar / (E_F \tau) = 1/(l k_F) = l_F / l \ll 1$, which are small if disorder is "weak", that we obtain quantities that are significant at diffusive length and time scales.  The only way to get this extra multiplicative factor is from the denominator of the Green's functions.  Moreover only terms in the denominator that are linear in $\bfq, \omega$  need to be considered, again because you can get only one compensating factor of $\Sigma/ E_F$ from the denominator. In summary, the numerator's dependence on $\bfq, \omega$ should be omitted, and the denominator should be expanded only to linear order in these quantities.

We compute the self-energy with the SCBA:
\begin{align}
\Sigma^{A,R}_{\alpha \beta}(E) &= \Gamma_{\alpha \nu \xi \beta} \int {d\bfk} \;G^{A,R}_{\nu \xi}(E,\bfk)= \gamma \int {d\bfk} \;G^{A,R}_{ \alpha \beta}(E,\bfk) \\
\Sigma^{A,R}(E) &=\gamma \int {d\bfk} \;  \frac{1}{E + H(\bfk) - \Sigma^{A,R}(E)\mp \imath \epsilon}
\end{align}
This equation must be solved iteratively until self consistency is obtained; otherwise the Ward-Takahashi identity will be broken.

If $\vec{b} = b_z \hat{z}$ then it is trivial to prove that $\Sigma_x = \Sigma_y = 0$, and the remaining integrals have rotational invariance resulting in one dimensional integrals:
\begin{align}
{\Sigma}^{A,R}(E) &=  \gamma  \;\frac{1}{2}\sum_s \int {d\bfk} \frac{ 1 +  s \sin \theta^{A,R} (E) \sigma_z}  {E +H_0(\bfk^2) -\Sigma_1^{A,R}(E) + s\sqrt{\Delta^2_{SO} \bfk^2 + (b_z - \Sigma_z^{A,R}(E))^2} \mp \imath \epsilon},  \\
\sin \theta^{A,R}(E) &= \frac{b_z - \Sigma_z^{A,R}(E)}{\sqrt{\Delta^2_{SO} \bfk^2 + (b_z - \Sigma_z^{A,R}(E))^2}}  \\
\Sigma_z &= \frac{1}{2}{Tr}(\Sigma \sigma_z), \; \Sigma_1 = \frac{1}{2}{Tr}(\Sigma )
\end{align}

We compute the main integral, assuming that the disorder has no spin structure:
\begin{align}
\overline{\Phi}_{\alpha \nu \eta \beta} & ( \bfq, \omega, E,\omega/2) = \frac{1}{4} \sum_{s,\acute{s}} \int {d\bfk} M_{\alpha \nu \eta \beta} \Delta \\
M &=( \delta_{\alpha \nu} + s (S^{A}(\bfk^2,E + \omega/2))^{-1} (\Delta_{SO} \bfk_y   \sigma_x - \Delta_{SO} \bfk_x   \sigma_y + b_z \sigma_z- \Sigma_z^{A}(E + \omega/2) \sigma_z)_{\alpha \nu}) \\
& \times (\delta_{\eta \beta} + \acute{s} (S^{R}(\bfk^2,E - \omega/2))^{-1} (\Delta_{SO} \bfk_y   \sigma_x - \Delta_{SO} \bfk_x   \sigma_y + b_z \sigma_z- \Sigma_z^{R}(E - \omega/2) \sigma_z)_{\eta \beta}) \\
\Delta &= ( {E +\omega/2+ H_0(\bfk^2) - \Sigma_1^{A} (E+\omega/2) + s S^{A}(\bfk^2,E+\omega/2) - \imath \epsilon  + \hbar \vec{V}^A(E+\omega/2) \cdot \bfq/2} )^{-1} \\
& \times ( {E -\omega/2+ H_0(\bfk^2) - \Sigma_1^{R} (E-\omega/2) + \acute{s} S^{R}(\bfk^2,E-\omega/2) + \imath \epsilon - \hbar \vec{V}^R(E-\omega/2) \cdot \bfq/2})^{-1} \\
& = ((-\omega/2- \Sigma_1^{R} (E) + \acute{s} S^{R}(\bfk^2,E-\omega/2) + \imath \epsilon - \hbar \vec{V}^R(E-\omega/2) \cdot \bfq/2) \\
& - (\omega/2- \Sigma_1^{A} (E+\omega/2) + s S^{A}(\bfk^2,E+\omega/2) - \imath \epsilon + \hbar \vec{V}^A(E+\omega/2) \cdot \bfq/2))^{-1}  \\
& \times (( {E +\omega/2 + H_0(\bfk^2) - \Sigma_1^{A} (E+\omega/2) + s S^{A}(\bfk^2,E+\omega/2) - \imath \epsilon  + \hbar \vec{V}^A(E+\omega/2) \cdot \bfq/2} )^{-1} \\
& - ({E -\omega/2+ H_0(\bfk^2) - \Sigma_1^{R} (E-\omega/2) + \acute{s} S^{R}(\bfk^2,E-\omega/2) + \imath \epsilon - \hbar \vec{V}^R(E-\omega/2) \cdot \bfq/2})^{-1} ) \\
\Delta & \approx \rho \;\; (\delta   - \hbar (\vec{V}^R(E-\omega/2) + \vec{V}^A(E+\omega/2)) \cdot \bfq/2)^{-1}   = (\delta   -V (\hbar \bfq \cdot \bfk))^{-1}\\
\Delta &\approx \rho \;\; (\frac{1}{\delta} + \hbar {\bfq} \cdot \hat{\bfk} \frac{V}{\delta^2} + (\hbar {\bfq} \cdot \hat{\bfk})^2 \frac{V^2}{\delta^3}) \\
\delta &= -\omega - \Sigma_1^{R} (E-\omega/2) + \Sigma_1^{A} (E+\omega/2) + \acute{s} S^{R}(\bfk^2,E-\omega/2) - s S^{A}(\bfk^2,E+\omega/2) \\
\rho &= (( {E +\omega/2+ H_0(\bfk^2) - \Sigma_1^{A} (E+\omega/2) + s S^{A}(\bfk^2,E+\omega/2) } )^{-1} - ({E -\omega/2+ H_0(\bfk^2) - \Sigma_1^{R} (E-\omega/2) + \acute{s} S^{R}(\bfk^2,E-\omega/2) })^{-1} ) 
\end{align}

At this point we can do all the angular integrals.  Writing $\overline{\Phi}$ now in terms of its spin and charge components, we obtain:  
\begin{align}
\overline{\Phi}_{00} &=  \mu_1 + \mu_{xy} + \mu_{zz} +  \bfq^2 (d_1 + d_{xy} + d_{zz}) \\
\overline{\Phi}_{11}  &= \mu_1 - \mu_{zz} +\bfq^2 (d_1 - d_{zz}) +  \frac{(\bfq_y^2 - \bfq_x^2)}{2} d_{xy} \\
\overline{\Phi}_{22}  &= \mu_1 - \mu_{zz} +\bfq^2 (d_1 - d_{zz}) - \frac{(\bfq_y^2 - \bfq_x^2)}{2} d_{xy}  \\
\overline{\Phi}_{33} &=  \mu_1 - \mu_{xy} + \mu_{zz} +  \bfq^2 (d_1 - d_{xy} + d_{zz})  \\
\overline{\Phi}_{12} &=  + \imath  \mu_{z-} - \bfq_x \bfq_y d_{xy}    + \imath  \bfq^2 d_{z-} \\
\overline{\Phi}_{21} &=  - \imath  \mu_{z-} - \bfq_x \bfq_y d_{xy}   - \imath  \bfq^2 d_{z-}  \\
\overline{\Phi}_{03} = \overline{\Phi}_{30}&=  \mu_{z+}   + \bfq^2 d_{z+}  \\
\overline{\Phi}_{13} & =\bfq_y  \gamma_{xyz+}  + \imath  \bfq_x  \gamma_{-}  \\
\overline{\Phi}_{31} & =\bfq_y  \gamma_{xyz+}  - \imath  \bfq_x  \gamma_{-} \\
\overline{\Phi}_{23}  &= - \bfq_x  \gamma_{xyz+} + \imath \bfq_y  \gamma_{-} \\
\overline{\Phi}_{32}  &= - \bfq_x  \gamma_{xyz+} - \imath \bfq_y  \gamma_{-}  \\
\overline{\Phi}_{01} &=  \bfq_y \gamma_+ + \imath  \bfq_x \gamma_{xyz-} \\
\overline{\Phi}_{10} &=  \bfq_y \gamma_+ - \imath \bfq_x \gamma_{xyz-} \\
\overline{\Phi}_{02} &= -  \bfq_x \gamma_+ + \imath \bfq_y \gamma_{xyz-} \\
\overline{\Phi}_{20} &= -  \bfq_x \gamma_+ - \imath \bfq_y \gamma_{xyz-} \\
\mu_1 &= \langle 1 \rangle , \; \mu_{xy} = \langle  s\acute{s}  X_{xy}^{A} \; X_{xy}^{R}   \rangle, \; \mu_{zz} =\langle s\acute{s} X_z^{A} \; X_z^{R}  \rangle, \; \mu_{z \pm} =\langle s X_z^{A} \pm \acute{s} X_z^{R}   \rangle \\
d_1 &= \langle \frac{\hbar^2 V^2}{2 \delta^2} \rangle , \; d_{xy} = \langle  s\acute{s} \frac{\hbar^2V^2}{2 \delta^2}  X_{xy}^{A} \; X_{xy}^{R}   \rangle, \; d_{zz} =\langle s\acute{s} \frac{\hbar^2V^2}{2 \delta^2} X_z^{A} \; X_z^{R}  \rangle, \; d_{z \pm} =\langle  \frac{\hbar^2V^2}{2 \delta^2} (s X_z^{A} \pm \acute{s} X_z^{R} )  \rangle  \\
\gamma_{\pm} &= \langle \frac{\hbar V}{2 \delta} ( s {X_{xy}^{A}} \pm \acute{s} {X_{xy}^{R}}) \rangle, \; \gamma_{xyz \pm} = \langle \frac{\hbar V}{2 \delta} s  \acute{s} (X_{z}^A X_{xy}^R \pm X_{z}^R X_{xy}^A)  \rangle \\
\langle  .... \rangle &=  \frac{1}{4} \sum_{s,\acute{s}} \int {d\bfk} \;\; \frac{\rho}{\delta}  \\
X_{xy}^{A,R} &= S^{A,R}(\bfk^2,E \pm \omega/2))^{-1} \Delta_{SO} | \bfk| \\
X_{z}^{A,R}  &= S^{A,R}(\bfk^2,E \pm \omega/2))^{-1} (b_z - \Sigma_z^{A,R}(E \pm \omega/2) )
\end{align}



In the next equations we redo the calculations with disorder that is purely magnetic, of the type where $M$ is proportional to $G^A \sigma_x \otimes G^R \sigma_x + G^A \sigma_y \otimes G^R \sigma_y$:
\begin{align}
\overline{\Phi}_{00} &=  \mu_1  - \mu_{zz} +  \bfq^2 (d_1 - d_{zz}) \\
\overline{\Phi}_{11}  &= \mu_{xy} +\bfq^2 d_{xy}  \\
\overline{\Phi}_{22}  &= \mu_{xy} +\bfq^2 d_{xy} \\
\overline{\Phi}_{33} &=  -\mu_1  + \mu_{zz} +  \bfq^2 (-d_1 + d_{zz})\\
\overline{\Phi}_{12} &=  0 \\
\overline{\Phi}_{21} &=  0 \\
\overline{\Phi}_{03} &=  \mu_{z-}   + \bfq^2 d_{z-}  \\
\overline{\Phi}_{30} &=  -\mu_{z-}   - \bfq^2 d_{z-}  \\
\overline{\Phi}_{13} & =\bfq_y   \langle  s \acute{s} X^A_z X^R_{xy} \frac{\hbar V}{2 \delta}\rangle  + \imath  \bfq_x  \langle  \acute{s} X^R_{xy} \frac{\hbar V}{2 \delta}\rangle =\bfq_y  (\gamma_{xyz+} + \gamma_{xyz-})/2 + \imath  \bfq_x  (\gamma_+ - \gamma_-)/2\\
\overline{\Phi}_{31} & =\bfq_y   \langle  s \acute{s} X^R_z X^A_{xy} \frac{\hbar V}{2 \delta}\rangle  + \imath  \bfq_x  \langle  s X^A_{xy} \frac{\hbar V}{2 \delta}\rangle  =\bfq_y   (\gamma_{xyz+} - \gamma_{xyz-})/2  + \imath  \bfq_x  (\gamma_+ + \gamma_-)/2 \\
\overline{\Phi}_{23} & = -\bfq_x   \langle  s \acute{s} X^A_z X^R_{xy} \frac{\hbar V}{2 \delta}\rangle  + \imath  \bfq_y  \langle  \acute{s} X^R_{xy} \frac{\hbar V}{2 \delta}\rangle  = -\bfq_x  (\gamma_{xyz+} + \gamma_{xyz-})/2  + \imath  \bfq_y  (\gamma_+ - \gamma_-)/2  \\
\overline{\Phi}_{32} & =-\bfq_x   \langle  s \acute{s} X^R_z X^A_{xy} \frac{\hbar V}{2 \delta}\rangle  + \imath  \bfq_y  \langle  s X^A_{xy} \frac{\hbar V}{2 \delta}\rangle =-\bfq_x   (\gamma_{xyz+} - \gamma_{xyz-})/2  + \imath  \bfq_y  (\gamma_+ + \gamma_-)/2  \\
\overline{\Phi}_{01} & = \imath \bfq_x   \langle  s \acute{s} X^R_z X^A_{xy} \frac{\hbar V}{2 \delta}\rangle  +  \bfq_y  \langle s X^A_{xy} \frac{\hbar V}{2 \delta}\rangle   = \imath \bfq_x  (\gamma_{xyz+} - \gamma_{xyz-})/2 +  \bfq_y  (\gamma_+ + \gamma_-)/2 \\
\overline{\Phi}_{10} & = \imath \bfq_x   \langle  s \acute{s} X^A_z X^R_{xy} \frac{\hbar V}{2 \delta}\rangle  +  \bfq_y  \langle \acute{s} X^R_{xy} \frac{\hbar V}{2 \delta}\rangle  = \imath \bfq_x   (\gamma_{xyz+} + \gamma_{xyz-})/2  +  \bfq_y  (\gamma_+ - \gamma_-)/2 \\
\overline{\Phi}_{02} & = \imath \bfq_y   \langle  s \acute{s} X^R_z X^A_{xy} \frac{\hbar V}{2 \delta}\rangle  -  \bfq_x  \langle s X^A_{xy} \frac{\hbar V}{2 \delta}\rangle  = \imath \bfq_y   (\gamma_{xyz+} - \gamma_{xyz-})/2  -  \bfq_x (\gamma_+ + \gamma_-)/2 \\
\overline{\Phi}_{20} & = \imath \bfq_y  \langle  s \acute{s} X^A_z X^R_{xy} \frac{\hbar V}{2 \delta}\rangle  -  \bfq_x  \langle \acute{s} X^R_{xy} \frac{\hbar V}{2 \delta}\rangle  = \imath \bfq_y  (\gamma_{xyz+} + \gamma_{xyz-})/2  -  \bfq_x  (\gamma_+ - \gamma_-)/2\\
\end{align}

It is possible to mix the disorder types, in which case $M$ will be a linear interpolation between the results with magnetic disorder and the results with spin-independent disorder.

There are some exotic terms in $\overline{\Phi}$ caused by  $\Sigma^A_z \neq \Sigma^R_z$:
\begin{itemize}
\item  The charge component $\overline{\Phi}_{00}$ and the charge-$S_z$ coupling $\overline{\Phi}_{03}=\overline{\Phi}_{30}$ acquire a part which is not proportional to $1 + s \acute{s}$, allowing $s= -\acute{s}$ physics to contribute.
\item The energy beating factor $\delta$, its derivative $W$, and the Fermi velocity $V$   acquire a part which is  proportional to $s+\acute{s}$, allowing $s= \acute{s}$ physics to couple to the spin-orbit interaction.
\item The charge-$S_x,S_y$ couplings become nonsymmetric, i.e. $\overline{\Phi}_{01}-\overline{\Phi}_{10}$, $\overline{\Phi}_{02}-\overline{\Phi}_{20}$ become nonzero.
\item The symmetric part of the charge-$S_x,S_y$ couplings, i.e. $\overline{\Phi}_{01}+\overline{\Phi}_{10}$ and $\overline{\Phi}_{02}+\overline{\Phi}_{20}$,  acquire a part which is not proportional to $s+\acute{s}$, allowing $s= -\acute{s}$ physics to contribute.
\item The spin-mixing terms $\overline{\Phi}_{12}-\overline{\Phi}_{21}$, $\overline{\Phi}_{23}-\overline{\Phi}_{32}$, $\overline{\Phi}_{13}-\overline{\Phi}_{31}$ acquire a part which is not proportional to $s-\acute{s}$, allowing $s= \acute{s}$ physics to contribute.
\end{itemize}

\section{Notes on the Bethe-Salpeter Equation and the Quantum Kinetic Equation} 
The Bethe-Salpeter equation is exact and offers a Ward-Takahashi identity, so we explored it extensively to see how it could help us with magnetized systems.  One of the most attractive things about this equation is that it offers a nice base for developing and understanding the Keldysh based approach of Mischenko Shytov and Halperin, which is closely tied to a Quantum Kinetic Equation.  Therefore  after exploring the Bethe-Salpeter equation  and its Ward-Takahashi identity we spent a lot of time trying to understand the Keldysh based approach and how it could be made fully general including for magnetized systems. Pletyukhov's PRB 75 155335 was particularly helpful here \cite{PhysRevB.75.155335}.  We were never able to bring this direction to a conclusion, and we still do not understand how to correctly generalize the Keldysh approach to magnetized systems.  

Eventually we decided that the Ward-Takahashi Identity offered by the Bethe-Salpeter equation is not powerful enough, after we realized that the Ward-Takahashi identity explained by Ramazashvili\cite{PhysRevB.66.220503} is more powerful.  We did not figure out how to conserve probability in a magnetized system until we had abandoned the Bethe-Salpeter approach in favor of Ramazashvili's approach.  

Probably the reason why the Bethe-Salpeter equation approach did not give us what we need is that it naturally results in a description of how the density matrix evolves; a quantum kinetic equation whose inputs and outputs are both densities.  This turns out to be a bit too ambitious.  What we have succeeded in doing is producing a formalism which, given a perturbing potential as an input, produces a response density.  We leave the problem of calculating the perturbing potential to another day.  This approach allows us to build on the foundation of linear response theory, which is rock solid and gives us what we need to obtain spin conservation.

It is clear that there is a unique way within linear response theory (and still unique after application of the SCBA) to obtain charge conservation in magnetized systems. Therefore a correct generalization of the Keldysh approach must necessarily give identical results to the approach we present in this paper.

We retain these notes because the Bethe-Salpeter equation and the Keldysh approach retain great interest in their own right.
\subsection{Density matrix and probability conservation}

Let us take an $N$-band system.
With an arbitrary basis $|\mu\bfk\rangle \ (\mu=1,\ldots,N)$, the (unperturbed) Hamiltonian can be written in the $N\times N$ matrix form, $\hat{H}(\bfk) = H_{\mu\nu}(\bfk)$.
Incorporating the disorder term $\hat{V}$,
the time evolution of the density matrix $\hat{\rho}(t)$ is given by
\begin{align}
\hat{\rho}(t) = \int \frac{dE}{2\pi} \int\frac{d\epsilon}{2\pi} e^{-i\epsilon t} \hat{G}^R(E+\tfrac{\epsilon}{2}) \hat{\rho}_0 \hat{G}^A(E-\tfrac{\epsilon}{2}),
\end{align}
where $\hat{\rho}_0$ is the density matrix at the initial time $(t=0)$.
The matrix element becomes
\begin{align}
& \rho_{\mu\nu}(\bfk,\bfq;t) \equiv \langle \mu,\bfk+\tfrac{\bfq}{2}|\hat{\rho}(t)|\nu,\bfk-\tfrac{\bfq}{2}\rangle \\
 &= \int \frac{dE}{2\pi} \int\frac{d\epsilon}{2\pi} e^{-i\epsilon t} \int d\bfk' \ G^R_{\mu\alpha}(\bfk+\tfrac{\bfq}{2},\bfk'+\tfrac{\bfq}{2},E+\tfrac{\epsilon}{2}) G^A_{\alpha'\nu}(\bfk'-\tfrac{\bfq}{2},\bfk-\tfrac{\bfq}{2},E-\tfrac{\epsilon}{2}) \rho^0_{\alpha\alpha'}(\bfk',\bfq)
\end{align}
Averaring over the disorder, we obtain
\begin{align}
\overline{\rho}_{\mu\nu}(\bfk,\bfq;t) &= \int \frac{dE}{2\pi} \int\frac{d\epsilon}{2\pi} e^{-i\epsilon t} \int d\bfk' \ \overline{G^R_{\mu\alpha}(\bfk+\tfrac{\bfq}{2},\bfk'+\tfrac{\bfq}{2},E+\tfrac{\epsilon}{2}) G^A_{\alpha'\nu}(\bfk'-\tfrac{\bfq}{2},\bfk-\tfrac{\bfq}{2},E-\tfrac{\epsilon}{2})} \rho^0_{\alpha\alpha'}(\bfk',\bfq) \\
 &\equiv \int \frac{dE}{2\pi} \int\frac{d\epsilon}{2\pi} e^{-i\epsilon t} \int d\bfk' \Phi^{\mu\alpha}_{\nu\alpha'}(\bfk,\bfk',\bfq;E,\epsilon) \rho^0_{\alpha\alpha'}(\bfk',\bfq). \label{eq:dis-density}
\end{align}
We need to calculate the tensor $\check{\Phi}(\bfk,\bfk',\bfq;E,\epsilon)$.

The total probability at time $t$ is given by
\begin{align}
P(t) = \sum_{\mu}\int d\bfk \langle \mu\bfk | \hat{\rho}(t) | \mu\bfk\rangle = \sum_{\mu}\int d\bfk \ \rho_{\mu\mu}(\bfk,0;t)
\end{align}
Using Eq.(\ref{eq:dis-density}), we obtain
\begin{align}
P(t) = \int\frac{d\epsilon}{2\pi} e^{-i\epsilon t} P(\epsilon),
\end{align}
where
\begin{align}
 P(\epsilon) = \int \frac{dE}{2\pi}  \int d\bfk d\bfk' \sum_\mu \Phi^{\mu\alpha}_{\mu\alpha'}(\bfk,\bfk',0;E,\epsilon) \rho^0_{\alpha\alpha'}(\bfk',0). \label{eq:Pe}
\end{align}
Thus, the probability conservation law is expressed by
\begin{align}
\frac{d}{dt}P(t) = -i \int\frac{d\epsilon}{2\pi} e^{-i\epsilon t} \epsilon P(\epsilon) \overset{!}{=}0. \label{eq:probability-conservation}
\end{align}
We need to show the last equality, i.e. $\epsilon P(\epsilon)$ is independent of $\epsilon$.

\subsection{Bethe-Salpeter equation and kinetic equation}
The disorder-averaged 4-point propagator $\Phi^{\mu\nu}_{\mu'\nu'}$ satisfies the Bethe-Salpeter equation,
\begin{align}
\Phi^{\mu\nu}_{\mu'\nu'}(\bfk,\bfk',E,\epsilon) = {\Phi_0}^{\mu\alpha}_{\mu'\alpha'}(\bfk,E,\epsilon) \left[ \delta^{\alpha\nu}\delta_{\alpha'\nu'}\delta(\bfk-\bfk') + \int d\bfk'' U^{\alpha\beta}_{\alpha'\beta'}(\bfk,\bfk'',E,\epsilon) \Phi^{\beta\nu}_{\beta'\nu'}(\bfk'',\bfk',E,\epsilon)\right],
\end{align}
where $U^{\alpha\beta}_{\alpha'\beta'}(\bfk,\bfk'',E,\epsilon)$ denotes the ``irreducible'' vertex function,
and the bare propagator $\Phi_0$ is defined by
\begin{align}
{\Phi_0}^{\mu\nu}_{\mu'\nu'}(\bfk,E,\epsilon) = G^R_{\mu\nu}(\bfk,E+\tfrac{\epsilon}{2}) G^A_{\nu'\mu'}(\bfk,E-\tfrac{\epsilon}{2}).
\end{align}
In the matrix form, this equation can be simplified as
\begin{align}
& \check{\Phi}(\bfk,\bfk',E,\epsilon) = \check{\Phi}_0(\bfk,E,\epsilon) \left[ \delta(\bfk-\bfk') + \int d\bfk'' \check{U}(\bfk,\bfk'',E,\epsilon) \check{\Phi}(\bfk'',\bfk',E,\epsilon)\right], \label{eq:Bethe-Salpeter} \\
& \check{\Phi}_0(\bfk,E,\epsilon) = \hat{G}^R(\bfk,E+\tfrac{\epsilon}{2}) \otimes \hat{G}^{AT}(\bfk,E-\tfrac{\epsilon}{2}).
\end{align}
We should be careful that the advanced Green's function is transposed.

Let us now fractionalize the bare propagator $\check{\Phi_0}$ as in the single-band case.
In the matrix form, it becomes
\begin{align}
\check{\Phi}_0(\bfk,E,\epsilon) = (\hat{G}^R \otimes \hat{1}) (\hat{1} \otimes \hat{G}^{AT}) = -\check{\mathcal{E}}^{-1}(\bfk,E,\epsilon) \check{\mathcal{G}}(\bfk,E,\epsilon),
\end{align}
where
\begin{align}
\check{\mathcal{E}}(\bfk,E,\epsilon) &= \left[(\hat{G}^R)^{-1} \otimes \hat{1} \right] - \left[\hat{1} \otimes (\hat{G}^{AT})^{-1} \right] \\
 &= \left[E+\frac{\epsilon}{2}-\hat{H}(\bfk)-\hat{\Sigma}^R(\bfk,E+\tfrac{\epsilon}{2}) \right] \otimes \hat{1} - \hat{1} \otimes  \left[E-\frac{\epsilon}{2}-\hat{H}(\bfk)-\hat{\Sigma}^A(\bfk,E-\tfrac{\epsilon}{2}) \right] \\
 &\equiv \epsilon - \delta \check{H}(\bfk) - \delta\check{\Sigma}(\bfk,E,\epsilon) \\
 & \quad \left( \delta\check{H}(\bfk) = [\hat{H}(\bfk)\otimes\hat{1}] - [\hat{1}\otimes\hat{H}(\bfk)], \quad  \delta\check{\Sigma}(\bfk,E,\epsilon) = [\hat{\Sigma}^R(\bfk,E+\tfrac{\epsilon}{2})\otimes\hat{1}] - [\hat{1}\otimes\hat{\Sigma}^A(\bfk,E-\tfrac{\epsilon}{2})]\right) \\
\check{\mathcal{G}}(\bfk,E,\epsilon) &= \left[\hat{G}^R(\bfk,E+\tfrac{\epsilon}{2}) \otimes \hat{1} \right] - \left[\hat{1} \otimes \hat{G}^A(\bfk,E-\tfrac{\epsilon}{2}) \right]
\end{align}

Multiplying $\check{\mathcal{E}}(\bfk,E,\epsilon)$ to Eq.(\ref{eq:Bethe-Salpeter}) from the left,
we obtain the kinetic equation
\begin{align}
\left[\epsilon - \delta \check{H}(\bfk) - \delta\check{\Sigma}(\bfk,E,\epsilon)\right]\check{\Phi}(\bfk,\bfk',E,\epsilon) = -\check{\mathcal{G}}(\bfk,E,\epsilon) \left[ \delta(\bfk-\bfk') + \int d\bfk'' \check{U}(\bfk,\bfk'',E,\epsilon) \check{\Phi}(\bfk'',\bfk',E,\epsilon)\right], \label{eq:kinetic-equation}
\end{align}
leading to
\begin{align}
\epsilon \check{\Phi}(\bfk,\bfk',E,\epsilon) = \left[ \delta \check{H}(\bfk) + \delta\check{\Sigma}(\bfk,E,\epsilon)\right]\check{\Phi}(\bfk,\bfk',E,\epsilon) -\check{\mathcal{G}}(\bfk,E,\epsilon) \left[ \delta(\bfk-\bfk') + \int d\bfk'' \check{U}(\bfk,\bfk'',E,\epsilon) \check{\Phi}(\bfk'',\bfk',E,\epsilon)\right].
\end{align}
Here, the left hand side is the form appearing in Eq.(\ref{eq:probability-conservation}).

\subsection{Ward-Takahashi identity and probability conservation}
Let us now substitute the right hand side of Eq.(\ref{eq:kinetic-equation}) to Eq.(\ref{eq:Pe}), and manipulate their contributions to $\epsilon P(\epsilon)$ one by one.

(A) \underline{$\delta\check{H} \check{\Phi}$ term:}
\begin{align}
\epsilon P^\mathrm{(A)}(\epsilon) &= \int \frac{dE}{2\pi}  \int d\bfk d\bfk' \sum_\mu \delta H^{\mu\alpha}_{\mu\alpha'}(\bfk) \Phi^{\alpha\beta}_{\alpha'\beta'}(\bfk,\bfk',E,\epsilon) \rho^0_{\beta\beta'}(\bfk') \\
 &= \int \frac{dE}{2\pi}  \int d\bfk d\bfk' \sum_\mu \left[H_{\mu\alpha}(\bfk)\delta_{\mu\alpha'} - \delta_{\mu\alpha}H_{\alpha'\mu}(\bfk) \right]\Phi^{\alpha\beta}_{\alpha'\beta'}(\bfk,\bfk',E,\epsilon) \rho^0_{\beta\beta'}(\bfk') \\
 &= \int \frac{dE}{2\pi}  \int d\bfk d\bfk'\left[ H_{\alpha'\alpha}(\bfk) - H_{\alpha'\alpha}(\bfk) \right] \Phi^{\alpha\beta}_{\alpha'\beta'}(\bfk,\bfk',E,\epsilon) \rho^0_{\beta\beta'}(\bfk') \\
 &= 0.
\end{align}

(B) \underline{$\delta\check{\Sigma} \check{\Phi}$ term:}
\begin{align}
\epsilon P^\mathrm{(B)}(\epsilon) &= \int \frac{dE}{2\pi}  \int d\bfk d\bfk' \sum_\mu \delta \Sigma^{\mu\alpha}_{\mu\alpha'}(\bfk) \Phi^{\alpha\beta}_{\alpha'\beta'}(\bfk,\bfk',E,\epsilon) \rho^0_{\beta\beta'}(\bfk') \\
 &= \int \frac{dE}{2\pi}  \int d\bfk d\bfk' \sum_\mu \left[\Sigma^R_{\mu\alpha}(\bfk,E+\tfrac{\epsilon}{2})\delta_{\mu\alpha'} - \delta_{\mu\alpha}\Sigma^A_{\alpha'\mu}(\bfk,E-\tfrac{\epsilon}{2}) \right]\Phi^{\alpha\beta}_{\alpha'\beta'}(\bfk,\bfk',E,\epsilon) \rho^0_{\beta\beta'}(\bfk') \\
 &= \int \frac{dE}{2\pi}  \int d\bfk d\bfk'\left[ \Sigma^R_{\alpha'\alpha}(\bfk,E+\tfrac{\epsilon}{2}) - \Sigma^A_{\alpha'\alpha}(\bfk,E-\tfrac{\epsilon}{2}) \right] \Phi^{\alpha\beta}_{\alpha'\beta'}(\bfk,\bfk',E,\epsilon) \rho^0_{\beta\beta'}(\bfk') \\
 &= \int \frac{dE}{2\pi}  \int d\bfk d\bfk' \ \Delta\Sigma_{\alpha'\alpha}(\bfk,E,\epsilon) \Phi^{\alpha\beta}_{\alpha'\beta'}(\bfk,\bfk',E,\epsilon) \rho^0_{\beta\beta'}(\bfk') \\
 &= \int \frac{dE}{2\pi}  \int d\bfk d\bfk' d\bfk'' \ U^{\alpha'\gamma'}_{\alpha\gamma}(\bfk,\bfk'',E,\epsilon) \Delta G_{\gamma'\gamma}(\bfk'',E,\epsilon) \Phi^{\alpha\beta}_{\alpha'\beta'}(\bfk,\bfk',E,\epsilon) \rho^0_{\beta\beta'}(\bfk')
\end{align}
In the last line we have used the Ward-Takahashi identity
\begin{align}
\Delta\Sigma_{\mu\mu'}(\bfk,E,\epsilon) = \int d\bfk'' U^{\mu\nu}_{\mu'\nu'}(\bfk,\bfk'',E,\epsilon) \Delta G_{\nu\nu'}(\bfk'',E,\epsilon).
\end{align}

(C) \underline{$\check{\mathcal{G}}$ term:}
\begin{align}
\epsilon P^\mathrm{(C)}(\epsilon) &= - \int \frac{dE}{2\pi}  \int d\bfk d\bfk' \sum_\mu \mathcal{G}^{\mu\alpha}_{\mu\alpha'}(\bfk,E,\epsilon)\delta(\bfk-\bfk')\rho^0_{\alpha\alpha'}(\bfk') \\
&= \int \frac{dE}{2\pi}  \int d\bfk' \sum_\mu \left[G^R_{\mu\alpha}(\bfk',E+\tfrac{\epsilon}{2})\delta_{\mu\alpha'} - \delta_{\mu\alpha} G^A_{\alpha'\mu}(\bfk',E-\tfrac{\epsilon}{2})\right]\rho^0_{\alpha\alpha'}(\bfk') \\
 &= \int \frac{dE}{2\pi}  \int d\bfk' \Delta G_{\alpha'\alpha}(\bfk',E) \rho^0_{\alpha\alpha'}(\bfk')
\end{align}
In the last line we have shifted the variable $E$ by $\pm \epsilon/2$ respectively.
Using
\begin{align}
\int dE \ \Delta G_{\mu\nu}(\bfk,E) = -2\pi i \delta_{\mu\nu},
\end{align}
we obtain
\begin{align}
\epsilon P^\mathrm{(C)}(\epsilon) = -i \int d\bfk' \delta_{\alpha'\alpha}  \rho^0_{\alpha\alpha'}(\bfk') = -i P_0,
\end{align}
which is independent of $\epsilon$.

(D) \underline{$\check{\mathcal{G}}\check{U}\check{\Phi}$ term:}
\begin{align}
\epsilon P^\mathrm{(D)}(\epsilon) &= - \int \frac{dE}{2\pi}  \int d\bfk d\bfk' d\bfk'' \sum_\mu \mathcal{G}^{\mu\alpha}_{\mu\alpha'}(\bfk,E,\epsilon)U^{\alpha\beta}_{\alpha'\beta'}(\bfk,\bfk'',E,\epsilon)\Phi^{\beta\gamma}_{\beta'\gamma'}(\bfk'',\bfk',E,\epsilon)\rho^0_{\gamma\gamma'}(\bfk') \\
 &= - \int \frac{dE}{2\pi}  \int d\bfk d\bfk' d\bfk'' \Delta G_{\alpha'\alpha}(\bfk,E,\epsilon) U^{\alpha\beta}_{\alpha'\beta'}(\bfk,\bfk'',E,\epsilon)\Phi^{\beta\gamma}_{\beta'\gamma'}(\bfk'',\bfk',E,\epsilon)\rho^0_{\gamma\gamma'}(\bfk') \\
 &= - \int \frac{dE}{2\pi}  \int d\bfk d\bfk' d\bfk'' U^{\beta'\alpha'}_{\beta\alpha}(\bfk'',\bfk,E,\epsilon) \Delta G_{\alpha'\alpha}(\bfk,E,\epsilon) \Phi^{\beta\gamma}_{\beta'\gamma'}(\bfk'',\bfk',E,\epsilon)\rho^0_{\gamma\gamma'}(\bfk'),
\end{align}
where we have used the time-reversal symmetry of the vertex function,
\begin{align}
U^{\alpha\beta}_{\alpha'\beta'}(\bfk,\bfk'',E,\epsilon) = U^{\beta'\alpha'}_{\beta\alpha}(\bfk'',\bfk,E,\epsilon).
\end{align}
Now we can see that $\epsilon P^\mathrm{(D)}(\epsilon)$ cancels with $\epsilon P^\mathrm{(B)}(\epsilon)$.

Therefore, we obtain
\begin{align}
\epsilon P(\epsilon) = \epsilon P^\mathrm{(C)}(\epsilon) = -i P_0,
\end{align}
leading to
\begin{align}
\frac{d}{dt} P(t) = 0 \quad \text{for} \ t>0.
\end{align}


\begin{acknowledgments}
We are greatly indebted to Y. Araki and A. H. Macdonald for their support throughout this work.  We gratefully acknowledge  stimulating discussions with S. Kettemann, P. Wenk, X. Dai, B. A. Bernevig,  A. Petrovic, A. Ferreira,  H. W. Kim, A. Ho, M. Titov, P. Ostrovsky,   J. Wunderlich, T. Ziman, M. Miron, M. Chshiev, B. Dieny, S. Parkin, J. Katine, P. Mavropoulous, K. Flensberg,  S. Blugel, P. Kelly, M. Sob, and D. Edwards.  We thank POSTECH, the University of York, the Institut Laue-Langevin, and the Forschungzentrum Juelich for hospitality.    Conversations with C. Mueller and C. Miniatura were very influential early in the project.  We acknowledge  support from EPSRC grant EP/M011038/1. 
  \end{acknowledgments}

\bibliography{Vincent1}

\end{document}